\documentclass[11pt]{cernrep}
\usepackage{graphicx}
\usepackage{here}

\def\ii{\'\i}

\newcommand{\rhor}{\rho_{\rm R}}

\def\lsim{\mathrel{\mathop
  {\hbox{\lower0.5ex\hbox{$\sim$}\kern-0.8em\lower-0.7ex\hbox{$<$}}}}}
\def\gsim{\mathrel{\mathop
  {\hbox{\lower0.5ex\hbox{$\sim$}\kern-0.8em\lower-0.7ex\hbox{$>$}}}}}

\begin{document}

\newcommand{\half}{{1\over2}}

\title{COSMOLOGY AND ASTROPHYSICS}
\author{J. Garc\'\i a-Bellido}
\institute{Departamento de F\'\i sica Te\'orica, Universidad Aut\'onoma
de Madrid, Cantoblanco 28049 Madrid, Spain}
\maketitle
\begin{abstract}
In these lectures I review the present status of the so-called
Standard Cosmological Model, based on the hot Big Bang Theory and the
Inflationary Paradigm. I will make special emphasis on the recent
developments in observational cosmology, mainly the acceleration of the
universe, the precise measurements of the microwave background
anisotropies, and the formation of structure like galaxies and clusters
of galaxies from tiny primodial fluctuations generated during inflation.
\end{abstract}

\section{INTRODUCTION}

The last five years have seen the coming of age of Modern Cosmology, 
a mature branch of science based on the hot Big Bang theory and the
Inflationary Paradigm. In particular, we can now define
rather precisely a Standard Model of Cosmology, where the basic
parameters are determined within small uncertainties, of just a few
percent, thanks to a host of experiments and observations.  This
precision era of cosmology has become possible thanks to important
experimental developments in all fronts, from measurements of
supernovae at high redshifts to the microwave background anisotropies,
as well as to the distribution of matter in galaxies and clusters of
galaxies.

In these lecture notes I will first introduce the basic concepts and
equations associated with hot Big Bang cosmology, defining the main
cosmological parameters and their corresponding relationships.  Then I
will address in detail the three fundamental observations that have
shaped our present knowledge: the recent acceleration of the universe,
the distribution of matter on large scales and the anisotropies in the
microwave background. Together these observations allow the precise
determination of a handful of cosmological parameters, in the context
of the inflationary plus cold dark matter paradigm.

\section{BIG BANG COSMOLOGY}

Our present understanding of the universe is based upon the successful
hot Big Bang theory, which explains its evolution from the first
fraction of a second to our present age, around 13.6 billion years
later. This theory rests upon four robust pillars, a theoretical
framework based on general relativity, as put forward by Albert
Einstein~\cite{Einstein} and Alexander A. Friedmann~\cite{Friedmann} in
the 1920s, and three basic observational facts: First, the expansion of
the universe, discovered by Edwin P. Hubble~\cite{Hubble} in the 1930s,
as a recession of galaxies at a speed proportional to their distance
from us. Second, the relative abundance of light elements, explained by
George Gamow~\cite{Gamow} in the 1940s, mainly that of helium, deuterium
and lithium, which were cooked from the nuclear reactions that took
place at around a second to a few minutes after the Big Bang, when the
universe was a few times hotter than the core of the sun. Third, the
cosmic microwave background (CMB), the afterglow of the Big Bang,
discovered in 1965 by Arno A. Penzias and Robert W. Wilson~\cite{Wilson}
as a very isotropic blackbody radiation at a temperature of about 3
degrees Kelvin, emitted when the universe was cold enough to form
neutral atoms, and photons decoupled from matter, approximately 380,000
years after the Big Bang. Today, these observations are confirmed to
within a few percent accuracy, and have helped establish the hot Big
Bang as the preferred model of the universe.

Modern Cosmology begun as a quantitative science with the advent of
Einstein's general relativity and the realization that the
geometry of space-time, and thus the general attraction of matter, is
determined by the energy content of the universe~\cite{Weinberg}
\begin{equation}\label{EinsteinEquations}
G_{\mu\nu}\equiv R_{\mu\nu} - {1\over2}g_{\mu\nu}R +
\Lambda\,g_{\mu\nu} = 8\pi G\,T_{\mu\nu}
\,.
\end{equation}
These non-linear equations are simply too difficult to solve without
invoking some symmetries of the problem at hand: the universe itself.

We live on Earth, just 8 light-minutes away from our star, the Sun,
which is orbiting at 8.5 kpc from the center of our
galaxy,\footnote{One parallax second (1 pc), {\em parsec} for short,
corresponds to a distance of about 3.26 light-years or
$3.09\times10^{18}$ cm.} the Milky Way, an ordinary galaxy within the
Virgo cluster, of size a few Mpc, itself part of a supercluster of
size a few 100 Mpc, within the visible universe, approximately 10,000
Mpc in size. Although at small scales the universe looks very
inhomogeneous and anisotropic, the deepest galaxy catalogs like 2dF
GRS and SDSS suggest that the universe on large scales (beyond the
supercluster scales) is very homogeneous and isotropic. Moreover, the
cosmic microwave background, which contains information about the
early universe, indicates that the deviations from homogeneity and
isotropy were just a few parts per million at the time of photon
decoupling. Therefore, we can safely impose those symmetries to the
univerge at large and determine the corresponding evolution equations.
The most general metric satisfying homogeneity and isotropy is the
Friedmann-Robertson-Walker (FRW) metric, written here in terms of the
invariant geodesic distance $ds^2=g_{\mu\nu}dx^\mu dx^\nu$ in four
dimensions~\cite{Weinberg} $\mu=0,1,2,3$,\footnote{I am using $c=1$
everywhere, unless specified, and a metric signature $(-,+,+,+)$.}
\begin{equation}\label{FRWmetric}
ds^2 = - dt^2 + a^2(t)\left[{dr^2\over1-K\,r^2} +
r^2(d\theta^2 + \sin^2\theta\,d\phi^2)\right]\,,
\end{equation}
characterized by just two quantities, a {\em scale factor} $a(t)$,
which determines the physical size of the universe, and a constant $K$,
which characterizes the {\em spatial} curvature of the universe,
\begin{equation}\label{SpatialCurvature}
{}^{(3)}\!R = {6K\over a^2(t)} \hspace{2cm}
\left\{\begin{array}{ll}K=-1\hspace{1cm}&{\rm OPEN}\\
K=0&{\rm FLAT}\\K=+1&{\rm CLOSED}
\end{array}\right.
\end{equation}
Spatially open, flat and closed universes have different three-geometries.
Light geodesics on these universes behave differently, and thus could in
principle be distinguished observationally, as we shall discuss later.
Apart from the three-dimensional spatial curvature, we can also compute
a four-dimensional {\em space-time} curvature,
\begin{equation}\label{SpacetimeCurvature}
{}^{(4)}\!R = 6{\ddot a\over a} + 6\left({\dot a\over a}\right)^2 + 
6{K\over a^2}\,. 
\end{equation}
Depending on the dynamics (and thus on the matter/energy content) of the
universe, we will have different possible outcomes of its evolution.
The universe may expand for ever, recollapse in the future or approach
an asymptotic state in between.

\subsection{The matter and energy content of the universe}

The most general matter fluid consistent with the assumption of
homogeneity and isotropy is a perfect fluid, one in which an observer
{\em comoving with the fluid} would see the universe around it as
isotropic. The energy momentum tensor associated with such a fluid can
be written as~\cite{Weinberg}
\begin{equation}\label{PerfectFluid}
T^{\mu\nu} = p\,g^{\mu\nu} + (p+\rho)\,U^\mu U^\nu\,,
\end{equation}
where $p(t)$ and $\rho(t)$ are the pressure and energy density of the
fluid at a given time in the expansion, as measured by this comoving
observer, and $U^\mu$ is the comoving four-velocity, satisfying $U^\mu
U_\mu=-1$. For such a comoving observer, the matter content looks
isotropic (in its rest frame),
\begin{equation}\label{PerfectFluidRestframe}
T^\mu_{\ \ \nu} = {\rm diag}(-\rho(t),\,p(t),\,p(t),\,p(t))\,.
\end{equation}
The conservation of energy ($T^{\mu\nu}_{\hspace{3mm};\nu} = 0$), a
direct consequence of the general covariance of the theory
($G^{\mu\nu}_{\hspace{3mm};\nu} = 0$), can be written in terms of the
FRW metric and the perfect fluid tensor (\ref{PerfectFluid}) as
\begin{equation}\label{EnergyConservation}
\dot\rho + 3{\dot a\over a}(p+\rho) = 0\,.
\end{equation}

In order to find explicit solutions, one has to supplement the
conservation equation with an {\em equation of state} 
relating the pressure and the density of the fluid, $p=p(\rho)$.
The most relevant fluids in cosmology are barotropic, i.e. fluids whose
pressure is linearly proportional to the density, $p=w\,\rho$, and
therefore the speed of sound is constant in those fluids.

We will restrict ourselves in these lectures to three main types of
barotropic fluids:

\begin{itemize}

\item {\em Radiation}, with equation of state $p_R=\rho_R/3$, associated with
relativistic degrees of freedom (i.e. particles with temperatures much
greater than their mass). In this case, the energy density of radiation
decays as $\rho_R \sim a^{-4}$ with the expansion of the universe.

\item {\em Matter}, with equation of state $p_M\simeq0$, associated with
nonrelativistic degrees of freedom (i.e. particles with temperatures
much smaller than their mass). In this case, the energy density of 
matter decays as $\rho_M \sim a^{-3}$ with the expansion of the universe.

\item {\em Vacuum energy}, with equation of state $p_V=-\rho_V$,
associated with quantum vacuum fluctuations. In this case, the vacuum
energy density remains constant with the expansion of the universe.

\end{itemize}

This is all we need in order to solve the Einstein equations.
Let us now write the equations of motion of observers comoving with
such a fluid in an expanding universe. According to general
relativity, these equations can be deduced from the Einstein equations
(\ref{EinsteinEquations}), by substituting the FRW metric
(\ref{FRWmetric}) and the perfect fluid tensor
(\ref{PerfectFluid}). The $\mu=i,\ \nu=j$ component of the Einstein
equations, together with the $\mu=0,\ \nu=0$ component constitute 
the so-called Friedmann equations,
\begin{eqnarray}
\label{FriedmannEquation}
\left({\dot a\over a}\right)^2 &=& {8\pi G\over3}\,\rho +
{\Lambda\over3} - {K\over a^2}\,,\\
{\ddot a\over a} &=& -\,{4\pi G\over3}\,(\rho + 3p) +
{\Lambda\over3} \,.\label{Evolution}
\end{eqnarray}
These equations contain all the relevant dynamics, since the energy
conservation equation (\ref{EnergyConservation}) can be obtained from these.

\subsection{The Cosmological Parameters}

I will now define the most important cosmological parameters. Perhaps
the best known is the {\em Hubble parameter} or rate of expansion today,
$H_0 = \dot a/a(t_0)$. We can write the Hubble parameter in units of
100 km\,s$^{-1}$Mpc$^{-1}$, which can be used to estimate the order of
magnitude for the present size and age of the universe,
\begin{eqnarray}
H_0&\equiv& 100\,h\ \ {\rm km\,s}^{-1}{\rm Mpc}^{-1}\,,\\
c\,H_0^{-1} &=& 3000\,h^{-1}\ {\rm Mpc}\,,\\
H_0^{-1} &=& 9.773\,h^{-1}\ {\rm Gyr}\,.
\end{eqnarray}
The parameter $h$ was measured to be in the range $0.4 < h< 1$ for
decades, and only in the last few years has it been found to lie within
4\% of $h=0.70$.  I will discuss those recent measurements in the next
Section.

Using the present rate of expansion, one can define a {\em critical}
density $\rho_c$, that which corresponds to a flat universe,
\begin{eqnarray}\label{CriticalDensity}
\rho_c\equiv{3H_0^2\over8\pi G}&=& 1.88\,h^2\,10^{-29}\ {\rm g/cm}^3\\
&=& 2.77\,h^{-1}\,10^{11}\ M_\odot/(h^{-1}\,{\rm Mpc})^3\\[1mm]
&=& 11.26\,h^2\ {\rm protons}/{\rm m}^3\,,
\end{eqnarray}
where $M_\odot=1.989\times10^{33}$ g \ is a solar mass unit. The
critical density $\rho_c$ corresponds to approximately 6 protons per
cubic meter, certainly a very dilute fluid! 

In terms of the critical density it is possible to define the 
{\rm density parameter}
\begin{equation}
\Omega_0\equiv{8\pi G\over3H_0^2}\,\rho(t_0)={\rho\over\rho_c}(t_0)\,,
\end{equation}
whose sign can be used to determine the spatial (three-)curvature.
Closed universes ($K=+1$) have $\Omega_0 > 1$, flat universes ($K=0$)
have $\Omega_0 = 1$, and open universes ($K=-1$) have $\Omega_0 < 1$,
no matter what are the individual components that sum up to the
density parameter.

In particular, we can define the individual ratios $\Omega_i \equiv
\rho_i/\rho_c$, for matter, radiation, cosmological constant and even
curvature, today, 
\begin{eqnarray}\label{Omega}
&&\Omega_M={8\pi G\,\rho_M\over3H_0^2}\hspace{2cm} 
\Omega_R={8\pi G\,\rho_R\over3H_0^2}\\
&&\,\Omega_\Lambda={\Lambda\over3H_0^2}\hspace{2.6cm}
\Omega_K=-\,{K\over a_0^2H_0^2}\,.
\end{eqnarray} 
For instance, we can evaluate today the radiation component
$\Omega_{\rm R}$, corresponding to relativistic particles, from the
density of microwave background photons, $\rho_{\rm CMB} =
\pi^2k^4T_{\rm CMB}^4/(15\hbar^3c^3) = 4.5\times 10^{-34}\
{\rm g/cm}^3$, which gives $\Omega_{\rm CMB} = 2.4\times 10^{-5}\
h^{-2}$. Three approximately massless neutrinos would contribute a
similar amount. Therefore, we can safely neglect the contribution of
relativistic particles to the total density of the universe today,
which is dominated either by non-relativistic particles (baryons, dark
matter or massive neutrinos) or by a cosmological constant, and write
the rate of expansion in terms of its value today, as
\begin{equation}\label{H2a}
H^2(a) = H_0^2\left(\Omega_R\,{a_0^4\over a^4} + 
\Omega_M\,{a_0^3\over a^3} + \Omega_\Lambda + 
\Omega_K\,{a_0^2\over a^2}\right)\,.
\end{equation}
An interesting consequence of these definitions is that one can now
write the Friedmann equation today, $a=a_0$, as a {\em cosmic sum rule}, 
\begin{equation}\label{CosmicSumRule}
1 = \Omega_M + \Omega_\Lambda + \Omega_K\,,
\end{equation}
where we have neglected $\Omega_{\rm R}$ today. That is, in the context
of a FRW universe, the total fraction of matter density, cosmological
constant and spatial curvature today must add up to one.  For instance,
if we measure one of the three components, say the spatial curvature, we
can deduce the sum of the other two. 

Looking now at the second Friedmann equation (\ref{Evolution}), we can
define another basic parameter, the {\em deceleration parameter},
\begin{equation}\label{Deceleration}
q_0 = - {a\,\ddot a\over\dot a^2}(t_0) = {4\pi G\over3H_0^2}\,
\Big[\rho(t_0)+3p(t_0)\Big]\,,
\end{equation}
defined so that it is positive for ordinary matter and radiation,
expressing the fact that the universe expansion should slow down due
to the gravitational attraction of matter. We can write this parameter
using the definitions of the density parameter for known and unknown
fluids (with density $\Omega_x$ and arbitrary equation of state $w_x$) as
\begin{equation}\label{DecelerationParameter}
q_0 = \Omega_R + \half\Omega_M - \Omega_\Lambda + \half
\sum_x (1+3w_x)\,\Omega_x\,.
\end{equation}
Uniform expansion corresponds to $q_0=0$ and requires a cancellation
between the matter and vacuum energies. For matter domination, $q_0 >0$,
while for vacuum domination, $q_0 < 0$. As we will see in a moment,
we are at present probing the time dependence of the deceleration
parameter and can determine with some accuracy the moment at which
the universe went from a decelerating phase, dominated by dark matter,
into an acceleration phase at present, which seems to indicate the
dominance of some kind of vacuum energy.

\subsection{The $(\Omega_M,\,\Omega_\Lambda)$ plane}

Now that we know that the universe is accelerating, one can parametrize
the matter/energy content of the universe with just two components:
the matter, characterized by $\Omega_M$, and the vacuum energy 
$\Omega_\Lambda$. Different values of these two parameters completely
specify the universe evolution. It is thus natural to plot the results
of observations in the plane ($\Omega_M,\ \Omega_\Lambda$), in order to
check whether we arrive at a consistent picture of the present universe
from several different angles (different sets of cosmological observations).

%%%%%%%%%%%%%%%%%%%%%%%%%
\begin{figure}
\begin{center}\hspace*{-3cm}
\includegraphics[width=10cm]{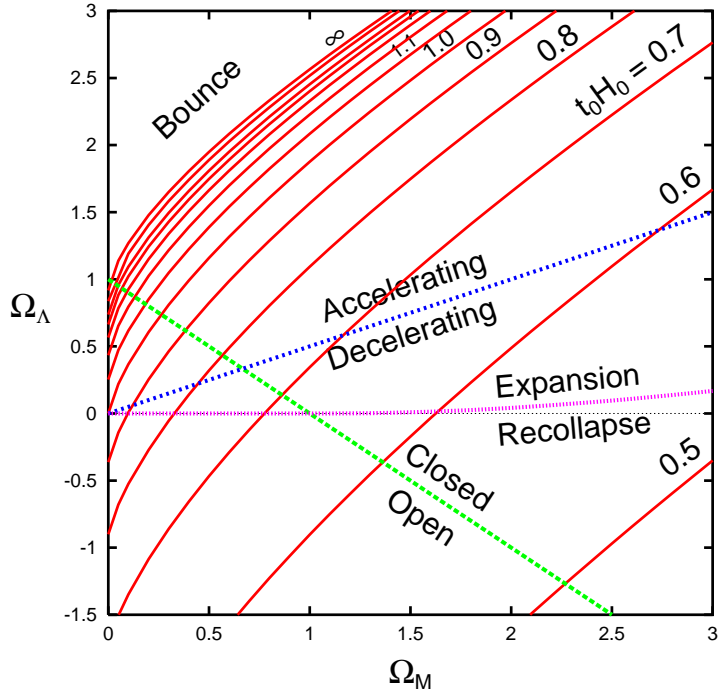} 
\vspace*{-1cm}
\caption{Parameter space $(\Omega_M,\,\Omega_\Lambda)$. The green
(dashed) line $\Omega_\Lambda=1-\Omega_M$ corresponds to a flat
universe, $\Omega_K=0$, separating open from closed universes. The blue
(dotted) line $\Omega_\Lambda=\Omega_M/2$ corresponds to uniform
expansion, $q_0 = 0$, separating accelerating from decelerating
universes. The violet (dot-dashed) line corresponds to critical
universes, separating eternal expansion from recollapse in the
future. Finally, the red (continuous) lines correspond to
$t_0H_0=0.5,\,0.6,\dots,\,\infty$, beyond which the universe has a
bounce.}
\end{center}
\end{figure}
%%%%%%%%%%%%%%%%%%%%%%%%%

Moreover, different regions of this plane specify different behaviors
of the universe. The boundaries between regions are well defined curves
that can be computed for a given model. I will now describe the various
regions and boundaries.

\begin{itemize}

\item {\em Uniform expansion} ($q_0=0$). Corresponds to the line
$\Omega_\Lambda=\Omega_M/2$. Points above this line correspond to
universes that are accelerating today, while those below correspond to
decelerating universes, in particular the old cosmological model of
Einstein-de Sitter (EdS), with $\Omega_\Lambda=0,\ \Omega_M=1$.  Since
1998, all the data from Supernovae of type Ia appear above this line, 
many standard deviations away from EdS universes.

\item {\em Flat universe} ($\Omega_K=0$). Corresponds to the line
$\Omega_\Lambda=1-\Omega_M$. Points to the right of this line
correspond to closed universes, while those to the left correspond to
open ones.  In the last few years we have mounting evidence that the
universe is spatially flat (in fact Euclidean).

\item {\em Bounce} ($t_0H_0=\infty$). Corresponds to a complicated
function of $\Omega_\Lambda(\Omega_M)$, normally expressed as an
integral equation, where
$$t_0H_0=\int_0^1 da\ [1+\Omega_M(1/a-1)+\Omega_\Lambda(a^2-1)]^{-1/2}$$
is the product of the age of the universe and the present rate of
expansion. Points above this line correspond to universes that have
contracted in the past and have later rebounced. At present, these 
universes are ruled out by observations of galaxies and quasars at
high redshift (up to $z=10$).

\item {\em Critical Universe} ($H=\dot H=0$). Corresponds to the
boundary between eternal expansion in the future and recollapse. For
$\Omega_M \leq1$, it is simply the line $\Omega_\Lambda=0$, but for
$\Omega_M >1$, it is a more complicated curve,
$$\Omega_\Lambda=4\Omega_M\sin^3\Big[{1\over3}\arcsin\Big({\Omega_M-
1\over\Omega_M}\Big)\Big]\simeq{4\over27}{(\Omega_M-1)^3\over\Omega_M^2}.$$
These critical solutions are asymptotic to the EdS model.

\end{itemize}

These boundaries, and the regions they delimit,
can be seen in Fig.~1, together with the lines of equal 
$t_0H_0$ values.

In summary, the basic cosmological parameters that are now been hunted 
by a host of cosmological observations are the following: the present
rate of expansion $H_0$; the age of the universe $t_0$; the
deceleration parameter $q_0$; the spatial curvature $\Omega_K$; the
matter content $\Omega_M$; the vacuum energy $\Omega_\Lambda$; the
baryon density $\Omega_B$; the neutrino density $\Omega_\nu$, and
many other that characterize the perturbations responsible for the
large scale structure (LSS) and the CMB anisotropies.

\subsection{The accelerating universe}

Let us first describe the effect that the expansion of the universe
has on the objects that live in it. In the absence of other forces
but those of gravity, the trajectory of a particle is given by
general relativity in terms of the geodesic equation
\begin{equation}\label{geodesic}
{du^\mu\over ds} + \Gamma^\mu_{\nu\lambda}\,u^\nu u^\lambda =0\,,
\end{equation}
where $u^\mu=(\gamma,\,\gamma v^i)$, with $\gamma^2=1-v^2$ and $v^i$ is
the peculiar velocity. Here $\Gamma^\mu_{\nu\lambda}$ is the Christoffel
connection~\cite{Weinberg}, whose only non-zero component is
$\Gamma^0_{ij} = (\dot a/a)\,g_{ij}$; substituting into the geodesic
equation, we obtain $|\vec u|\propto 1/a$, and thus the particle's
momentum decays with the expansion like $p\propto 1/a$. In the case of a
photon, satisfying the de Broglie relation $p=h/\lambda$, one obtains
the well known {\em photon redshift}
\begin{equation}\label{redshift}
{\lambda_1\over\lambda_0} = {a(t_1)\over a(t_0)} \ \ \Rightarrow \ 
\ z\equiv{\lambda_0-\lambda_1\over\lambda_1} = {a_0\over a_1} - 1\,,
\end{equation}
where $\lambda_0$ is the wavelength measured by an observer at time
$t_0$, while $\lambda_1$ is the wavelength emitted when the universe was
younger $(t_1<t_0)$. Normally we measure light from stars in distant
galaxies and compare their observed spectra with our laboratory
(restframe) spectra. The fraction (\ref{redshift}) then gives the redshift
$z$ of the object. We are assuming, of course, that both the emitted and
the restframe spectra are identical, so that we can actually measure the
effect of the intervening expansion, i.e. the growth of the scale factor
from $t_1$ to $t_0$, when we compare the two spectra. Note that if the
emitting galaxy and our own participated in the expansion, i.e. if our
measuring rods (our rulers) also expanded with the universe, we would
see no effect!  The reason we can measure the redshift of light from a
distant galaxy is because our galaxy is a gravitationally bounded
object that has decoupled from the expansion of the universe. It is the
distance between galaxies that changes with time, not the sizes of
galaxies, nor the local measuring rods.

We can now evaluate the relationship between physical distance and
redshift as a function of the rate of expansion of the universe.
Because of homogeneity we can always choose our position to be at the
origin $r=0$ of our spatial section. Imagine an object (a star) emitting
light at time $t_1$, at coordinate distance $r_1$ from the
origin. Because of isotropy we can ignore the angular coordinates
$(\theta,\phi)$. Then the physical distance, to first order, will be
$d=a_0\,r_1$. Since light travels along null geodesics~\cite{Weinberg},
we can write $0 = -dt^2+a^2(t)\,dr^2/(1-Kr^2)$, and therefore,
\begin{equation}\label{distance}
\int_{t_1}^{t_0}{dt\over a(t)} = \int_0^{r_1}{dr\over\sqrt{1-Kr^2}} 
\equiv f(r_1) = \left\{\begin{array}{ll}\arcsin r_1 \hspace{1cm} & K=1\\
r_1 & K=0\\{\rm arcsinh}\, r_1 & K=-1\end{array}\right.
\end{equation}
If we now Taylor expand the scale factor to first order,
\begin{equation}\label{Taylor}
{1\over1+z} = {a(t)\over a_0} = 1 + H_0(t-t_0)+{\cal O}(t-t_0)^2\,,
\end{equation}
we find, to first approximation,
$$r_1 \approx f(r_1) = {1\over a_0}(t_0 - t_1) + \dots = 
{z\over a_0H_0} + \dots$$
Putting all together we find the famous Hubble law
\begin{equation}\label{HubbleLaw}
H_0\,d = a_0H_0r_1 = z \simeq vc\,,
\end{equation}
which is just a kinematical effect (we have not included yet
any dynamics, i.e. the matter content of the universe). Note that at
low redshift $(z\ll1)$, one is tempted to associate the observed
change in wavelength with a Doppler effect due to a hypothetical
recession velocity of the distant galaxy. This is only an
approximation. In fact, the redshift cannot be ascribed to the
relative velocity of the distant galaxy because in general relativity
(i.e. in curved spacetimes) one cannot compare velocities through
parallel transport, since the value depends on the path! If the
distance to the galaxy is small, i.e. $z\ll1$, the physical spacetime
is not very different from Minkowsky and such a comparison is
approximately valid. As $z$ becomes of order one, such a relation is
manifestly false: galaxies cannot travel at speeds greater than the
speed of light; it is the stretching of spacetime which is responsible
for the observed redshift.

%%%%%%%%%%%%%%%%%%%%%%%%%
\begin{figure}[htb]
\begin{center}\label{fig:stretchfactor}
\includegraphics[width=7.8cm]{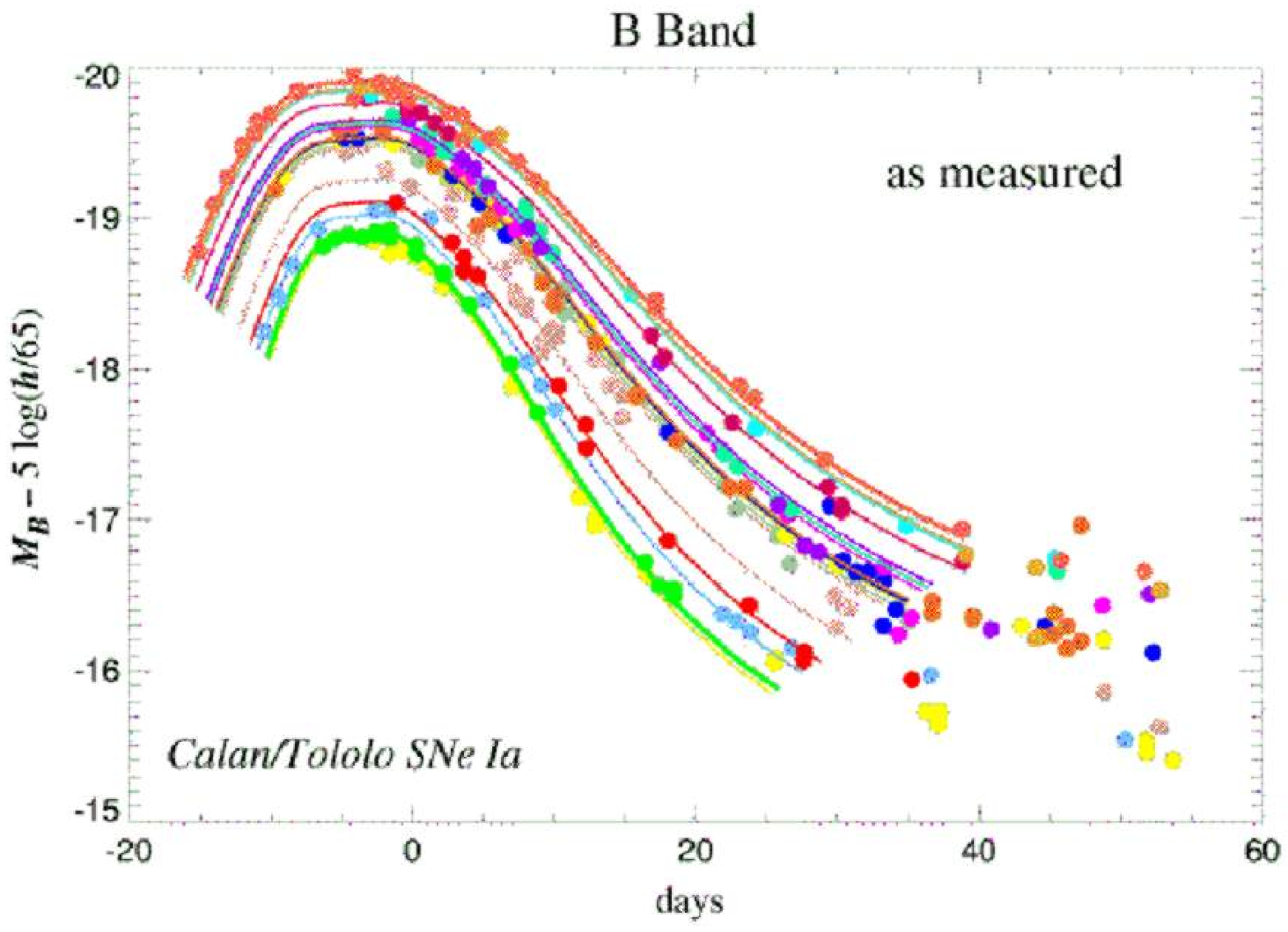} 
\includegraphics[width=7.8cm]{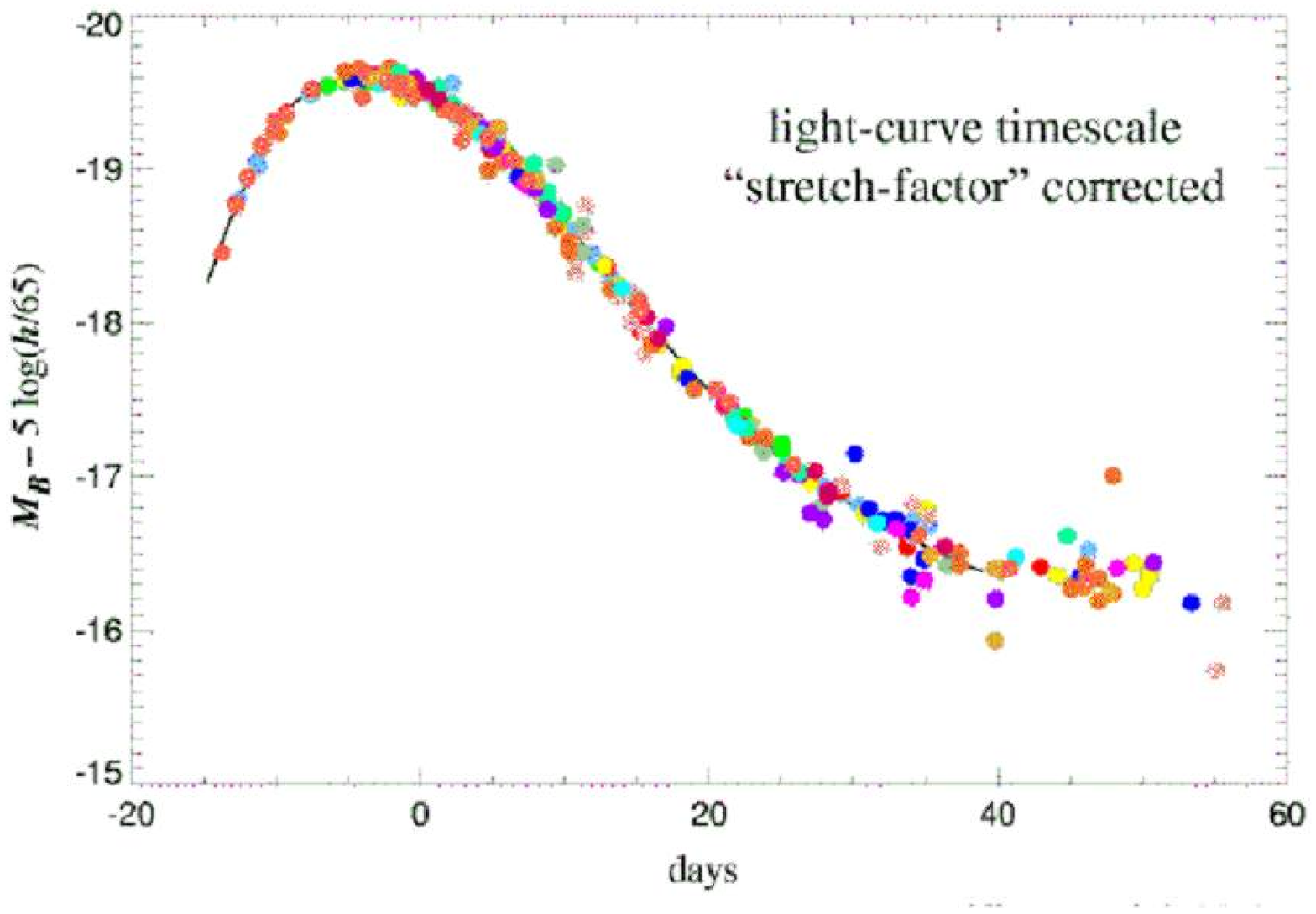} 
\caption{The Type Ia supernovae observed nearby show a relationship
between their absolute luminosity and the timescale of their light
curve: the brighter supernovae are slower and the fainter ones are
faster. A simple linear relation between the absolute magnitude and a
``stretch factor'' multiplying the light curve timescale fits the data
quite well. From Ref.~\cite{SCP}.}
\end{center}
\end{figure}
%%%%%%%%%%%%%%%%%%%%%%%%%

Hubble's law has been confirmed by observations ever since the 1920s,
with increasing precision, which have allowed cosmologists to determine
the Hubble parameter $H_0$ with less and less systematic errors.
Nowadays, the best determination of the Hubble parameter was made by the
Hubble Space Telescope Key Project~\cite{HSTKP}, $H_0 = 72 \pm 8$
km/s/Mpc. This determination is based on objects at distances up to 500
Mpc, corresponding to redshifts $z\leq0.1$.

Nowadays, we are beginning to probe much greater distances,
corresponding to $z\simeq1$, thanks to type Ia supernovae.  These are
white dwarf stars at the end of their life cycle that accrete matter
from a companion until they become unstable and violently explode in a
natural thermonuclear explosion that out-shines their progenitor
galaxy. The intensity of the distant flash varies in time, it takes
about three weeks to reach its maximum brightness and then it declines
over a period of months.  Although the maximum luminosity varies from
one supernova to another, depending on their original mass, their
environment, etc., there is a pattern: brighter explosions last longer
than fainter ones. By studying the characteristic light curves, see
Fig.~2, of a reasonably large statistical sample, cosmologists from
the Supernova Cosmology Project~\cite{SCP} and the High-redshift
Supernova Project~\cite{HRS}, are now quite confident that they can
use this type of supernova as a standard candle. Since the light
coming from some of these rare explosions has travelled a large
fraction of the size of the universe, one expects to be able to infer
from their distribution the spatial curvature and the rate of
expansion of the universe. 

%%%%%%%%%%%%%%%%%%%%%%%%%
\begin{figure}[htb]
\label{fig:Knop03}
\begin{center}%\vspace*{-3cm}\vspace*{1cm}
\includegraphics[width=9cm]{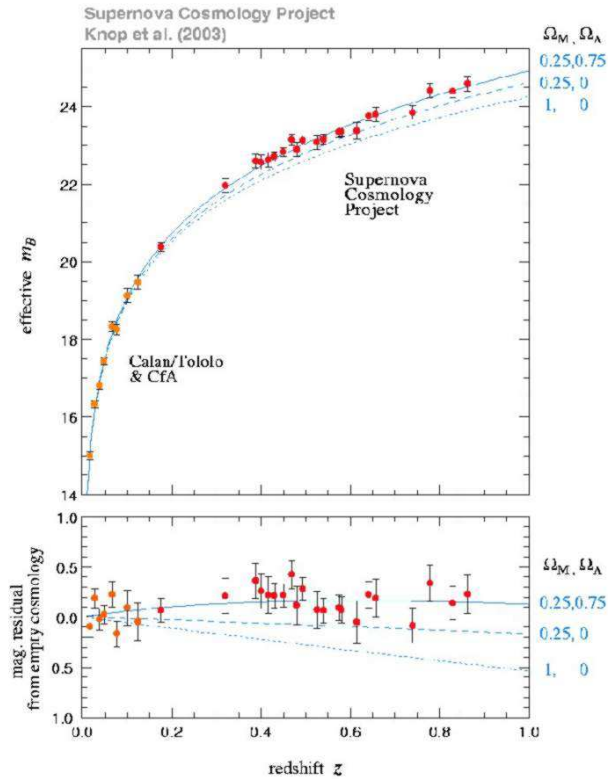}
%\vspace*{-7cm}
\caption{Upper panel: The Hubble diagram in linear redshift
scale. Supernovae with $\Delta z < 0.01$ of eachother have been
weighted-averaged binned.  The solid curve represents the best-fit
flat universe model, $(\Omega_M=0.25,\,\Omega_\Lambda=0.75)$. Two
other cosmological models are shown for comparison, $(\Omega_M=0.25,
\,\Omega_\Lambda=0)$ and $(\Omega_M=1,\,\Omega_\Lambda=0)$. Lower
panel: Residuals of the averaged data relative to an empty
universe. From Ref.~\cite{SCP}.}
\end{center}
\end{figure}
%%%%%%%%%%%%%%%%%%%%%%%%%

The connection between observations of high
redshift supernovae and cosmological parameters is done via the
luminosity distance, defined as the distance $d_L$ at which a source
of absolute luminosity (energy emitted per unit time) ${\cal L}$ gives
a flux (measured energy per unit time and unit area of the detector)
${\cal F}={\cal L}/4\pi\,d_L^2$. One can then evaluate, within a given
cosmological model, the expression for $d_L$ as a function of
redshift~\cite{JGB},
\begin{equation}\label{LuminosityDistance}
H_0\,d_L(z) = {(1+z)\over|\Omega_K|^{1/2}}\,{\rm sinn}\left[
\int_0^z {|\Omega_K|^{1/2}\ dz'\over\sqrt{(1+z')^2(1+z'\Omega_M) - 
z'(2+z')\Omega_\Lambda}}\right]\,,
\end{equation}
where ${\rm sinn}(x)= x\ {\rm if}\ K=0;\ \sin(x)\ {\rm if}\ K=+1\ {\rm
and}\ \sinh(x)\ {\rm if}\ K=-1$, and we have used the cosmic sum rule
(\ref{CosmicSumRule}). 

Astronomers measure the relative luminosity of a distant object in
terms of what they call the effective magnitude, which has a peculiar
relation with distance,
\begin{equation}\label{EffectiveMagnitude}
m(z) \equiv M + 5\,\log_{10}\Big[{d_L(z)\over{\rm Mpc}}\Big] + 25
= \bar M + 5\,\log_{10}[H_0\,d_L(z)]\,.
\end{equation}
Since 1998, several groups have obtained serious evidence that high
redshift supernovae appear fainter than expected for either an open
$(\Omega_M < 1)$ or a flat $(\Omega_M = 1)$ universe, see Fig.~3.  In
fact, the universe appears to be accelerating instead of decelerating,
as was expected from the general attraction of matter, see
Eq.~(\ref{DecelerationParameter}); something seems to be acting as a
repulsive force on very large scales. The most natural explanation for
this is the presence of a cosmological constant, a diffuse vacuum
energy that permeates all space and, as explained above, gives the
universe an acceleration that tends to separate gravitationally bound
systems from each other. The best-fit results from the Supernova
Cosmology Project~\cite{SCP2003} give a linear combination
$$0.8\,\Omega_M - 0.6\,\Omega_\Lambda = - 0.16 \pm 0.05 \ \ (1\sigma),$$
which is now many sigma away from an EdS model with $\Lambda=0$.
In particular, for a flat universe this gives 
$$\Omega_\Lambda = 0.71 \pm 0.05 \hspace{1cm} {\rm and} \hspace{1cm} 
\Omega_M = 0.29 \pm 0.05 \ \ (1\sigma).$$ 
Surprising as it may seem, arguments for a significant dark energy
component of the universe where proposed long before these
observations, in order to accommodate the ages of globular clusters,
as well as a flat universe with a matter content below critical, which
was needed in order to explain the observed distribution of galaxies,
clusters and voids.

Taylor expanding the scale factor to third order,
\begin{equation}
{a(t)\over a_0} = 1 + H_0(t-t_0) - {q_0\over2!}H_0^2(t-t_0)^2 + 
{j_0\over3!}H_0^3(t-t_0)^3 + {\cal O}(t-t_0)^4\,,
\end{equation}
where
\begin{eqnarray}
&&q_0=-\,{\ddot a\over a H^2}(t_0) = \half\sum_i(1+3w_i)\Omega_i=
\half\Omega_M-\Omega_\Lambda\,, \\
&&j_0=+\,{\stackrel{\dots}{a}\over a H^3}(t_0) = \half\sum_i(1+3w_i)
(2+3w_i)\Omega_i=\Omega_M+\Omega_\Lambda\,,\label{JerkParameter}
\end{eqnarray}
are the deceleration and ``jerk'' parameters. Substituting into
Eq.~(\ref{LuminosityDistance}) we find
\begin{equation}\label{DLZ}
H_0\,d_L(z) = z + {1\over2}(1-q_0)\,z^2 - 
{1\over6}(1-q_0-3q_0^2+j_0)\,z^3 + {\cal O}(z^4)\,.
\end{equation}
This expression goes beyond the leading linear term, corresponding to
the Hubble law, into the second and third order terms, which are
sensitive to the cosmological parameters $\Omega_M$ and
$\Omega_\Lambda$. It is only recently that cosmological observations
have gone far enough back into the early universe that we can begin to
probe these terms, see Fig.~4.

%%%%%%%%%%%%%%%%%%%%%%%%%
\begin{figure}\label{fig:Riess04}
\begin{center}\hspace*{-4cm}
\includegraphics[width=9cm]{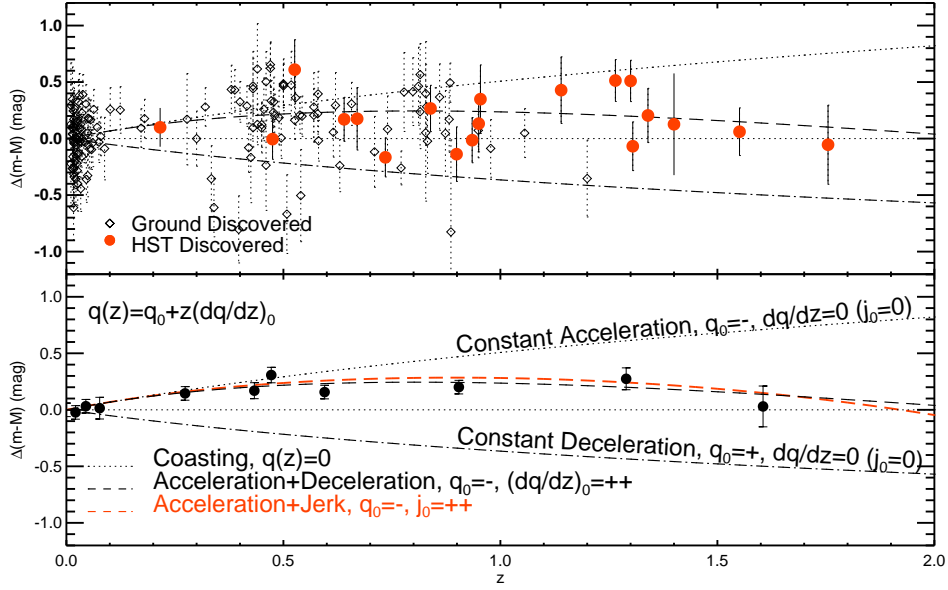}
\vspace*{-4cm}
\caption{The Supernovae Ia residual Hubble diagram. Upper panel:
Ground-based discoveries are represented by diamonds, HST-discovered
SNe Ia are shown as filled circles. Lower panel: The same but with
weighted averaged in fixed redshift bins. Kinematic models of the
expansion history are shown relative to an eternally coasting model
$q(z)=0$. From Ref.~\cite{Riess2004}.}
\end{center}
\end{figure}
%%%%%%%%%%%%%%%%%%%%%%%%%

This extra component of the critical density would have to resist
gravitational collapse, otherwise it would have been detected already
as part of the energy in the halos of galaxies.  However, if most of
the energy of the universe resists gravitational collapse, it is
impossible for structure in the universe to grow. This dilemma can be
resolved if the hypothetical dark energy was negligible in the past
and only recently became the dominant component. According to general
relativity, this requires that the dark energy have {\rm negative}
pressure, since the ratio of dark energy to matter density goes like
$a(t)^{-3p/\rho}$. This argument would rule out almost all of the
usual suspects, such as cold dark matter, neutrinos, radiation, and
kinetic energy, since they all have zero or positive pressure. Thus,
we expect something like a cosmological constant, with a negative
pressure, $p\approx-\rho$, to account for the missing energy.

However, if the universe was dominated by dark matter in the past, in
order to form structure, and only recently became dominated by dark
energy, we must be able to see the effects of the transition from the
deceleration into the acceleration phase in the luminosity of distant
type Ia supernovae. This has been searched for since 1998, when the
first convincing results on the present acceleration appeared.
However, only recently~\cite{Riess2004} do we have clear evidence of
this transition point in the evolution of the universe.  This {\em
coasting point} is defined as the time, or redshift, at which the
deceleration parameter vanishes,
\begin{equation}
q(z) = -1 + (1+z)\,{d\over dz}\ln H(z) = 0\,,
\end{equation}
where
\begin{equation}
H(z) = H_0\Big[\Omega_M (1+z)^3 + \Omega_x\,e^{3\int_0^z(1+w_x(z'))
{dz'\over1+z'}}\ + \Omega_K (1+z)^2\Big]^{1/2}\,,
\end{equation}
and we have assumed that the dark energy is parametrized by a
density $\Omega_x$ today, with a redshift-dependent equation of 
state, $w_x(z)$, not necessarily equal to $-1$. Of course, in the
case of a true cosmological constant, this reduces to the usual 
expression.

Let us suppose for a moment that the barotropic parameter $w$ is
constant, then the coasting redshift can be determined from
\begin{eqnarray}
&&q(z) = \half\Big[{\Omega_M + (1+3w)\,\Omega_x\,(1+z)^{3w}\over
\Omega_M + \Omega_x\,(1+z)^{3w} + \Omega_K (1+z)^{-1}}\Big] = 0\,,\\
&&\hspace{1cm}\Rightarrow z_c = \left({(3|w|-1)\Omega_x\over\Omega_M}
\right)^{1\over3|w|}-1\,,
\label{DecelerationParamz}
\end{eqnarray}
which, in the case of a true cosmological constant, reduces to
\begin{equation}\label{CoastingPoint}
z_c = \Big({2\Omega_\Lambda\over\Omega_M}\Big)^{1/3}-1\,.
\end{equation}
When substituting $\Omega_\Lambda\simeq0.7$ and $\Omega_M\simeq0.3$,
one obtains $z_c \simeq 0.6$, in excellent agreement with recent
observations~\cite{Riess2004}. The plane $(\Omega_M,\,\Omega_\Lambda)$
can be seen in Fig.~5, which shows a significant improvement with
respect to previous data.

%%%%%%%%%%%%%%%%%%%%%%%%%
\begin{figure}[htb]
\label{fig:RiessOMOL}
\begin{center}%\hspace*{-4cm}%\vspace*{-1cm}
\includegraphics[width=9cm]{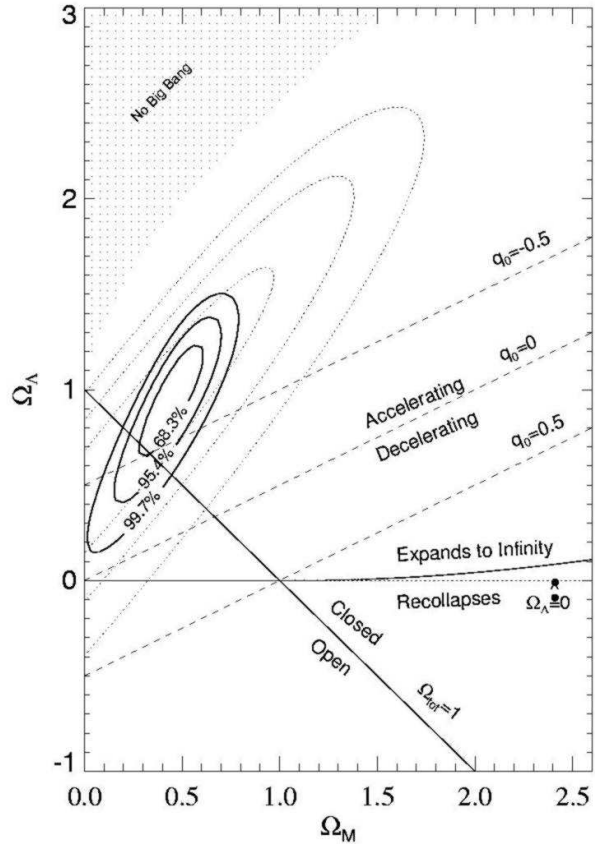}
%\vspace*{5mm}
\caption{The recent supernovae data on the
$(\Omega_M,\,\Omega_\Lambda)$ plane. Shown are the 
1-, 2- and 3-$\sigma$ contours, as well as the data from 1998,
for comparison. It is clear that the old EdS cosmological model at 
$(\Omega_M=1,\,\Omega_\Lambda=0)$ is many standard deviations away
from the data. From Ref.~[12].}
\end{center}
\end{figure}
%%%%%%%%%%%%%%%%%%%%%%%%%

Now, if we have to live with this vacuum energy, we might as well try
to comprehend its origin. For the moment it is a complete mystery,
perhaps the biggest mystery we have in physics
today~\cite{WeinbergCosmo}. We measure its value but we don't
understand why it has the value it has. In fact, if we naively predict
it using the rules of quantum mechanics, we find a number that is many
(many!) orders of magnitude off the mark. Let us describe this
calculation in some detail. In non-gravitational physics, the
zero-point energy of the system is irrelevant because forces arise
from gradients of potential energies. However, we know from general
relativity that even a constant energy density gravitates.  Let us
write down the most general energy momentum tensor compatible with the
symmetries of the metric and that is covariantly conserved.  This is
precisely of the form $T_{\mu\nu}^{(vac)} = p_V\,g_{\mu\nu} =
-\,\rho_V\,g_{\mu\nu}$, see Fig.~6. Substituting into the Einstein
equations (\ref{EinsteinEquations}), we see that the cosmological
constant and the vacuum energy are completely equivalent, $\Lambda =
8\pi G\,\rho_V$, so we can measure the vacuum energy with the
observations of the acceleration of the universe, which tells us that
$\Omega_\Lambda\simeq0.7$.

On the other hand, we can estimate the contribution to the vacuum
energy coming from the quantum mechanical zero-point energy of the
quantum oscillators associated with the fluctuations of all quantum
fields,
\begin{equation}\label{ZeroPointEnergy}
\rho_V^{th} = \sum_i \int_0^{\Lambda_{UV}}\!\!{d^2k\over(2\pi)^3}\,
\half\hbar\omega_i(k)={\hbar\Lambda_{UV}^4\over16\pi^2}\sum_i
(-1)^F N_i + {\cal O}(m_i^2\Lambda_{UV}^2)\,,
\end{equation}
where $\Lambda_{UV}$ is the ultraviolet cutoff signaling the scale of
new physics. Taking the scale of quantum gravity, $\Lambda_{UV} =
M_{Pl}$, as the cutoff, and barring any fortuituous cancellations,
then the theoretical expectation (\ref{ZeroPointEnergy}) appears to be
120 orders of magnitude larger than the observed vacuum energy
associated with the acceleration of the universe,
\begin{eqnarray}\label{vacuumenergyth}
&&\rho_V^{th} \simeq 1.4\times 10^{74} \ {\rm GeV}^4  =
3.2\times 10^{91} \ {\rm g/cm}^3\,,\\
&&\rho_V^{obs} \simeq 0.7\,\rho_c = 0.66\times 10^{-29} \ {\rm g/cm}^3 =
2.9\times 10^{-11} \ {\rm eV}^4\,.\label{vacuumenergyobs}
\end{eqnarray}
Even if we assumed that the ultraviolet cutoff associated with quantum
gravity was as low as the electroweak scale (and thus around the
corner, liable to be explored in the LHC), the theoretical expectation
would still be 60 orders of magnitude too big. This is by far the
worst mismatch between theory and observations in all of science.
There must be something seriously wrong in our present understanding
of gravity at the most fundamental level. Perhaps we don't understand
the vacuum and its energy does not gravitate after all, or perhaps we
need to impose a new principle (or a symmetry) at the quantum gravity
level to accommodate such a flagrant mismatch.

%%%%%%%%%%%%%%%%%%%%%%%%%
\begin{figure}[htb]
%\vspace*{-5mm}
\begin{center}%\hspace*{-1cm}
\includegraphics[width=10cm]{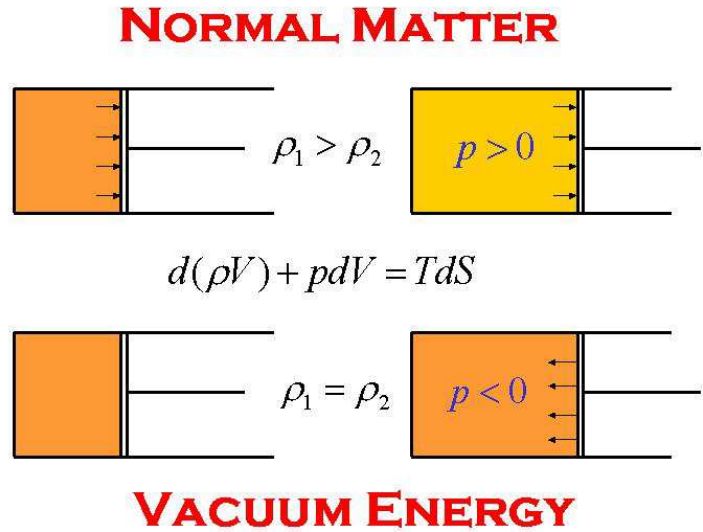}
%\vspace*{-5mm}
\caption{Ordinary matter dilutes as it expands. According to the
second law of Thermodynamics, its pressure on the walls should be
positive, which excerts a force, and energy is lost in the expansion.
On the other hand, vacuum energy is always the same, independent of
the volume of the region, and thus, according to the second law, its
pressure must be negative and of the same magnitude as the energy
density. This negative pressure means that the volume tends to increase
more and more rapidly, which explains the exponential expansion of the
universe dominated by a cosmological constant.}
\end{center}
\label{fig:VacuumPressure}
\end{figure}
%%%%%%%%%%%%%%%%%%%%%%%%%

In the meantime, one can at least parametrize our ignorance by making
variations on the idea of a {\em constant} vacuum energy. Let us
assume that it actually evolves slowly with time. In that case, we do
not expect the equation of state $p=-\rho$ to remain true, but instead
we expect the barotropic parameter $w(z)$ to depend on redshift.  Such
phenomenological models have been proposed, and until recently
produced results that were compatible with $w=-1$ today, but with
enough uncertainty to speculate on alternatives to a truly constant
vacuum energy.  However, with the recent supernovae
results~\cite{Riess2004}, there seems to be little space for
variations, and models of a time-dependent vacuum energy are less and
less favoured. In the near future, the SNAP satellite~\cite{SNAP} will
measure several thousand supernovae at high redshift and therefore map
the redshift dependence of both the dark energy density and its
equation of state with great precision. This will allow a much better
determination of the cosmological parameters $\Omega_M$ and
$\Omega_\Lambda$.

\subsection{Thermodynamics of an expanding plasma}

In this section I will describe the main concepts associated with
ensembles of particles in thermal equilibrium and the brief periods in
which the universe fell out of equilibrium. To begin with, let me make
contact between the covariant energy conservation law
(\ref{EnergyConservation}) and the second law of thermodynamics,
\begin{equation}\label{SecondLawThermodynamics}
T\,dS = dU + p\,dV\,,
\end{equation}
where $U = \rho\,V$ is the total energy of the  fluid, and $p=w\,\rho$
is its barotropic pressure. Taking a comoving volume for the universe,
$V=a^3$, we find
\begin{equation}
T\,{dS\over dt} = {d\over dt}(\rho\,a^3) + p\,{d\over dt}(a^3) = 0\,,
\end{equation}
where we have used (\ref{EnergyConservation}). Therefore, entropy is
conserved during the expansion of the universe, $dS=0$; i.e., the
expansion is adiabatic even in those epochs in which the equation of
state changes, like in the matter-radiation transition (not a proper
phase transition). Using (\ref{EnergyConservation}), we can write
\begin{equation}
{d\over dt}\ln(\rho\,a^3) = - 3H\,w\,.
\end{equation}
Thus, our universe expands like a gaseous fluid in thermal equilibrium
at a temperature $T$. This temperature decreases like that of any
expanding fluid, in a way that is inversely proportional to the cubic
root of the volume. This implies that in the past the universe was
necessarily denser and hotter. As we go back in time we reach higher
and higher temperatures, which implies that the mean energy of plasma
particles is larger and thus certain fundamental reactions are now
possible and even common, giving rise to processes that today we can
only attain in particle physics accelerators. That is the reason why
it is so important, for the study of early universe, to know the
nature of the fundamental interactions at high energies, and the basic
connection between cosmology and high energy particle physics.
However, I should clarify a misleading statement that is often used:
``high energy particle physics colliders reproduce the early
universe'' by inducing collisions among relativistic particles.
Although the energies of some of the interactions at those collisions
reach similar values as those attained in the early universe, the
physical conditions are rather different. The interactions within the
detectors of the great particle physics accelerators occur typically
in the perturbative regime, locally, and very far from thermal
equilibrium, lasting a minute fraction of a second; on the other hand,
the same interactions occurred within a hot plasma in equilibrium in
the early universe while it was expanding adiabatically and its
duration could be significantly larger, with a distribution in energy
that has nothing to do with those associated with particle
accelerators.  What is true, of course, is that the fundamental
parameters corresponding to those interactions $-$ masses and
couplings $-$ are assumed to be the same, and therefore present
terrestrial experiments can help us imagine what it could have been
like in the early universe, and make predictions about the evolution
of the universe, in the context of an expanding plasma a high
temperatures and high densities, and in thermal equilibrium.

\subsubsection{Fluids in thermal equilibrium}

In order to understand the thermodynamical behaviour of a plasma of
different species of particles at high temperatures we will consider a
gas of particles with $g$ internal degrees of freedom weakly
interacting. The degrees of freedom corresponding to the different
particles can be seen in Table~1. For example, leptons and quarks have
4 degrees of freedom since they correspond to the two helicities for
both particle and antiparticle. However, the nature of neutrinos
is still unknown. If they happen to be Majorana fermions, then they
would be their own antiparticle and the number of degrees of freedom
would reduce to 2. For photons and gravitons (without mass) their 2
d.o.f. correspond to their states of polarization. The 8 gluons (also
without mass) are the gauge bosons responsible for the strong
interaction betwen quarks, and also have 2 d.o.f. each. The vector
bosons $W^\pm$ and $Z^0$ are massive and thus, apart from the
transverse components of the polarization, they also have longitudinal
components.

\begin{table}[htb]\label{table8}
\begin{center}
\begin{tabular}{|l|c|c|l|}
\hline\hline
\noalign{\smallskip}
Particle & Spin & Degrees of freedom ($g$) & Nature \\
\hline\hline
\noalign{\smallskip}
Higgs & 0 & 1 & Massive scalar \\[1mm]
photon & 1 & 2 & Massless vector \\[1mm]
graviton & 2 & 2 & Massless tensor \\[1mm]
gluon & 1 & 2 & Massless vector \\[1mm]
$W$ y $Z$ & 1 & 3 & Massive vector  \\[1mm]
leptons \& quarks & 1/2 & 4 & Dirac Fermion \\[1mm]
neutrinos & 1/2 & 4 (2) & Dirac (Majorana) Fermion \\[1mm]
\hline\hline
\end{tabular}
\caption{The internal degrees of freedom of various fundamental
particles.}
\end{center}
\end{table}

For each of these particles we can compute the number density $n$, the
energy density $\rho$ and the pressure $p$, in thermal equilibrium at
a given temperature $T$,
\begin{eqnarray}
&&n = g\,\int{d^3{\bf p}\over(2\pi)^3}\,f({\bf p})\,,\\[2mm]
&&\rho = g\,\int{d^3{\bf p}\over(2\pi)^3}\,E({\bf p})\,f({\bf p})\,,\\[2mm]
&&p = g\,\int{d^3{\bf p}\over(2\pi)^3}\,{|{\bf p}|^2\over3E}\,
f({\bf p})\,,
\end{eqnarray}
where the energy is given by  $\,E^2 = |{\bf p}|^2+m^2\,$ and the
momentum distribution in thermal (kinetic) equilibrium is
\begin{equation}\label{EquilibriumDistribution}
f({\bf p}) = {1\over e^{(E-\mu)/T} \pm 1} \hspace{1cm} \left\{
\begin{array}{ll}-1 & \hspace{1cm} {\rm Bose-Einstein}\\[1mm]
+1 & \hspace{1cm} {\rm Fermi-Dirac}\end{array}\right.
\end{equation}
The chemical potential $\mu$ is conserved in these reactions if they
are in thermal equilibrium. For example, for reactions of the type
$\,i + j \longleftrightarrow k + l\,$, we have $\,\mu_i + \mu_j =
\mu_k + \mu_l$. For example, the chemical potencial of the photon
vanishes $\mu_\gamma = 0$, and thus particles and antiparticles have
opposite chemical potentials.

From the equilibrium distributions one can obtain the number density
$n$, the energy $\rho$ and the pressure $p$, of a particle of mass
$m$ with chemical potential $\mu$ at the temperature $T$,
\begin{eqnarray}\label{nT}
&&n = {g\over2\pi^2}\,\int_m^\infty dE\,{E(E^2-m^2)^{1/2}\over
e^{(E-\mu)/T} \pm 1}\,,\\[2mm]
\label{rhoT}
&&\rho = {g\over2\pi^2}\,\int_m^\infty dE\,{E^2(E^2-m^2)^{1/2}\over
e^{(E-\mu)/T} \pm 1}\,,\\[2mm]
&&p = {g\over6\pi^2}\,\int_m^\infty dE\,{(E^2-m^2)^{3/2}\over
e^{(E-\mu)/T} \pm 1}\,.\label{pT}
\end{eqnarray}
For a non-degenerate ($\mu\ll T$) relativistic gas ($m\ll T$), we find 
\begin{eqnarray}\label{nTrel}
&&n = {g\over2\pi^2}\,\int_0^\infty {E^2\,dE\over
e^{E/T} \pm 1} = \left\{\begin{array}{ll}{\displaystyle
{\zeta(3)\over\pi^2}\,g\,T^3} & \hspace{1cm} {\rm Bosons}\\[3mm]
{\displaystyle {3\over4}\,{\zeta(3)\over\pi^2}\,g\,T^3}
& \hspace{1cm} {\rm Fermions}\end{array}\right.\,,\\[2mm]
\label{rhoTrel}
&&\rho = {g\over2\pi^2}\,\int_0^\infty {E^3\,dE\over
e^{E/T} \pm 1} = \left\{\begin{array}{ll} {\displaystyle
{\pi^2\over30}\,g\,T^4} & \hspace{1cm} {\rm Bosons}\\[3mm]
{\displaystyle {7\over8}\,{\pi^2\over30}\,g\,T^4}
& \hspace{1cm} {\rm Fermions}\end{array}\right.\,,\\[2mm]
&&p = {1\over3}\rho\,,\label{pTrel}
\end{eqnarray}
where $\zeta(3)=1.20206\dots$ is the Riemann Zeta function. For
relativistic fluids, the energy density per particle is
\begin{equation}\label{AverageEnergyRel}
\langle E\rangle \equiv {\rho\over n} = \left\{
\begin{array}{ll} {\displaystyle {\pi^4\over30\zeta(3)}\,T 
\simeq 2.701\,T} & \hspace{1cm} {\rm Bosons}\\[3mm]
{\displaystyle {7\pi^2\over180\zeta(3)}\,T \simeq 3.151\,T}
& \hspace{1cm} {\rm Fermions}\end{array}\right.
\end{equation}
For relativistic bosons or fermions with $\mu < 0$ and $|\mu| < T$,
we have
\begin{eqnarray}\label{nTmu}
&&n = {g\over\pi^2}\,T^3\,e^{\mu/T}\,,\\[2mm]
\label{rhoTmu}
&&\rho = {3g\over\pi^2}\,T^4\,e^{\mu/T}\,,\\[2mm]
&&p = {1\over3}\rho\,.\label{pTmu}
\end{eqnarray}
For a bosonic particle, a positive chemical potential, $\mu > 0$, 
indicates the presence of a Bose-Einstein condensate, and should
be treated separately from the rest of the modes.

On the other hand, for a non-relativistic gas ($m\gg T$), with
arbitrary chemical potential $\mu$, we find
\begin{eqnarray}\label{nTnorel}
&&n = g\,\left({mT\over2\pi}\right)^{3/2}\,e^{-(m-\mu)/T}\,,\\[2mm]
\label{rhoTnorel}
&&\rho = m\,n\,,\\[2mm]
&&p = n\,T \ll \rho\,.\label{pTnorel}
\end{eqnarray}
The average energy density per particle is
\begin{equation}\label{AverageEnergyNoRel}
\langle E\rangle \equiv {\rho\over n} = m + {3\over2}\,T\,.
\end{equation}
Note that, at any given temperature $T$, the contribution to the
energy density of the universe coming from non-relativistic particles
in thermal equilibrium is exponentially supressed with respect to that
of relativistic particles, therefore we can write
\begin{eqnarray}
&&\rhor = {\pi^2\over30}\,g_*\,T^4\,, \hspace{1cm}
p_{\rm R} = {1\over3}\,\rhor\,,\\[2mm]
&&g_*(T) = \sum_{\rm bosons} g_i\,\left({T_i\over T}\right)^4
+ {7\over8}\,\sum_{\rm fermions} g_i\,\left({T_i\over T}\right)^4\,,
\end{eqnarray}
where the factor $7/8$ takes into account the difference between the
Fermi and Bose statistics; $g_*$ is the total number of light d.o.f.
($m\ll T$), and we have also considered the possibility that particle
species $i$ (bosons or fermions) have an equilibrium distribution at a
temperature $T_i$ different from that of photons, as happens for
example when a given relativistic species decouples from the thermal
bath, as we will discuss later. This number, $g_*$, strongly depends
on the temperature of the universe, since as it expands and cools,
different particles go out of equilibrium or become non-relativistic
($m\gg T$) and thus become exponentially suppressed from that moment on.
A plot of the time evolution of $g_*(T)$ can be seen in Fig.~7.

%%%%%%%%%%%%%%%%%%%%%%%%%
\begin{figure}[htb]
%\vspace*{-1.5cm}
\begin{center}%\hspace*{-1cm}
\includegraphics[width=10cm]{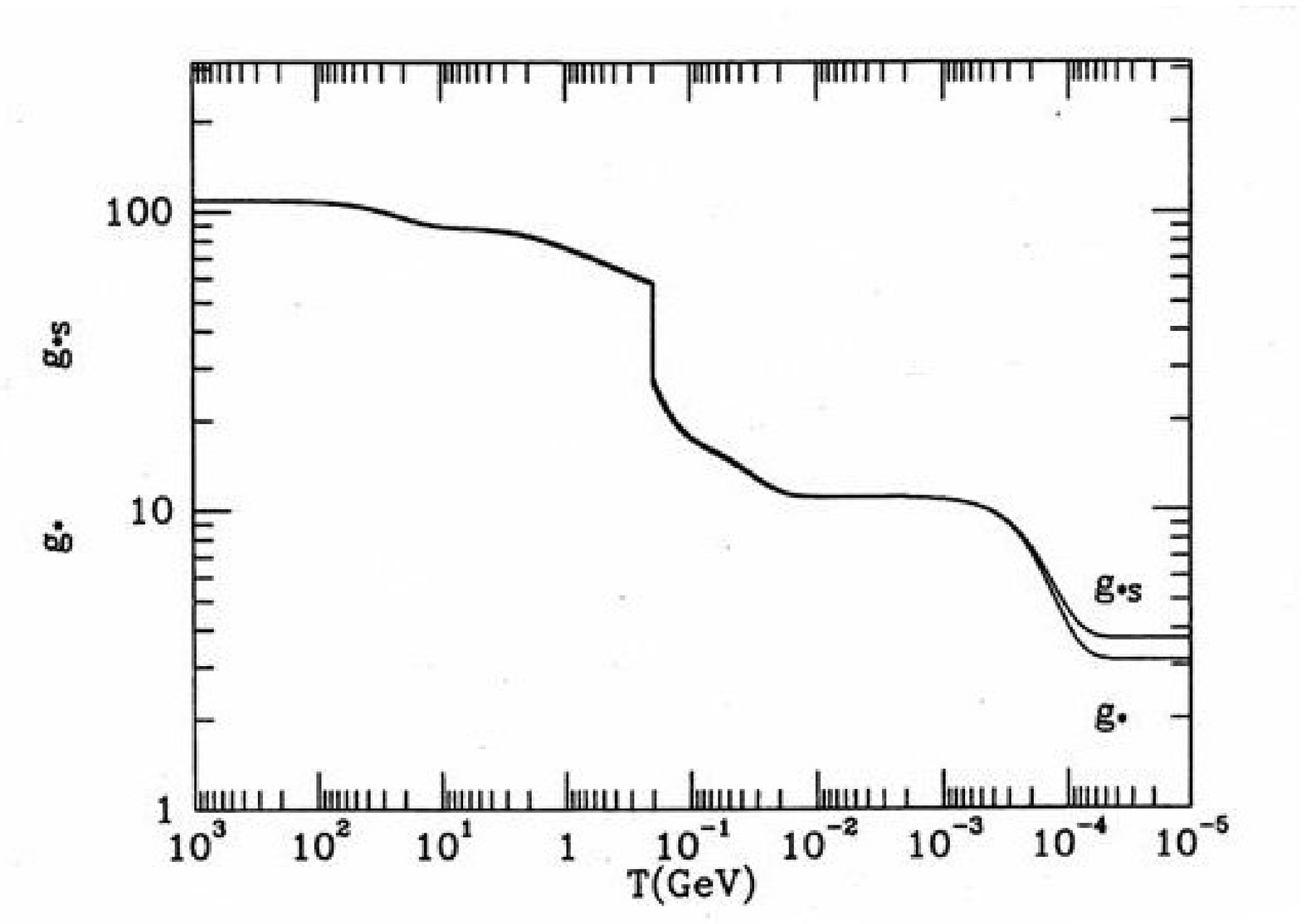}
%\vspace*{-5mm}
\caption{the light degrees of freedom $g_*$ and $g_{*S}$ as a function of
the temperature of the universe. From Ref.~\cite{KT}.}
\end{center}
\label{fig49}
\end{figure}
%%%%%%%%%%%%%%%%%%%%%%%%%

For example, for $T\ll 1$ MeV, i.e. after the time of primordial Big
Bang Nucleosynthesis (BBN) and neutrino decoupling, the only
relativistic species are the 3 light neutrinos and the photons; since
the temperature of the neutrinos is $T_\nu = (4/11)^{1/3} T_\gamma =
1.90$ K, see below, we have $g_* = 2 + 3\times {7\over4} \times
\Big({4\over11} \Big)^{4/3} = 3.36$, while $g_{*S} = 2 + 3 \times
{7\over4} \times\Big({4\over11}\Big) = 3.91$.

For 1 MeV $\ll T\ll 100$ MeV, i.e. between BBN and the phase
transition from a quark-gluon plasma to hadrons and mesons, we have,
as relativistic species, apart from neutrinos and photons, also the
electrons and positrons, so $g_* = 2+3\times{7\over4}+2\times{7\over4}
= 10.75$.

For $T\gg 250$ GeV, i.e. above the electroweak (EW) symmetry breaking
scale, we have one photon (2 polarizations), 8 gluons (massless), the
$W^\pm$ and $Z^0$ (massive), 3 families of quarks \& leptones, a Higgs
(still undiscovered), with which one finds $g_* = {427\over4} =
106.75$.

At temperatures well above the electroweak transition we ignore the
number of d.o.f. of particles, since we have never explored those
energies in particle physics accelerators. Perhaps in the near future,
with the results of the Large Hadron Collider (LHC) at CERN, we may
may predict the behaviour of the universe at those energy scales. For
the moment we even ignore whether the universe was in thermal
equilibrium at those temperatures. The highest energy scale at which
we can safely say the universe was in thermal equilibrium is that of
BBN, i.e. 1 MeV, due to the fact that we observe the present relative
abundances of the light element produced at that time. For instance,
we can't even claim that the universe went through the quark-gluon
phase transtion, at $\sim$ 200 MeV, since we have not observed yet any
signature of such an event, not to mention the electroweak phase
transition, at $\sim$ 1 TeV.

Let us now use the relation between the rate of expansion and the
temperature of relativistic particles to obtain the time scale of the
universe as a function of its temperature,
\begin{equation}
H = 1.66\,g_*^{1/2}\,{T^2\over M_P} = {1\over 2t} \hspace{1cm}
\Longrightarrow \hspace{1cm} t = 0.301\,g_*^{-1/2}\,{M_P\over T^2}
\sim \left({T\over {\rm MeV}}\right)^{-2} \ {\rm s}\,,
\end{equation}
thus, e.g. at the EW scale (100 GeV) the universe was just $10^{-10}$ s
old, while during the primordial BBN ($1 - 0.1$ MeV), it was 1 s to 
3 min old.

\subsubsection{The entropy of the universe}

During most of the history of the universe, the rates of reaction,
$\Gamma_{\rm int}$, of particles in the thermal bath are much bigger
than the rate of expansion of the universe, $H$, so that local thermal
equilibrium was mantained. In this case, the entropy per comoving
volume remained constant. In an expanding universe, the second law of
thermodynamics, applied to the element of comoving volume, of unit
coordinate volume and physical volume $V=a^3$, can be written as, see
(\ref{SecondLawThermodynamics}),
\begin{equation}
T\,dS = d(\rho\,V) + p\,dV = d[(\rho+p)V] - V\,dp\,.
\end{equation}
Using the Maxwell condition of integrability,
${\displaystyle {\partial^2S\over\partial T\partial V} =
{\partial^2S\over\partial V\partial T}}$, we find that
$dp = (\rho+p)dT/T$, so that
\begin{equation}
dS = d\left[(\rho+p){V\over T} + {\rm const.}\right]\,,
\end{equation}
i.e. the entropy in a comoving volume is $\ S = (\rho+p){V\over T}$,
except for a constant. Using now the first law, the covariant
conservation of energy, $T^{\mu\nu}_{\ \ ;\nu}=0$, we have
\begin{equation}
d\Big[(\rho+p)V\Big] = V\,dp \hspace{1cm}
\Longrightarrow \hspace{1cm} d\Big( (\rho+p){V\over T}\Big) = 0\,,
\end{equation}
and thus, in thermal equilibrium, the total entropy in a comoving
volume, $S=a^3(\rho+p)/T$, is conserved. During most of the evolution
of the universe, this entropy was dominated by the contribution from
relativistic particles,
\begin{eqnarray}\label{ConstantEntropy}
&&S = {2\pi^2\over45}\,g_{*S}\,(aT)^3 = {\rm const.}\,,\\[2mm]
&&g_*(T) = \sum_{\rm bosons} g_i\,\left({T_i\over T}\right)^3
+ {7\over8}\,\sum_{\rm fermions} g_i\,\left({T_i\over T}\right)^3\,,
\end{eqnarray}
where $g_{*S}$ is the number of ``entropic'' degrees of freedom, as we
can see in Fig.~7. Above the electron-positron annihilation, all
relativistic particles had the same temperature and thus $g_{*S}=g_*$.
It may be also useful to realize that the entropy density, $s=S/a^3$,
is propotional to the number density of relativistic particles, and in
particular to the number density of photons, $s=1.80g_{*S}\,n_\gamma$;
today, $s=7.04\,n_\gamma$. However, since $g_{*S}$ in general is a
function of temperature, we can't always interchange $s$ and $n_\gamma$.

The conservation of $S$ implies that the entropy density satisfies
$s\propto a^{-3}$, and thus the physical size of the comoving volume
is $a^3 \propto s^{-1}$; therefore, the number of particles of a given
species in a comoving volume, $N = a^3 n$, is proportional to the
number density of that species over the entropy density $s$,
\begin{equation}\label{NumberDensityEntropy}
N \sim {n\over s} = \left\{
\begin{array}{ll} {\displaystyle {45\zeta(3)\,g\over2\pi^4\,g_{*S}} } 
& \hspace{1cm} T \gg  m,\ \mu\\[4mm]
{\displaystyle {45\,g\over4\pi^5\sqrt{2}\,g_{*S}}\,
\Big({m\over T}\Big)^{3/2}
\,e^{-{m-\mu\over T}} } & \hspace{1cm} T \ll  m\end{array}\right.
\end{equation}
If this number does not change, i.e. if those particles are neither
created nor destroyed, then $\,n/s\,$ remains constant. As a useful
example, we will consider the barionic number in a comoving volume,
\begin{equation}
{n_B\over s} \equiv {n_b - n_{\bar b}\over s}\,.
\end{equation}
As long as the interactions that violate barion number occur
sufficiently slowly, the barionic number per comoving volume,
$n_B/s$, will remain constant. Although
\begin{equation}
\eta \equiv {n_B\over n_\gamma} = 1.80\,g_{*S}\,{n_B\over s}\,,
\end{equation}
the ratio between barion and photon numbers it does not remain
constant during the whole evolution of the universe since $g_{*S}$
varies; e.g. during the annihilation of electrons and positrons, the
number of photons per comoving volume, $N_\gamma = a^3\,n_\gamma$,
grows a factor $11/4$, and $\eta$ decreases by the same factor. After
this epoch, however, $g_*$ is constant so that $\eta \simeq 7 n_B/s$
and $n_B/s$ can be used indistinctly.

Another consequence of Eq,~(\ref{ConstantEntropy}) is that $S=$
const. implies that the temperature of the universe evolves as
\begin{equation}\label{Tredshift}
T \propto g_{*S}^{-1/3}\,a^{-1}\,.
\end{equation}
As long as $g_{*S}$ remains constant, we recover the well known result
that the universe cools as it expands according to $T\propto 1/a$. The
factor $g_{*S}^{-1/3}$ appears because when a species becomes
non-relativistic (when $T\leq m$), and effectively disappears from the
energy density of the universe, its entropy is transferred to the rest
of the relativistic particles in the plasma, making $T$ decrease not
as quickly as $1/a$, until $g_{*S}$ again becomes constant.

From the observational fact that the universe expands today one can
deduce that in the past it must have been hotter and denser, and that
in the future it will be colder and more dilute. Since the ratio of
scale factors is determined by the redshift parameter $z$, we can
obtain (to very good approximation) the temperature of the universe
in the past with
\begin{equation}\label{TemperatureRedshift}
T = T_0\,(1+z)\,.
\end{equation}
This expression has been spectacularly confirmed thanks to the
absorption spectra of distant quasars~\cite{QSO}. These spectra
suggest that the radiation background was acting as a thermal bath
for the molecules in the interstellar medium with a temperature of
9 K at a redshift $z\sim2$, and thus that in the past the photon
background was hotter than today. Furthermore, observations of the
anisotropies in the microwave background confirm that the universe
at a redshift $z=1089$ had a temperature of $0.3$ eV, in
agreement with Eq.~(\ref{TemperatureRedshift}).

\subsection{The thermal evolution of the universe}

In a strict mathematical sense, it is impossible for the universe to
have been always in thermal equilibrium since the FRW model does not
have a timelike Killing vector. In practice, however, we can say that
the universe has been most of its history very close to thermal
equilibrium. Of course, those periods in which there were deviations
from thermal equilibrium have been crucial for its evolution
thereafter (e.g. baryogenesis, QCD transition, primordial
nucleosynthesis, recombination, etc.); without these the universe
today would be very different and probably we would not be here to
tell the story.

The key to understand the thermal history of the universe is the
comparison between the rates of interaction between particles
(microphysics) and the rate of expansion of the universe
(macrophysics). Ignoring for the moment the dependence of $g_*$ on
temperature, the rate of change of $T$ is given directly by the rate
of expansion, $\dot T/T = -H$. As long as the local interactions
$-$ necessary in order that the particle distribution function
adjusts {\em adiabatically}  to the change of temperature $-$
are sufficiently fast compared with the rate of expansion of the
universe, the latter will evolve as a succession of states very close
to thermal equilibrium, with a temperature proportional to $a^{-1}$.
If we evaluate the interaction rates as
\begin{equation}\label{Gammaint}
\Gamma_{\rm int} \equiv \langle n\,\sigma\,|v|\rangle\,,
\end{equation}
where $n(t)$ is the number density of target particles, $\sigma$ is
the cross section on the interaction and $v$ is the relative velocity
of the reaction, all averaged on a thermal distribution; then a rule
of thumb for ensuring that thermal equilibrium is maintained is
\begin{equation}\label{ThermalEq}
\Gamma_{\rm int} \gsim H\,.
\end{equation}
This criterium is understandable. Suppose, as often occurs, that the
interaction rate in thermal equilibrium is $\Gamma_{\rm int}
\propto T^n$, with $n>2$; then, the number of interactions of a
particle after time $t$ is
\begin{equation}\label{Nint}
N_{\rm int} = \int_t^\infty \Gamma_{\rm int}(t')dt' = {1\over n-2}\,
{\Gamma_{\rm int}\over H}(t)\,,
\end{equation}
therefore the particle interacts less than once from the moment in
which $\Gamma_{\rm int}\approx H$. If $\Gamma_{\rm int}\gsim H$, the
species remains coupled to the thermal plasma. This doesn't mean that,
necessarily, the particle is out of local thermal equilibrium, since
we have seen already that relativistic particles that have decoupled
retain their equilibrium distribution, only at a different temperature
from that of the rest of the plasma.

In order to obtain an approximate description of the decoupling of a
particle species in an expanding universe, let us consider two types
of interaction:

\noindent i) interactions mediated by massless gauge bosons, like for
example the photon. In this case, the cross section for particles with
significant momentum transfer can be written as $\sigma \sim
\alpha^2/T^2$, with $\alpha = g^2/4\pi$ the coupling constant of the
interaction. Assuming local thermal equilibrium, $n(t) \sim T^3$ and
thus the interaction rate becomes $\Gamma \sim n\,\sigma\,|v| \sim
\alpha^2\,T$. Therefore, 
\begin{equation}\label{Gintsinmasa}
{\Gamma\over H}\sim \alpha^2\,{M_P\over T}\,,
\end{equation}
so that for temperatures of the universe $T \lsim \alpha^2\,M_P \sim
10^{16}$ GeV, the reactions are fast enough and the plasma is in
equilibrium, while for $T \gsim 10^{16}$ GeV, reactions are too slow
to maintain equilibrium and it is said that they are ``frozen-out''.
An important consequence of this result is that the universe could
never have been in thermal equilibrium above the grand unification
(GUT) scale.

\noindent ii) interactions mediated by massive gauge bosons, e.g. like
the $W^\pm$ and $Z^0$, or those responsible for the GUT interactions,
$X$ and $Y$. We will generically call them $X$ bosons. The cross
section depends rather strongly on the temperature of the plasma,
\begin{equation}\label{sigmaconmasa}
\sigma \sim \left\{
\begin{array}{ll} G_X^2 T^2 & \hspace{1cm} T \ll  M_X \\[4mm]
{\displaystyle {\alpha^2\over T^2} } & \hspace{1cm} T \gg  M_X
\end{array}\right.
\end{equation}
where $G_X\sim \alpha/M_X^2$ is the effective coupling constant of the
interaction at energies well below the mass of the vector boson,
analogous to the Fermi constant of the electroweak interaction, $G_F =
g^2/(4\sqrt{2}M_W^2)$ at tree level. Note that for $T \gg M_X$ we
recover the result for massless bosons, so we will concentrate here
on the other case. For $T \leq M_X$, the rate of thermal interactions
is $\Gamma \sim n\,\sigma\,|v| \sim G_X^2\,T^5$. Therefore,
\begin{equation}\label{Gintconmasa}
{\Gamma\over H}\sim G_X^2\,M_P\,T^3\,,
\end{equation}
such that at temperatures in the range
\begin{equation}
M_X \gsim T \gsim G_X^{-2/3}\, M_P^{-1/3} \sim \left({M_X\over100\
{\rm GeV}}\right)^{4/3}\ {\rm MeV}\,,
\end{equation}
reactions occur so fast that the plasma is in thermal
equilibrium, while for $T \lsim (M_X/100\ {\rm GeV})^{4/3}$ MeV, those
reactions are too slow for maintaining equilibrium and they effective
freeze-out, see Eq.~(\ref{Nint}).

\subsubsection{The decoupling of relativistic particles}

Those relativistic particles that have decoupled from the thermal bath
do not participate in the transfer of entropy when the temperature of
the universe falls below the mass thershold of a given species
$T\simeq m$; in fact, the temperature of the decoupled relativistic
species falls as $T\propto 1/a$, as we will now show. Suppose that a
relativistic particle is initially in local thermal equilibrium, and
that it decoples at a temperature $T_D$ and time $t_D$. The phase
space distribution at the time of decoupling is given by the equilibrium
distribution,
\begin{equation}
f({\bf p}, t_D) = {1\over e^{E/T_D}\pm1}\,.
\end{equation}
After decoupling, the energy of each massless particle suffers
redshift, $E(t) = E_D\,(a_D/a(t))$. The number density of particles
also decreases, $n(t) = n_D\, (a_D/a(t))^3$. Thus, the phase
space distribution at a time $t>t_D$ is
\begin{equation}
f({\bf p}, t) = {d^3n\over d^3{\bf p}} = f({\bf p}{a\over a_D}, t_D)
= {1\over e^{Ea/a_DT_D}\pm1} = {1\over e^{E/T}\pm1}\,,
\end{equation}
so that we conclude that the distribution function of a particle
that has decoupled while being relativistic remains self-similar
as the universe expands, with a temperature that decreases as
\begin{equation}
T = T_D\,{a_D\over a} \propto a^{-1}\,,
\end{equation}
and {\em not} as $g_{*S}^{-1/3}\,a^{-1}$, like the rest of the plasma
in equilibrium (\ref{Tredshift}).

\subsubsection{The decoupling of non-relativistic particles}

Those particles that decoupled from the thermal bath when they were
non-relativistic ($m\gg T$) behave differently. Let us study the
evolution of the distribution function of a non-relativistic particle
that was in local thermal equilibrium at a time $t_D$, when the
universe had a temperature $T_D$. The moment of each particle suffers
redshift as the universe expands, $|{\bf p}| = |{\bf p}_D|\,(a_D/a)$,
see Eq.~(\ref{redshift}). Therefore, their kinetic energy satisfies $E
= E_D\,(a_D/a)^2$. On the other hand, the particle number density also
varies, $n(t) = n_D\, (a_D/a(t))^3$, so that a decoupled
non-relativistic particle will have an equilibrium distribution
function characterized by a temperature
\begin{equation}
T = T_D\,{a_D^2\over a^2} \propto a^{-2}\,,
\end{equation}
and a chemical potential
\begin{equation}
\mu(t) = m + (\mu_D - m)\,{T\over T_D}\,,
\end{equation}
whose variation is precisely that which is needed for the number
density of particle to decrease as $a^{-3}$.

In summary, a particle species that decouples from the thermal bath
follows an equilibrium distribution function with a temperature that
decreases like $T_R\propto a^{-1}$ for relativistic particles ($T_D\gg
m$) or like $T_{NR}\propto a^{-2}$ for non-relativistic particles
($T_D\ll m$). On the other hand, for semi-relativistic particles
($T_D\sim m$), its phase space distribution {\em does not maintain} an
equilibrium distribution function, and should be computed case by
case.

\subsubsection{Brief thermal history of the universe}

I will briefly summarize here the thermal history of the
universe, from the Planck era to the present. As we go back in time, the
universe becomes hotter and hotter and thus the amount of energy
available for particle interactions increases. As a consequence, the
nature of interactions goes from those described at low energy by long
range gravitational and electromagnetic physics, to atomic physics,
nuclear physics, all the way to high energy physics at the electroweak
scale, gran unification (perhaps), and finally quantum gravity. The last
two are still uncertain since we do not have any experimental evidence
for those ultra high energy phenomena, and perhaps Nature has followed a
different path.

The way we know about the high energy interactions of matter is via
particle accelerators, which are unravelling the details of those
fundamental interactions as we increase in energy. However, one should
bear in mind that the physical conditions that take place in our high
energy colliders are very different from those that occurred in the
early universe. These machines could never reproduce the conditions of
density and pressure in the rapidly expanding thermal plasma of the
early universe. Nevertheless, those experiments are crucial in
understanding the nature and {\em rate} of the local fundamental
interactions available at those energies. What interests cosmologists is
the statistical and thermal properties that such a plasma should have,
and the role that causal horizons play in the final outcome of the early
universe expansion. For instance, of crucial importance is the time at
which certain particles {\em decoupled} from the plasma, i.e. when their
interactions were not quick enough compared with the expansion of the
universe, and they were left out of equilibrium with the plasma.

One can trace the evolution of the universe from its origin till today.
There is still some speculation about the physics that took place in the
universe above the energy scales probed by present colliders.
Nevertheless, the overall layout presented here is a plausible and
hopefully testable proposal. According to the best accepted view, the
universe must have originated at the Planck era ($10^{19}$ GeV,
$10^{-43}$ s) from a quantum gravity fluctuation. Needless to say, we
don't have any experimental evidence for such a statement: Quantum
gravity phenomena are still in the realm of physical speculation.
However, it is plausible that a primordial era of cosmological {\em
inflation} originated then. Its consequences will be discussed below.
Soon after, the universe may have reached the Grand Unified Theories
(GUT) era ($10^{16}$ GeV, $10^{-35}$ s). Quantum fluctuations of the
inflaton field most probably left their imprint then as tiny
perturbations in an otherwise very homogenous patch of the universe. At
the end of inflation, the huge energy density of the inflaton field was
converted into particles, which soon thermalized and became the origin
of the hot Big Bang as we know it. Such a process is called {\em
reheating} of the universe. Since then, the universe became radiation
dominated. It is probable (although by no means certain) that the
asymmetry between matter and antimatter originated at the same time as
the rest of the energy of the universe, from the decay of the
inflaton. This process is known under the name of {\em baryogenesis}
since baryons (mostly quarks at that time) must have originated then,
from the leftovers of their annihilation with antibaryons. It is a
matter of speculation whether baryogenesis could have occurred at
energies as low as the electroweak scale ($100$ GeV, $10^{-10}$ s).
Note that although particle physics experiments have reached energies as
high as 100 GeV, we still do not have observational evidence that the
universe actually went through the EW phase transition. If confirmed,
baryogenesis would constitute another ``window'' into the early
universe. As the universe cooled down, it may have gone through the
quark-gluon phase transition ($10^2$ MeV, $10^{-5}$ s), when baryons
(mainly protons and neutrons) formed from their constituent quarks.

The furthest window we have on the early universe at the moment is that
of {\em primordial nucleosynthesis} ($1 - 0.1$ MeV, 1 s -- 3 min), when
protons and neutrons were cold enough that bound systems could form,
giving rise to the lightest elements, soon after {\em neutrino
decoupling}: It is the realm of nuclear physics. The observed relative
abundances of light elements are in agreement with the predictions of
the hot Big Bang theory. Immediately afterwards, electron-positron
annihilation occurs (0.5 MeV, 1 min) and all their energy goes into
photons. Much later, at about (1 eV, $\sim 10^5$ yr), matter and
radiation have equal energy densities. Soon after, electrons become
bound to nuclei to form atoms (0.3 eV, $3\times10^5$ yr), in a process
known as {\em recombination}: It is the realm of atomic physics.
Immediately after, photons decouple from the plasma, travelling freely
since then. Those are the photons we observe as the cosmic microwave
background. Much later ($\sim 1-10$ Gyr), the small inhomogeneities
generated during inflation have grown, via gravitational collapse, to
become galaxies, clusters of galaxies, and superclusters, characterizing
the epoch of {\em structure formation}. It is the realm of long range
gravitational physics, perhaps dominated by a vacuum energy in the form
of a cosmological constant. Finally (3K, 13 Gyr), the Sun, the Earth,
and biological life originated from previous generations of stars, and
from a primordial soup of organic compounds, respectively.

I will now review some of the more robust features of the Hot Big Bang
theory of which we have precise observational evidence.

\subsubsection{Primordial nucleosynthesis and light element abundance}

In this subsection I will briefly review Big Bang nucleosynthesis and
give the present observational constraints on the amount of baryons in
the universe. In 1920 Eddington suggested that the sun might derive its
energy from the fusion of hydrogen into helium. The detailed reactions
by which stars burn hydrogen were first laid out by Hans Bethe in 1939.
Soon afterwards, in 1946, George Gamow realized that similar processes
might have occurred also in the hot and dense early universe and gave
rise to the first light elements~\cite{Gamow}. These processes could
take place when the universe had a temperature of around $T_{_{\rm NS}}
\sim 1-0.1$ MeV, which is about 100 times the temperature in the core of
the Sun, while the density is $\rho_{_{\rm NS}} =
{\pi^2\over30}g_*T_{_{\rm NS}}^4\sim 82$ g\,cm$^{-3}$, about the same
density as the core of the Sun. Note, however, that although both
processes are driven by identical thermonuclear reactions, the physical
conditions in star and Big Bang nucleosynthesis are very different. In
the former, gravitational collapse heats up the core of the star and
reactions last for billions of years (except in supernova explosions,
which last a few minutes and creates all the heavier elements beyond
iron), while in the latter the universe expansion cools the hot and
dense plasma in just a few minutes. Nevertheless, Gamow reasoned that,
although the early period of cosmic expansion was much shorter than the
lifetime of a star, there was a large number of free neutrons at that
time, so that the lighter elements could be built up quickly by
succesive neutron captures, starting with the reaction \ $n + p
\rightarrow D + \gamma$. The abundances of the light elements would then
be correlated with their neutron capture cross sections, in rough
agreement with observations~\cite{Weinberg,Burles}.

%%%%%%%%%%%%%%%%%%%%%%%%%
\begin{figure}[htb]
\begin{center}
\includegraphics[width=8.5cm]{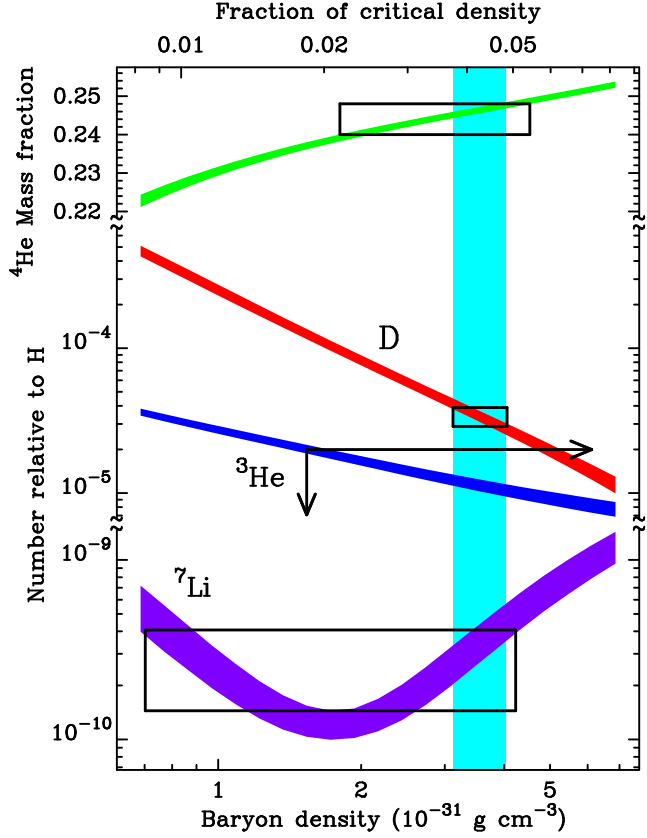} 
\caption{The relative abundance of light elements to Hidrogen. Note the
large range of scales involved. From Ref.~\cite{Burles}.}
\label{fig3}
\end{center}
\end{figure}
%%%%%%%%%%%%%%%%%%%%%%%%%

Nowadays, Big Bang nucleosynthesis (BBN) codes compute a chain of
around 30 coupled nuclear reactions~\cite{BBN}, to produce all the
light elements up to beryllium-7.~\footnote{The rest of nuclei, up to
iron (Fe), are produced in heavy stars, and beyond Fe in novae and
supernovae explosions.} Only the first four or five elements can be
computed with accuracy better than 1\% and compared with cosmological
observations. These light elements are $H, {}^4\!He, D, {}^3\!He,
{}^7\!Li$, and perhaps also ${}^6\!Li$. Their observed relative
abundance to hydrogen is
$[1:0.25:3\cdot10^{-5}:2\cdot10^{-5}:2\cdot10^{-10}]$ with various
errors, mainly systematic. The BBN codes calculate these abundances
using the laboratory measured nuclear reaction rates, the decay rate
of the neutron, the number of light neutrinos and the homogeneous FRW
expansion of the universe, as a function of {\em only} one variable,
the number density fraction of baryons to photons, $\eta\equiv n_{\rm
B}/n_\gamma$.  In fact, the present observations are only consistent,
see Fig.~8 and Ref.~\cite{Burles,BBN,PDG}, with a very narrow
range of values of
\begin{equation}\label{EtaBaryon}
\eta_{10} \equiv 10^{10}\,\eta = 6.2 \pm 0.6 \,.
\end{equation}
Such a small value of $\eta$ indicates that there is about one baryon
per $10^9$ photons in the universe today. Any acceptable theory of
baryogenesis should account for such a small number. Furthermore, the
present baryon fraction of the critical density can be calculated from
$\eta_{10}$ as
\begin{equation}\label{OmegaBaryon}
\Omega_{\rm B}h^2 = 3.6271\times 10^{-3}\,\eta_{10} = 
0.0224 \pm 0.0022 \hspace{5mm} (95\%\ {\rm c.l.})
\end{equation}
Clearly, this number is well below closure density, so baryons cannot
account for all the matter in the universe, as I shall discuss below.

\subsubsection{Neutrino decoupling}

Just before the nucleosynthesis of the lightest elements in the early
universe, weak interactions were too slow to keep neutrinos in thermal
equilibrium with the plasma, so they decoupled. We can estimate the
temperature at which decoupling occurred from the weak interaction
cross section, $\sigma_{\rm w} \simeq G_F^2 T^2$ at finite temperature
$T$, where $G_F=1.2\times10^{-5}$ GeV$^{-2}$ is the Fermi constant.  The
neutrino interaction rate, via W boson exchange in \ $n+\nu
\leftrightarrow p+e^-$ and \ $p+\bar\nu\leftrightarrow n+e^+$, can
be written as~\cite{KT}
\begin{equation}\label{NeutrinoInteraction}
\Gamma_\nu = n_\nu\langle\sigma_{\rm w}|v|\rangle\simeq G_F^2T^5\,,
\end{equation}
while the rate of expansion of the universe at that time ($g_*=10.75$)
was $H\simeq5.4\ T^2/M_{\rm P}$, where $M_{\rm P} = 1.22\times10^{19}$
GeV is the Planck mass. Neutrinos decouple when their interaction rate
is slower than the universe expansion, $\Gamma_\nu \leq H$ or,
equivalently, at $T_{\nu-{\rm dec}} \simeq 0.8$ MeV. Below this
temperature, neutrinos are no longer in thermal equilibrium with the
rest of the plasma, and their temperature continues to decay inversely
proportional to the scale factor of the universe. Since neutrinos
decoupled before $e^+e^-$ annihilation, the cosmic background of
neutrinos has a temperature today lower than that of the microwave
background of photons. Let us compute the difference. At temperatures
above the the mass of the electron, $T>m_e = 0.511$ MeV, and below 0.8
MeV, the only particle species contributing to the entropy of the
universe are the photons ($g_*=2$) and the electron-positron pairs
($g_*=4\times {7\over8}$); total number of degrees of freedom
$g_*={11\over2}$. At temperatures $T\simeq m_e$, electrons and positrons
annihilate into photons, heating up the plasma (but not the neutrinos,
which had decoupled already). At temperatures $T< m_e$, only photons
contribute to the entropy of the universe, with $g_*=2$ degrees of
freedom. Therefore, from the conservation of entropy, we find that the
ratio of $T_\gamma$ and $T_\nu$ today must be
\begin{equation}\label{NeutrinoTemperature}
{T_\gamma\over T_\nu} = \Big({11\over4}\Big)^{1/3} = 1.401
\hspace{5mm} \Rightarrow \hspace{5mm} T_\nu = 1.945\ {\rm K}\,,
\end{equation}
where I have used $T_{_{\rm CMB}} = 2.725\pm0.002$ K. We still have not
measured such a relic background of neutrinos, and probably will remain
undetected for a long time, since they have an average energy of order
$10^{-4}$ eV, much below that required for detection by present
experiments (of order GeV), precisely because of the relative
weakness of the weak interactions. Nevertheless, it would be fascinating
if, in the future, ingenious experiments were devised to detect such a
background, since it would confirm one of the most robust features of
Big Bang cosmology. 

\subsubsection{Matter-radiation equality}

Relativistic species have energy densities proportional to the quartic
power of temperature and therefore scale as $\rho_{\rm R}\propto
a^{-4}$, while non-relativistic particles have essentially zero pressure
and scale as $\rho_{\rm M}\propto a^{-3}$. Therefore, there will be a
time in the evolution of the universe in which both energy densities are
equal $\rho_{\rm R}(t_{\rm eq})=\rho_{\rm M}(t_{\rm eq})$.  Since then
both decay differently, and thus
\begin{equation}\label{Equality}
1+z_{\rm eq} = {a_0\over a_{\rm eq}} = {\Omega_{\rm M}\over\Omega_{\rm
R}} = 3.1\times10^4\ \Omega_{\rm M} h^2\,,
\end{equation}
where I have used $\Omega_{\rm R} h^2= \Omega_{_{\rm CMB}} h^2 +
\Omega_\nu h^2 = 3.24\times10^{-5}$ for three massless neutrinos at
$T=T_\nu$.  As I will show later, the matter content of the universe
today is below critical, $\Omega_{\rm M} \simeq 0.3$, while
$h\simeq0.71$, and therefore $(1+z_{\rm eq}) \simeq 3400$, or about
$t_{\rm eq} = 1308\, (\Omega_{\rm M} h^2)^{-2} {\rm yr} \simeq
61,000$ years after the origin of the universe. Around the time of
matter-radiation equality, the rate of expansion (\ref{H2a}) can be
written as ($a_0\equiv1$)
\begin{equation}\label{RateEquality}
H(a) = H_0\,\Big(\Omega_{\rm R}\,a^{-4} + \Omega_{\rm M}\,a^{-3}\Big)^{1/2} =
H_0\,\Omega_{\rm M}^{1/2}\,a^{-3/2}\Big(1+{a_{\rm eq}\over a}\Big)^{1/2}\,.
\end{equation}
The {\em horizon size} is the coordinate distance travelled by a
photon since the beginning of the universe, $d_H\sim H^{-1}$, i.e.
the size of causally connected regions in the universe. The
{\em comoving} horizon size is then given by
\begin{equation}\label{HorizonSize}
d_H={c\over aH(a)}=c\,H_0^{-1}\Omega_{\rm M}^{-1/2}\,a^{1/2}
\Big(1+{a_{\rm eq}\over a}\Big)^{-1/2}\,.
\end{equation}
Thus the horizon size at matter-radiation equality ($a=a_{\rm eq}$) is
\begin{equation}\label{HorizonSizeEquality}
d_H(a_{\rm eq})={c\,H_0^{-1}\over\sqrt2}\,\Omega_{\rm M}^{-1/2}\,
a_{\rm eq}^{1/2} \simeq 12\,(\Omega_{\rm M} h)^{-1}\,h^{-1}{\rm Mpc}\,.
\end{equation}
This scale plays a very important role in theories of structure
formation.

\subsubsection{Recombination and photon decoupling}

As the temperature of the universe decreased, electrons could eventually
become bound to protons to form neutral hydrogen. Nevertheless, there is
always a non-zero probability that a rare energetic photon ionizes
hydrogen and produces a free electron. The {\em ionization fraction} of
electrons in equilibrium with the plasma at a given temperature is given
by the Saha equation~\cite{KT}
\begin{equation}\label{IonizationFraction}
{1-X_e^{\rm eq}\over X_e^{\rm eq}} = {4\sqrt2\zeta(3)\over\sqrt\pi}\,
\eta\,\left({T\over m_e}\right)^{3/2}\,e^{E_{\rm ion}/T}\,,
\end{equation}
where $E_{\rm ion} = 13.6$ eV is the ionization energy of hydrogen, and
$\eta$ is the baryon-to-photon ratio (\ref{EtaBaryon}). If we now use
Eq.~(\ref{TemperatureRedshift}), we can compute the ionization fraction
$X_e^{\rm eq}$ as a function of redshift $z$. Note that the huge number
of photons with respect to electrons (in the ratio $^{4}\!He: H: \gamma
\simeq 1:4:10^{10}$) implies that even at a very low temperature, the
photon distribution will contain a sufficiently large number of
high-energy photons to ionize a significant fraction of hydrogen. In
fact, {\em defining} recombination as the time at which $X_e^{\rm
eq}\equiv0.1$, one finds that the recombination temperature is $T_{\rm
rec} = 0.31\ {\rm eV} \ll E_{\rm ion}$, for $\eta_{10}\simeq
6.2$. Comparing with the present temperature of the microwave
background, we deduce the corresponding redshift at recombination,
$(1+z_{\rm rec}) \simeq 1331$.

Photons remain in thermal equilibrium with the plasma of baryons and
electrons through elastic Thomson scattering, with cross section
\begin{equation}\label{ThomsonCrossSection}
\sigma_{_T} = {8\pi\alpha^2\over3m_e^2} = 6.65\times 10^{-25} \
{\rm cm}^2 =0.665\ {\rm barn}\,,
\end{equation}
where $\alpha=1/137.036$ \ is the dimensionless electromagnetic
coupling constant. The mean free path of photons $\lambda_\gamma$ in
such a plasma can be estimated from the photon interaction rate,
$\lambda_\gamma^{-1}\sim \Gamma_\gamma = n_e\sigma_{_T}$. For
temperatures above a few eV, the mean free path is much smaller that
the causal horizon at that time and photons suffer multiple
scattering: the plasma is like a dense fog. Photons will decouple from
the plasma when their interaction rate cannot keep up with the
expansion of the universe and the mean free path becomes larger than
the horizon size: the universe becomes transparent. We can estimate
this moment by evaluating $\Gamma_\gamma = H$ at photon decoupling.
Using $n_e=X_e\,\eta\,n_\gamma$, one can compute the decoupling
temperature as $T_{\rm dec} = 0.26$ eV, and the corresponding redshift
as $1+z_{\rm dec} \simeq 1100$. Recently, WMAP measured this redshift
to be $1+z_{\rm dec} \simeq 1089\pm1$~\cite{WMAP}. This redshift
defines the so called {\em last scattering surface}, when photons last
scattered off protons and electrons and travelled freely ever
since. This decoupling occurred when the universe was approximately
$t_{\rm dec} = 1.5\times 10^5\,(\Omega_{\rm M} h^2)^{-1/2} \simeq
380,000$ years old.

%%%%%%%%%%%%%%%%%%%%%%%%%
\begin{figure}[htb]
\begin{center}\hspace{-1.8cm}
\includegraphics[width=6cm,angle=90]{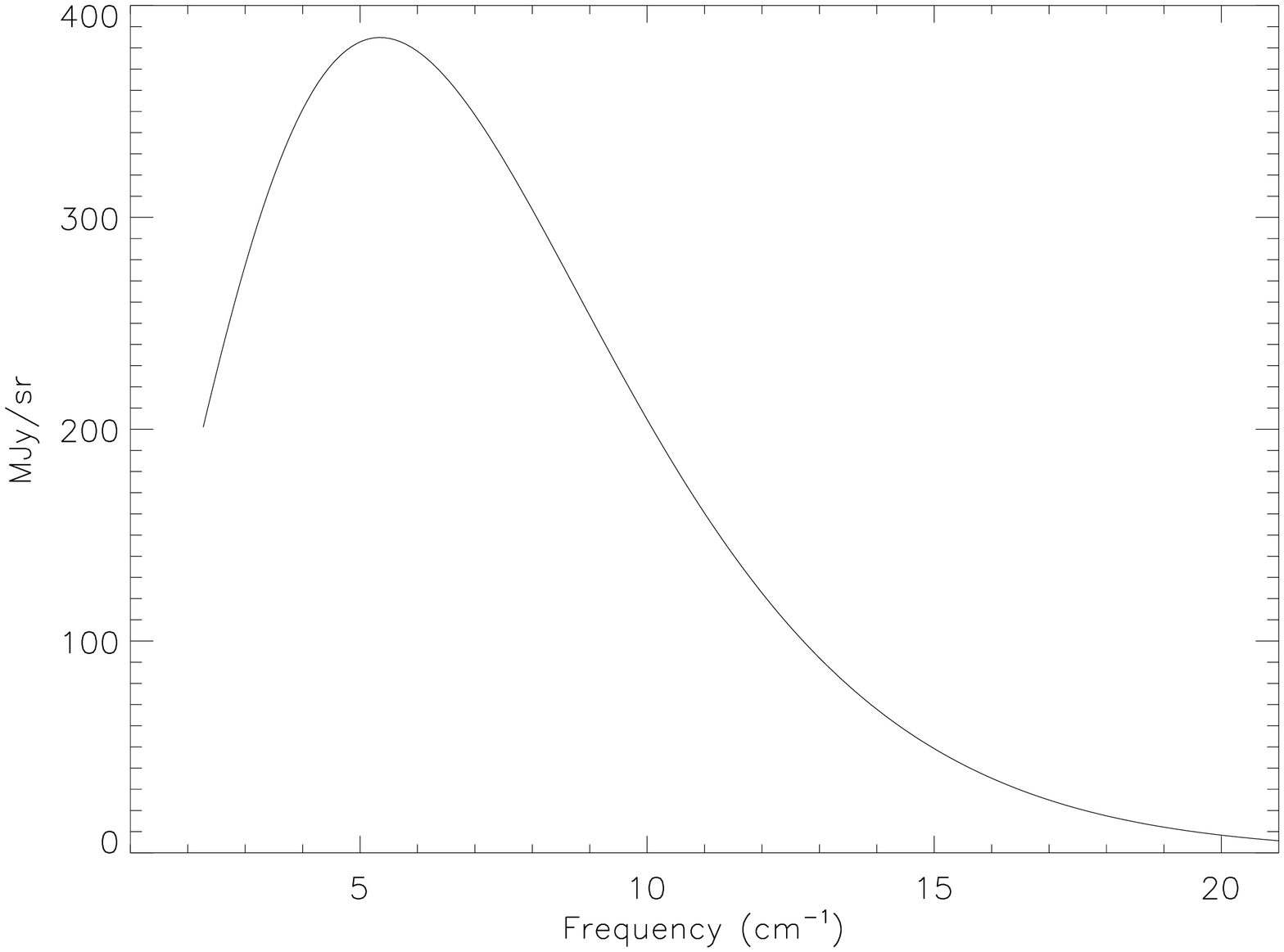}
\includegraphics[width=6cm,angle=90]{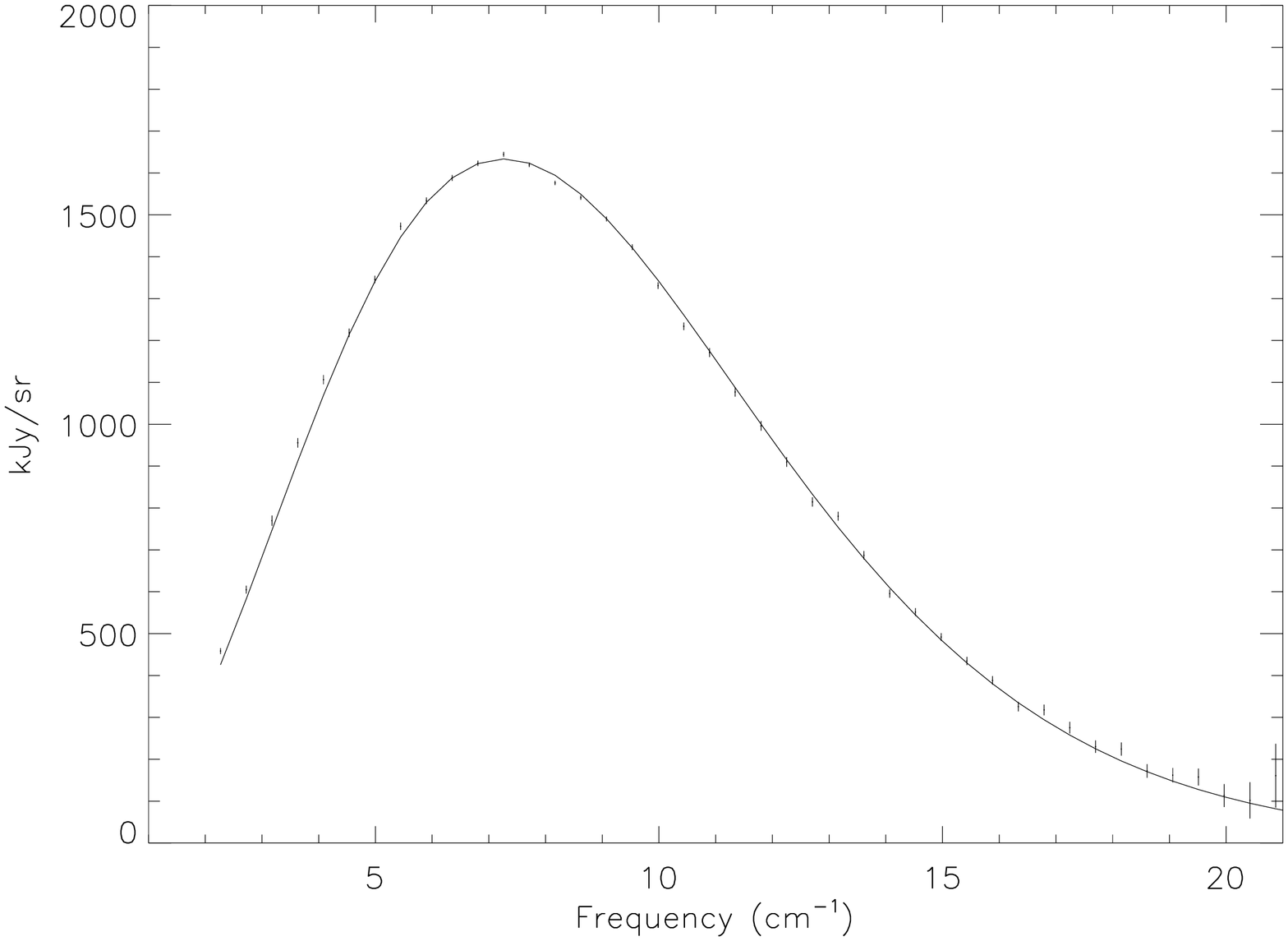}
\end{center}\vspace*{-5mm}
\caption{The Cosmic Microwave Background Spectrum seen by the FIRAS
instrument on COBE. The left panel corresponds to the monopole spectrum,
$T_0 = 2.725\pm0.002$ K, where the error bars are smaller than the 
line width. The right panel shows the dipole spectrum, $\delta T_1 =
3.372\pm0.014$ mK. From Ref.~\cite{FIRAS}.}
\label{fig5}
\end{figure}
%%%%%%%%%%%%%%%%%%%%%%%%%

\subsubsection{The microwave background}

One of the most remarkable observations ever made my mankind is the
detection of the relic background of photons from the Big Bang. This
background was predicted by George Gamow and collaborators in the 1940s,
based on the consistency of primordial nucleosynthesis with the observed
helium abundance. They estimated a value of about 10 K, although a
somewhat more detailed analysis by Alpher and Herman in 1950 predicted
$T_\gamma \approx 5$ K. Unfortunately, they had doubts whether the
radiation would have survived until the present, and this remarkable
prediction slipped into obscurity, until Dicke, Peebles, Roll and
Wilkinson~\cite{Dicke} studied the problem again in 1965. Before they
could measure the photon background, they learned that Penzias and
Wilson had observed a weak isotropic background signal at a radio
wavelength of 7.35 cm, corresponding to a blackbody temperature of
$T_\gamma=3.5\pm1$ K. They published their two papers back to back, with
that of Dicke et al.  explaining the fundamental significance of their
measurement~\cite{Weinberg}.

Since then many different experiments have confirmed the existence of
the microwave background. The most outstanding one has been the Cosmic
Background Explorer (COBE) satellite, whose FIRAS instrument measured
the photon background with great accuracy over a wide range of
frequencies ($\nu = 1-97$ cm$^{-1}$), see Ref.~\cite{FIRAS}, with a
spectral resolution ${\Delta\nu\over\nu}=0.0035$. Nowadays,
the photon spectrum is confirmed to be a blackbody spectrum with a
temperature given by~\cite{FIRAS}
\begin{equation}\label{T0}
T_{_{\rm CMB}} = 2.725 \pm 0.002 \ {\rm K} \ 
({\rm systematic}, \ 95\%\ {\rm c.l.}) \ 
\pm 7 \ \mu{\rm K} \ (1\sigma\ {\rm statistical})
\end{equation}
In fact, this is the best blackbody spectrum ever measured, see
Fig.~\ref{fig5}, with spectral distortions below the level of 10 parts
per million (ppm).

%%%%%%%%%%%%%%%%%%%%%%%%%
\begin{figure}[htb]
\vspace*{-5mm}
\begin{center}
\includegraphics[width=9cm,angle=0]{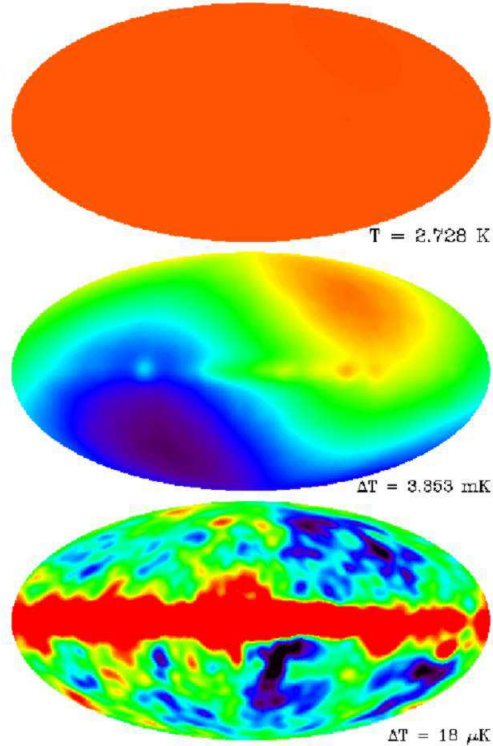}
\end{center}
\vspace*{-1cm}
\caption{The Cosmic Microwave Background Spectrum seen by the DMR
instrument on COBE. The top figure corresponds to the monopole, $T_0 =
2.725\pm0.002$ K. The middle figure shows the dipole, $\delta T_1 =
3.372\pm0.014$ mK, and the lower figure shows the quadrupole and higher
multipoles, $\delta T_2 = 18\pm2 \ \mu$K. The central region corresponds
to foreground by the galaxy. From Ref.~\cite{DMR}.}
\label{fig6}
\end{figure}
%%%%%%%%%%%%%%%%%%%%%%%%%

Moreover, the differential microwave radiometer (DMR) instrument on
COBE, with a resolution of about $7^\circ$ in the sky, has also
confirmed that it is an extraordinarily isotropic background. The
deviations from isotropy, i.e. differences in the temperature of the
blackbody spectrum measured in different directions in the sky, are of
the order of 20\,$\mu$K on large scales, or one part in $10^5$, see
Ref.~\cite{DMR}.  There is, in fact, a dipole anisotropy of one part in
$10^3$, $\delta T_1 = 3.372\pm0.007$ mK (95\% c.l.), in the direction of
the Virgo cluster, $(l,b) = (264.14^\circ \pm 0.30, 48.26^\circ \pm
0.30)$ (95\% c.l.). Under the assumption that a Doppler effect is
responsible for the entire CMB dipole, the velocity of the Sun with
respect to the CMB rest frame is $v_\odot=371\pm0.5$ km/s, see
Ref.~\cite{FIRAS}.\footnote{COBE even determined the annual variation
due to the Earth's motion around the Sun -- the ultimate proof of
Copernicus' hypothesis.} When subtracted, we are left with a whole
spectrum of anisotropies in the higher multipoles (quadrupole, octupole,
etc.), $\delta T_2 = 18\pm2 \ \mu$K (95\% c.l.), see Ref.~\cite{DMR} and
Fig.~\ref{fig6}.

Soon after COBE, other groups quickly confirmed the detection of
temperature anisotropies at around 30\,$\mu$K and above, at higher
multipole numbers or smaller angular scales. As I shall discuss below,
these anisotropies play a crucial role in the understanding of the
origin of structure in the universe.

\subsubsection{Large-scale structure formation}

Although the isotropic microwave background indicates that the universe
in the {\em past} was extraordinarily homogeneous, we know that the
universe {\em today} is not exactly homogeneous: we observe galaxies,
clusters and superclusters on large scales. These structures are
expected to arise from very small primordial inhomogeneities that grow
in time via gravitational instability, and that may have originated from
tiny ripples in the metric, as matter fell into their troughs. Those
ripples must have left some trace as temperature anisotropies in the
microwave background, and indeed such anisotropies were finally
discovered by the COBE satellite in 1992. The reason why they took so
long to be discovered was that they appear as perturbations in
temperature of only one part in $10^5$.

While the predicted anisotropies have finally been seen in the CMB, not
all kinds of matter and/or evolution of the universe can give rise to
the structure we observe today. If we define the density contrast
as~\cite{Peebles}
\begin{equation}\label{DensityContrast}
\delta(\vec x,a) \equiv {\rho(\vec x,a)-\bar\rho(a)\over\bar\rho(a)}
= \int d^3\vec{k}\ \delta_k(a)\ e^{i\vec{k}\cdot\vec{x}}\,,
\end{equation}
where $\bar\rho(a)=\rho_0\,a^{-3}$ is the average cosmic density, we need a
theory that will grow a density contrast with amplitude $\delta \sim
10^{-5}$ at the last scattering surface ($z=1100$) up to density
contrasts of the order of $\delta \sim 10^2$ for galaxies at redshifts
$z\ll1$, i.e. today. This is a {\em necessary} requirement for any
consistent theory of structure formation~\cite{Padmanabhan}.

Furthermore, the anisotropies observed by the COBE satellite correspond
to a small-amplitude scale-invariant primordial power spectrum of
inhomogeneities
\begin{equation}\label{HarrisonZeldovich}
P(k) = \langle|\delta_k|^2\rangle \propto k^n\,, \hspace{5mm} {\rm with}
\hspace{5mm} n=1\,,
\end{equation}
where the brackets $\langle\cdot\rangle$ represent integration over an
ensemble of different universe realizations. These inhomogeneities are
like waves in the space-time metric. When matter fell in the troughs of
those waves, it created density perturbations that collapsed
gravitationally to form galaxies and clusters of galaxies, with a
spectrum that is also scale invariant.  Such a type of spectrum was
proposed in the early 1970s by Edward R. Harrison, and independently by
the Russian cosmologist Yakov B. Zel'dovich, see Ref.~\cite{HZ}, to
explain the distribution of galaxies and clusters of galaxies on very
large scales in our observable universe.

%%%%%%%%%%%%%%%%%%%%%%%%%
\begin{figure}[htb]
%\vspace*{-1cm}
\begin{center}
\includegraphics[width=10cm,angle=0]{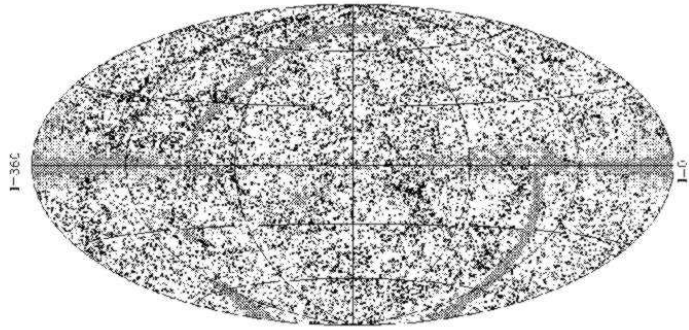}
\end{center}%\vspace*{-1.2cm}
\caption{The IRAS Point Source Catalog redshift survey contains some
15,000 galaxies, covering over 83\% of the sky up to redshifts of
$z\leq0.05$. We show here the projection of the galaxy distribution in
galactic coordinates. From Ref.~\cite{PSCz}.}
\label{fig7}
\end{figure}
%%%%%%%%%%%%%%%%%%%%%%%%%

Today various telescopes -- like the Hubble Space Telescope, the twin
Keck telescopes in Hawaii and the European Southern Observatory
telescopes in Chile -- are exploring the most distant regions of the
universe and discovering the first galaxies at large distances. The
furthest galaxies observed so far are at redshifts of $z\simeq10$ (at
a distance of 13.7 billion light years from Earth), whose light was
emitted when the universe had only about 3\% of its present age.  Only
a few galaxies are known at those redshifts, but there are at present
various catalogs like the CfA and APM galaxy catalogs, and more
recently the IRAS Point Source redshift Catalog, see Fig.~\ref{fig7},
and Las Campanas redshift surveys, that study the spatial distribution
of hundreds of thousands of galaxies up to distances of a billion
light years, or $z<0.1$, or the 2 degree Field Galaxy Redshift Survey
(2dFGRS) and the Sloan Digital Sky Survey (SDSS), which reach $z<0.5$
and study millions of galaxies.  These catalogs are telling us about
the evolution of clusters and superclusters of galaxies in the
universe, and already put constraints on the theory of structure
formation. From these observations one can infer that most galaxies
formed at redshifts of the order of $2 - 6$; clusters of galaxies
formed at redshifts of order 1, and superclusters are forming
now. That is, cosmic structure formed from the bottom up: from
galaxies to clusters to superclusters, and not the other way
around. This fundamental difference is an indication of the type of
matter that gave rise to structure.

We know from Big Bang nucleosynthesis that all the baryons in the
universe cannot account for the observed amount of matter, so there
must be some extra matter (dark since we don't see it) to account for
its gravitational pull. Whether it is relativistic (hot) or
non-relativistic (cold) could be inferred from observations:
relativistic particles tend to diffuse from one concentration of
matter to another, thus transferring energy among them and preventing
the growth of structure on small scales. This is excluded by
observations, so we conclude that most of the matter responsible for
structure formation must be cold. How much there is is a matter of
debate at the moment. Some recent analyses suggest that there is not
enough cold dark matter to reach the critical density required to make
the universe flat. If we want to make sense of the present
observations, we must conclude that some other form of energy
permeates the universe. In order to resolve this issue, 2dFGRS and
SDSS started taking data a few years ago. The first has already been
completed, but the second one is still taking data up to redshifts
$z\simeq5$ for quasars, over a large region of the sky. These important
observations will help astronomers determine the nature of the dark
matter and test the validity of the models of structure formation.

Before COBE discovered the anisotropies of the microwave background
there were serious doubts whether gravity alone could be responsible for
the formation of the structure we observe in the universe today. It
seemed that a new force was required to do the job. Fortunately, the
anisotropies were found with the right amplitude for structure to be
accounted for by gravitational collapse of primordial inhomogeneities
under the attraction of a large component of non-relativistic dark
matter. Nowadays, the standard theory of structure formation is a cold
dark matter model with a non vanishing cosmological constant in a
spatially flat universe. Gravitational collapse amplifies the density
contrast initially through linear growth and later on via non-linear
collapse. In the process, overdense regions decouple from the Hubble
expansion to become bound systems, which start attracting eachother to
form larger bound structures. In fact, the largest structures,
superclusters, have not yet gone non-linear.

The primordial spectrum (\ref{HarrisonZeldovich}) is reprocessed by
gravitational instability after the universe becomes matter dominated
and inhomogeneities can grow. Linear perturbation theory shows that the
growing mode~\footnote{The decaying modes go like $\delta(t)\sim
t^{-1}$, for all $\omega$.} of small density contrasts go
like~\cite{Peebles,Padmanabhan}
\begin{equation}\label{LinearGrowth}
\delta(a) \propto a^{1+3\omega} = \left\{\begin{array}{ll}
a^2\,,&\hspace{1cm}a<a_{\rm eq}\\
a\,,&\hspace{1cm}a>a_{\rm eq}\end{array}\right.
\end{equation}
in the Einstein-de Sitter limit ($\omega = p/\rho = 1/3$ and 0, for
radiation and matter, respectively). There are slight deviations for
$a\gg a_{\rm eq}$, if $\Omega_{\rm M}\neq1$ or $\Omega_\Lambda\neq0$,
but we will not be concerned with them here. The important observation
is that, since the density contrast at last scattering is of order
$\delta \sim 10^{-5}$, and the scale factor has grown since then only a
factor $z_{\rm dec} \sim 10^3$, one would expect a density contrast
today of order $\delta_0 \sim 10^{-2}$. Instead, we observe structures
like galaxies, where $\delta\sim10^2$. So how can this be possible? The
microwave background shows anisotropies due to fluctuations in the
baryonic matter component only (to which photons couple,
electromagnetically). If there is an additional matter component that
only couples through very weak interactions, fluctuations in that
component could grow as soon as it decoupled from the plasma, well
before photons decoupled from baryons. The reason why baryonic
inhomogeneities cannot grow is because of photon pressure: as baryons
collapse towards denser regions, radiation pressure eventually halts the
contraction and sets up acoustic oscillations in the plasma that prevent
the growth of perturbations, until photon decoupling. On the other hand,
a weakly interacting cold dark matter component could start
gravitational collapse much earlier, even before matter-radiation
equality, and thus reach the density contrast amplitudes observed today.
The resolution of this mismatch is one of the strongest arguments for
the existence of a weakly interacting cold dark matter component of the
universe.

%%%%%%%%%%%%%%%%%%%%%%%%%
\begin{figure}[htb]
\vspace*{-2mm}
\begin{center}
\includegraphics[width=9cm]{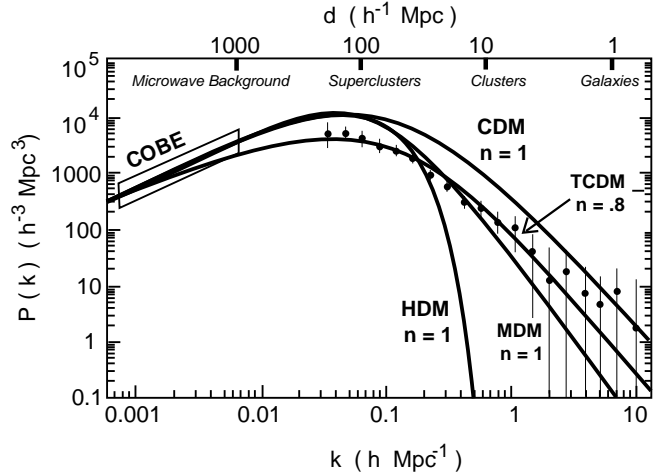}
\end{center}\vspace*{-5mm}
\caption{The power spectrum for cold dark matter (CDM), tilted cold dark
matter (TCDM), hot dark matter (HDM), and mixed hot plus cold dark
matter (MDM), normalized to COBE, for large-scale structure formation.
From Ref.~\cite{PJS}.}
\label{fig9}
\end{figure}
%%%%%%%%%%%%%%%%%%%%%%%%%

How much dark matter there is in the universe can be deduced from the
actual power spectrum (the Fourier transform of the two-point
correlation function of density perturbations) of the observed large
scale structure. One can decompose the density contrast in Fourier
components, see Eq.~(\ref{DensityContrast}). This is very convenient
since in linear perturbation theory individual Fourier components evolve
independently.  A comoving wavenumber $k$ is said to ``enter the
horizon'' when $k=d_H^{-1}(a)=aH(a)$. If a certain perturbation, of
wavelength $\lambda =k^{-1}<d_H(a_{\rm eq})$, enters the horizon before
matter-radiation equality, the fast radiation-driven expansion prevents
dark-matter perturbations from collapsing. Since light can only cross
regions that are smaller than the horizon, the suppression of growth due
to radiation is restricted to scales smaller than the horizon, while
large-scale perturbations remain unaffected. This is the reason why the
horizon size at equality, Eq.~(\ref{HorizonSizeEquality}), sets an
important scale for structure growth,
\begin{equation}\label{EqualityScale}
k_{\rm eq} = d_H^{-1}(a_{\rm eq})\simeq 0.083\,(\Omega_{\rm M} h)\,h\ 
{\rm Mpc}^{-1}\,.
\end{equation}
The suppression factor can be easily computed from (\ref{LinearGrowth})
as $f_{\rm sup} = (a_{\rm enter}/a_{\rm eq})^2 = (k_{\rm eq}/k)^2$.
In other words, the processed power spectrum $P(k)$ will have the form:
\begin{equation}\label{PowerSpectrum}
P(k) \propto \left\{\begin{array}{ll}
k\,,&\hspace{1cm}k\ll k_{\rm eq}\\[2mm]
k^{-3}\,,&\hspace{1cm}k\gg k_{\rm eq}\end{array}\right.
\end{equation}
This is precisely the shape that large-scale galaxy catalogs are bound
to test in the near future, see Fig.~\ref{fig9}. Furthermore, since
relativistic Hot Dark Matter (HDM) transfer energy between clumps of
matter, they will wipe out small scale perturbations, and this should be
seen as a distinctive signature in the matter power spectra of future
galaxy catalogs. On the other hand, non-relativistic Cold Dark Matter
(CDM) allow structure to form on {\em all} scales via gravitational
collapse. The dark matter will then pull in the baryons, which will
later shine and thus allow us to see the galaxies.

Naturally, when baryons start to collapse onto dark matter potential
wells, they will convert a large fraction of their potential energy into
kinetic energy of protons and electrons, ionizing the medium. As a
consequence, we expect to see a large fraction of those baryons
constituting a hot ionized gas surrounding large clusters of galaxies.
This is indeed what is observed, and confirms the general picture of
structure formation.

\section{DETERMINATION OF COSMOLOGICAL PARAMETERS}

In this Section, I will restrict myself to those recent measurements of
the cosmological parameters by means of standard cosmological
techniques, together with a few instances of new results from recently
applied techniques. We will see that a large host of observations are
determining the cosmological parameters with some reliability of the
order of 10\%. However, the majority of these measurements are dominated
by large systematic errors. Most of the recent work in observational
cosmology has been the search for virtually systematic-free observables,
like those obtained from the microwave background anisotropies, and
discussed in Section 4.4. I will devote, however, this Section to the
more `classical' measurements of the following cosmological parameters:
The rate of expansion $H_0$; the matter content $\Omega_{\rm M}$; the
cosmological constant $\Omega_\Lambda$; the spatial curvature
$\Omega_K$, and the age of the universe $t_0$.

\subsection{The rate of expansion $H_0$}

Over most of last century the value of $H_0$ has been a constant source
of disagreement~\cite{Freedman}. Around 1929, Hubble measured the rate
of expansion to be $H_0 = 500$ km\,s$^{-1}$Mpc$^{-1}$, which implied an
age of the universe of order $t_0\sim2$ Gyr, in clear conflict with
geology.  Hubble's data was based on Cepheid standard candles that were
incorrectly calibrated with those in the Large Magellanic Cloud. Later
on, in 1954 Baade recalibrated the Cepheid distance and obtained a lower
value, $H_0 = 250$ km\,s$^{-1}$Mpc$^{-1}$, still in conflict with ratios
of certain unstable isotopes. Finally, in 1958 Sandage realized that the
brightest stars in galaxies were ionized HII regions, and the Hubble
rate dropped down to $H_0 = 60$ km\,s$^{-1}$ Mpc$^{-1}$, still with
large (factor of two) systematic errors. Fortunately, in the past 15
years there has been significant progress towards the determination of
$H_0$, with systematic errors approaching the 10\% level. These
improvements come from two directions. First, technological, through the
replacement of photographic plates (almost exclusively the source of
data from the 1920s to 1980s) with charged couple devices (CCDs), i.e.
solid state detectors with excellent flux sensitivity per pixel, which
were previously used successfully in particle physics detectors. Second,
by the refinement of existing methods for measuring extragalactic
distances (e.g. parallax, Cepheids, supernovae, etc.). Finally, with the
development of completely new methods to determine $H_0$, which fall
into totally independent and very broad categories: a) Gravitational
lensing; b) Sunyaev-Zel'dovich effect; c) Extragalactic distance scale,
mainly Cepheid variability and type Ia Supernovae; d) Microwave
background anisotropies. I will review here the first three, and leave
the last method for Section~4.4, since it involves knowledge about the
primordial spectrum of inhomogeneities.

\subsubsection{Gravitational lensing}

Imagine a quasi-stellar object (QSO) at large redshift ($z\gg1$) whose
light is lensed by an intervening galaxy at redshift $z\sim1$ and
arrives to an observer at $z=0$. There will be at least two different
images of the same background {\em variable} point source. The arrival
times of photons from two different gravitationally lensed images of the
quasar depend on the different path lengths and the gravitational
potential traversed. Therefore, a measurement of the time delay and the
angular separation of the different images of a variable quasar can be
used to determine $H_0$ with great accuracy. This method, proposed in
1964 by Refsdael~\cite{Refsdael}, offers tremendous potential because it
can be applied at great distances and it is based on very solid physical
principles~\cite{Blandford}.

Unfortunately, there are very few systems with both a favourable
geometry (i.e. a known mass distribution of the intervening galaxy) and
a variable background source with a measurable time delay. That is the
reason why it has taken so much time since the original proposal for the
first results to come out. Fortunately, there are now very powerful
telescopes that can be used for these purposes. The best candidate
to-date is the QSO\ $0957+561$, observed with the 10m Keck telescope,
for which there is a model of the lensing mass distribution that is
consistent with the measured velocity dispersion. Assuming a flat space
with $\Omega_{\rm M}=0.25$, one can determine~\cite{Grogin}
\begin{equation}
H_0 = 72 \pm 7 \ (1\sigma\ {\rm statistical})\ 
\pm 15\%\ ({\rm systematic})\ \ {\rm km\,s}^{-1}{\rm Mpc}^{-1}\,.
\end{equation}
The main source of systematic error is the degeneracy between the mass
distribution of the lens and the value of $H_0$. Knowledge of the
velocity dispersion within the lens as a function of position helps
constrain the mass distribution, but those measurements are very
difficult and, in the case of lensing by a cluster of galaxies, the dark
matter distribution in those systems is usually unknown, associated with
a complicated cluster potential. Nevertheless, the method is just
starting to give promising results and, in the near future, with the
recent discovery of several systems with optimum properties, the
prospects for measuring $H_0$ and lowering its uncertainty with this
technique are excellent.

\subsubsection{Sunyaev-Zel'dovich effect}

\begin{figure}[htb]
\vspace*{1mm}
\begin{center}
\includegraphics[width=5cm,angle=0]{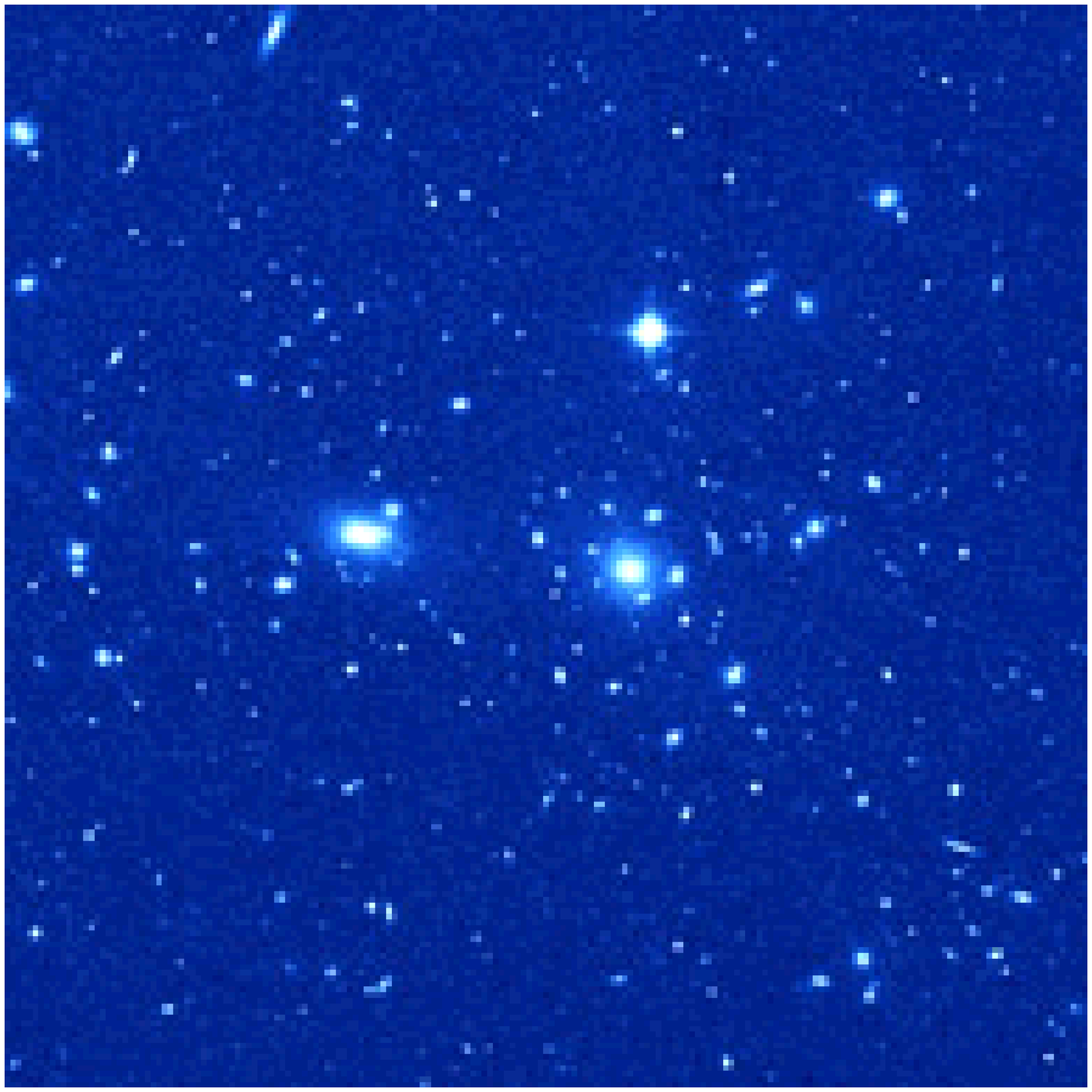}
\includegraphics[width=5cm,angle=0]{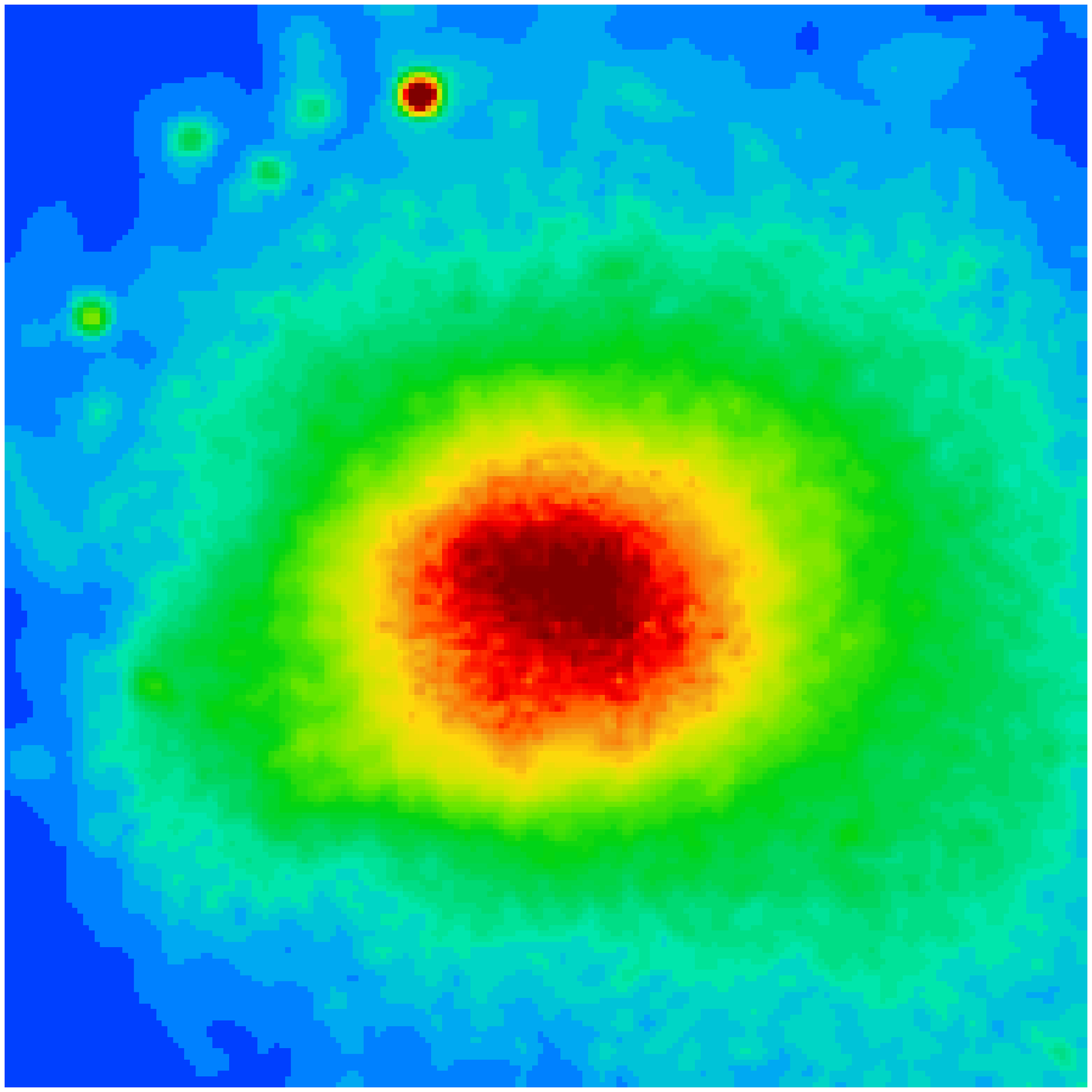}
\end{center}
\vspace*{-1mm}
\caption{The Coma cluster of galaxies, seen here in an optical image
(left) and an X-ray image (right), taken by the recently launched 
Chandra X-ray Observatory. From Ref.~\cite{Chandra}.}
\label{fig12}
\end{figure}

As discussed in the previous Section, the gravitational collapse of
baryons onto the potential wells generated by dark matter gave rise to
the reionization of the plasma, generating an X-ray halo around rich
clusters of galaxies, see Fig.~\ref{fig12}. The inverse-Compton
scattering of microwave background photons off the hot electrons in the
X-ray gas results in a measurable distortion of the blackbody spectrum
of the microwave background, known as the Sunyaev-Zel'dovich (SZ)
effect. Since photons acquire extra energy from the X-ray electrons, we
expect a shift towards higher frequencies of the spectrum,
$(\Delta\nu/\nu) \simeq (k_{\rm B}T_{\rm gas}/m_e c^2) \sim
10^{-2}$. This corresponds to a {\em decrement} of the microwave
background temperature at low frequencies (Rayleigh-Jeans region) and an
increment at high frequencies, see Ref.~\cite{Birkinshaw}.  

%%%%%%%%%%%%%%%%%%%%%%%%%
%%%%%%%%%%%%%%%%%%%%%%%%%

Measuring the {\em spatial} distribution of the SZ effect (3 K
spectrum), together with a high resolution X-ray map ($10^8$ K spectrum)
of the cluster, one can determine the density and temperature
distribution of the hot gas. Since the X-ray flux is distance-dependent
(${\cal F}={\cal L}/4\pi d_L^2$), while the SZ decrement is not (because
the energy of the CMB photons increases as we go back in redshift,
$\nu=\nu_0(1+z)$, and exactly compensates the redshift in energy of the
photons that reach us), one can determine from there the distance to the
cluster, and thus the Hubble rate $H_0$.

The advantages of this method are that it can be applied to large
distances and it is based on clear physical principles. The main
systematics come from possible clumpiness of the gas (which would reduce
$H_0$), projection effects (if the clusters are prolate, $H_0$ could be
larger), the assumption of hydrostatic equilibrium of the X-ray gas,
details of models for the gas and electron densities, and possible
contaminations from point sources. Present measurements give the
value~\cite{Birkinshaw}
\begin{equation}
H_0 = 60 \pm 10 \ (1\sigma\ {\rm statistical})\ 
\pm 20\%\ ({\rm systematic})\ \ {\rm km\,s}^{-1}{\rm Mpc}^{-1}\,,
\end{equation}
compatible with other determinations. A great advantage of this
completely new and independent method is that nowadays more and more
clusters are observed in the X-ray, and soon we will have
high-resolution 2D maps of the SZ decrement from several balloon
flights, as well as from future microwave background satellites,
together with precise X-ray maps and spectra from the Chandra X-ray
observatory recently launched by NASA, as well as from the European
X-ray satellite XMM launched a few months ago by ESA, which will deliver
orders of magnitude better resolution than the existing Einstein X-ray
satellite.

\subsubsection{Cepheid variability}

Cepheids are low-mass variable stars with a period-luminosity relation
based on the helium ionization cycles inside the star, as it contracts
and expands. This time variability can be measured, and the star's
absolute luminosity determined from the calibrated relationship. From
the observed flux one can then deduce the luminosity distance, see
Eq.~(\ref{LuminosityDistance}), and thus the Hubble rate $H_0$. The
Hubble Space Telescope (HST) was launched by NASA in 1990 (and repaired
in 1993) with the specific project of calibrating
the extragalactic distance scale and thus determining the Hubble rate
with 10\% accuracy. The most recent results from HST are the
following~\cite{HST}
\begin{equation}
H_0 = 71 \pm 4 \ ({\rm random})\ \pm 7\ ({\rm systematic})\ \ 
{\rm km\,s}^{-1}{\rm Mpc}^{-1}\,.
\end{equation}
The main source of systematic error is the distance to the Large
Magellanic Cloud, which provides the fiducial comparison for Cepheids in
more distant galaxies. Other systematic uncertainties that affect the
value of $H_0$ are the internal extinction correction method used, a
possible metallicity dependence of the Cepheid period-luminosity
relation and cluster population incompleteness bias, for a set of 21
galaxies within 25 Mpc, and 23 clusters within $z\lsim0.03$.

With better telescopes already taking data, like the Very Large
Telescope (VLT) interferometer of the European Southern Observatory
(ESO) in the Chilean Atacama desert, with 8 synchronized telescopes,
and others coming up soon, like the Next Generation Space Telescope
(NGST) proposed by NASA for 2008, and the Gran TeCan of the European
Northern Observatory in the Canary Islands, for 2010, it is expected
that much better resolution and therefore accuracy can be obtained for
the determination of $H_0$.

\subsection{Dark Matter}

In the 1920s Hubble realized that the so called nebulae were actually
distant galaxies very similar to our own. Soon afterwards, in 1933,
Zwicky found dynamical evidence that there is possibly ten to a hundred
times more mass in the Coma cluster than contributed by the luminous
matter in galaxies~\cite{Zwicky}. However, it was not until the 1970s
that the existence of dark matter began to be taken more seriously. At
that time there was evidence that rotation curves of galaxies did not
fall off with radius and that the dynamical mass was increasing with
scale from that of individual galaxies up to clusters of galaxies. Since
then, new possible extra sources to the matter content of the universe
have been accumulating:
\begin{eqnarray}\label{MatterContent}
\Omega_M &=& \Omega_{B,\ {\rm lum}} \hspace{0.9cm} 
({\rm stars\ in\ galaxies})\\
&+& \Omega_{B,\ {\rm dark}} \hspace{0.8cm} ({\rm MACHOs?})\\
&+& \Omega_{CDM} \hspace{1cm} ({\rm weakly\ interacting:\ 
axion,\ neutralino?})\\
&+& \Omega_{HDM} \hspace{1cm} ({\rm massive\ neutrinos?})
\end{eqnarray}

The empirical route to the determination of $\Omega_M$ is nowadays one
of the most diversified of all cosmological parameters. The matter
content of the universe can be deduced from the mass-to-light ratio of
various objects in the universe; from the rotation curves of galaxies;
from microlensing and the direct search of Massive Compact Halo Objects
(MACHOs); from the cluster velocity dispersion with the use of the
Virial theorem; from the baryon fraction in the X-ray gas of clusters;
from weak gravitational lensing; from the observed matter distribution
of the universe via its power spectrum; from the cluster abundance and
its evolution; from direct detection of massive neutrinos at
SuperKamiokande; from direct detection of Weakly Interacting Massive
Particles (WIMPs) at CDMS, DAMA or UKDMC, and finally from microwave
background anisotropies. I will review here just a few of them.

\subsubsection{Rotation curves of spiral galaxies}

The flat rotation curves of spiral galaxies provide the most direct
evidence for the existence of large amounts of dark matter. Spiral
galaxies consist of a central bulge and a very thin disk, stabilized
against gravitational collapse by angular momentum conservation, and
surrounded by an approximately spherical halo of dark matter. One can
measure the orbital velocities of objects orbiting around the disk as
a function of radius from the Doppler shifts of their spectral lines.

The rotation curve of the Andromeda galaxy was first measured by
Babcock in 1938, from the stars in the disk. Later it became possible
to measure galactic rotation curves far out into the disk, and a trend
was found~\cite{Freeman}. The orbital velocity rose linearly from the
center outward until it reached a typical value of 200 km/s, and then
remained flat out to the largest measured radii. This was completely
unexpected since the observed surface luminosity of the disk falls off
exponentially with radius~\cite{Freeman}, $I(r) = I_0
\exp(-r/r_D)$. Therefore, one would expect that most of the galactic
mass is concentrated within a few disk lengths $r_D$, such that the
rotation velocity is determined as in a Keplerian orbit, $v_{\rm rot}
= (GM/r)^{1/2} \propto r^{-1/2}$. No such behaviour is observed. In
fact, the most convincing observations come from radio emission (from
the 21 cm line) of neutral hydrogen in the disk, which has been
measured to much larger galactic radii than optical tracers. A typical
case is that of the spiral galaxy NGC 6503, where $r_D = 1.73$ kpc,
while the furthest measured hydrogen line is at $r=22.22$ kpc, about
13 disk lengths away.  Nowadays, thousands of galactic rotation curves
are known, see Fig.~14, and all suggest the existence of about ten
times more mass in the halos of spiral galaxies than in the stars of
the disk. Recent numerical simulations of galaxy formation in a CDM
cosmology~\cite{Frenk} suggest that galaxies probably formed by the
infall of material in an overdense region of the universe that had
decoupled from the overall expansion.  

%%%%%%%%%%%%%%%%%%%%%%%%%
\begin{figure}[htb]
\label{fig:NGC6503}
\begin{center}%\vspace*{-1cm}
\includegraphics[width=8cm]{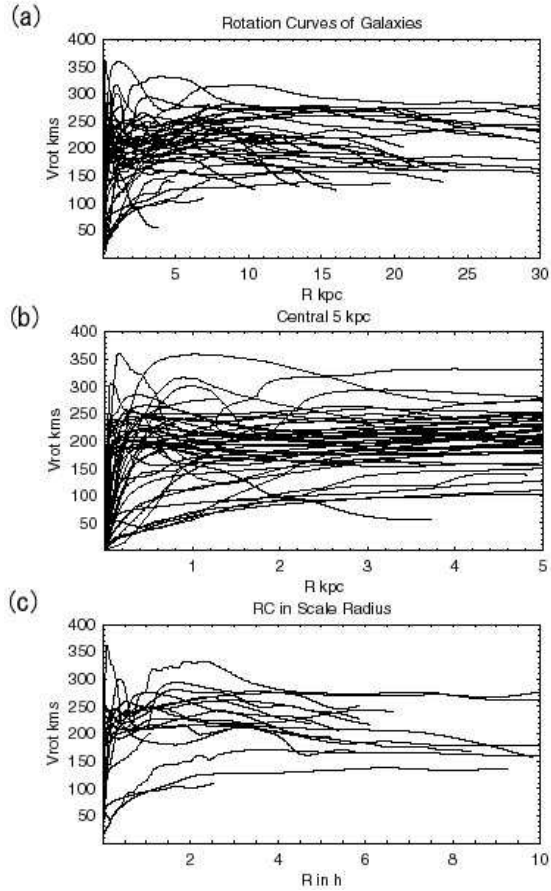}
\caption{The rotation curves of several hundred galaxies. Upper panel:
As a function of their radii in kpc. Middle panel: The central 5 kpc.
Lower panel: As a function of scale radius.}
\end{center}
\end{figure}
%%%%%%%%%%%%%%%%%%%%%%%%%

The dark matter is supposed to
undergo violent relaxation and create a virialized system, i.e. in
hydrostatic equilibrium. This picture has led to a simple model of
dark-matter halos as isothermal spheres, with density profile $\rho(r)
= \rho_c/(r_c^2 + r^2)$, where $r_c$ is a core radius and $\rho_c =
v_\infty^2/4\pi G$, with $v_\infty$ equal to the plateau value of the
flat rotation curve. This model is consistent with the universal
rotation curves seen in Fig.~6. At large radii the dark matter
distribution leads to a flat rotation curve. The question is for how
long. In dense galaxy clusters one expects the galactic halos to
overlap and form a continuum, and therefore the rotation curves should
remain flat from one galaxy to another. However, in field galaxies,
far from clusters, one can study the rotation velocities of
substructures (like satellite dwarf galaxies) around a given galaxy,
and determine whether they fall off at sufficiently large distances
according to Kepler's law, as one would expect, once the edges of the
dark matter halo have been reached. These observations are rather
difficult because of uncertainties in distinguishing between true
satellites and interlopers.  Recently, a group from the Sloan Digital
Sky Survey Collaboration claim that they have seen the edges of the
dark matter halos around field galaxies by confirming the fall-off at
large distances of their rotation curves~\cite{Klypin}. These results, if
corroborated by further analysis, would constitute a tremendous
support to the idea of dark matter as a fluid surrounding galaxies and
clusters, while at the same time eliminates the need for modifications
of Newtonian of even Einstenian gravity at the scales of galaxies, to
account for the flat rotation curves.

That's fine, but how much dark matter is there at the galactic scale?
Adding up all the matter in galactic halos up to a maximum radii, one 
finds
\begin{equation}\label{OmegaHalo}
\Omega_{\rm halo} \simeq 10\ \Omega_{\rm lum} \geq 0.03 - 0.05\,.
\end{equation}
Of course, it would be extraordinary if we could confirm, through
direct detection, the existence of dark matter in our own galaxy. For
that purpose, one should measure its rotation curve, which is much
more difficult because of obscuration by dust in the disk, as well as
problems with the determination of reliable galactocentric distances
for the tracers. Nevertheless, the rotation curve of the Milky Way has
been measured and conforms to the usual picture, with a plateau value
of the rotation velocity of 220 km/s. For dark matter searches, the
crucial quantity is the dark matter density in the solar
neighbourhood, which turns out to be (within a factor of two
uncertainty depending on the halo model) $\rho_{\rm DM} = 0.3$
GeV/cm$^3$. We will come back to direct searched of dark matter in a
later subsection.

\subsubsection{Baryon fraction in clusters}

Since large clusters of galaxies form through gravitational collapse,
they scoop up mass over a large volume of space, and therefore the ratio
of baryons over the total matter in the cluster should be representative
of the entire universe, at least within a 20\% systematic error. Since
the 1960s, when X-ray telescopes became available, it is known that
galaxy clusters are the most powerful X-ray sources in the
sky~\cite{Sarazin}. The emission extends over the whole cluster and
reveals the existence of a hot plasma with temperature $T\sim 10^7 -
10^8$ K, where X-rays are produced by electron bremsstrahlung. Assuming
the gas to be in hydrostatic equilibrium and applying the virial theorem
one can estimate the total mass in the cluster, giving general agreement
(within a factor of 2) with the virial mass estimates. From these
estimates one can calculate the baryon fraction of clusters
\begin{equation}\label{BaryonFraction}
f_{\rm B}h^{3/2} = 0.08 \hspace{5mm} \Rightarrow
\hspace{5mm} {\Omega_B\over\Omega_M} \approx 0.14\,,
\hspace{5mm} {\rm for} \hspace{3mm} h=0.70\,.
\end{equation}
Since $\Omega_{\rm lum} \simeq 0.002 - 0.006$, the previous expression
suggests that clusters contain far more baryonic matter in the form of
hot gas than in the form of stars in galaxies. Assuming this fraction
to be representative of the entire universe, and using the Big Bang
nucleosynthesis value of $\Omega_B = 0.04 \pm 0.01$, for $h=0.7$, we
find
\begin{equation}\label{OmegaXray}
\Omega_M = 0.3 \pm 0.1\ ({\rm statistical})\ 
\pm 20\%\ ({\rm systematic})\,.
\end{equation}
This value is consistent with previous determinations of $\Omega_M$.
If some baryons are ejected from the cluster during gravitational
collapse, or some are actually bound in nonluminous objects like planets,
then the actual value of $\Omega_M$ is smaller than this estimate.

\subsubsection{Weak gravitational lensing}

Since the mid 1980s, deep surveys with powerful telescopes have
observed huge arc-like features in galaxy clusters. The spectroscopic
analysis showed that the cluster and the giant arcs were at very
different redshifts. The usual interpretation is that the arc is the
image of a distant background galaxy which is in the same line of
sight as the cluster so that it appears distorted and magnified by the
gravitational lens effect: the giant arcs are essentially partial
Einstein rings. From a systematic study of the cluster mass
distribution one can reconstruct the shear field responsible for the
gravitational distortion~\cite{Bartelmann}. This analy\-sis shows that
there are large amounts of dark matter in the clusters, in rough
agreement with the virial mass estimates, although the lensing masses
tend to be systematically larger. At present, the estimates indicate
$\Omega_M = 0.2 - 0.3$ on scales $\lsim 6\,h^{-1}$ Mpc.

\subsubsection{Large scale structure formation and the matter power spectrum}

Although the isotropic microwave background indicates that the
universe in the {\em past} was extraordinarily homogeneous, we know
that the universe {\em today} is far from homogeneous: we observe
galaxies, clusters and superclusters on large scales. These structures
are expected to arise from very small primordial inhomogeneities that
grow in time via gravitational instability, and that may have
originated from tiny ripples in the metric, as matter fell into their
troughs. Those ripples must have left some trace as temperature
anisotropies in the microwave background, and indeed such anisotropies
were finally discovered by the COBE satellite in 1992. However, not
all kinds of matter and/or evolution of the universe can give rise to
the structure we observe today. If we define the density contrast
as
\begin{equation}\label{DensityContrast2}
\delta(\vec x,a) \equiv {\rho(\vec x,a)-\bar\rho(a)\over\bar\rho(a)}
= \int d^3\vec{k}\ \delta_k(a)\ e^{i\vec{k}\cdot\vec{x}}\,,
\end{equation}
where $\bar\rho(a)=\rho_0\,a^{-3}$ is the average cosmic density, we need a
theory that will grow a density contrast with amplitude $\delta \sim
10^{-5}$ at the last scattering surface ($z=1100$) up to density
contrasts of the order of $\delta \sim 10^2$ for galaxies at redshifts
$z\ll1$, i.e. today. This is a {\em necessary} requirement for any
consistent theory of structure formation.

Furthermore, the anisotropies observed by the COBE satellite correspond
to a small-amplitude scale-invariant primordial power spectrum of
inhomogeneities
\begin{equation}\label{HarrisonZeldovich2}
P(k) = \langle|\delta_k|^2\rangle \propto k^n\,, \hspace{5mm} {\rm with}
\hspace{5mm} n=1\,,
\end{equation}
These inhomogeneities are like waves in the space-time metric. When
matter fell in the troughs of those waves, it created density
perturbations that collapsed gravitationally to form galaxies and
clusters of galaxies, with a spectrum that is also scale invariant.
Such a type of spectrum was proposed in the early 1970s by Edward
R. Harrison, and independently by the Russian cosmologist Yakov
B. Zel'dovich~\cite{HZ}, to explain the distribution of galaxies and
clusters of galaxies on very large scales in our observable universe,
see Fig.~15.

%%%%%%%%%%%%%%%%%%%%%%%%%
\begin{figure}\label{2dF}
\begin{center}
\includegraphics[width=10cm]{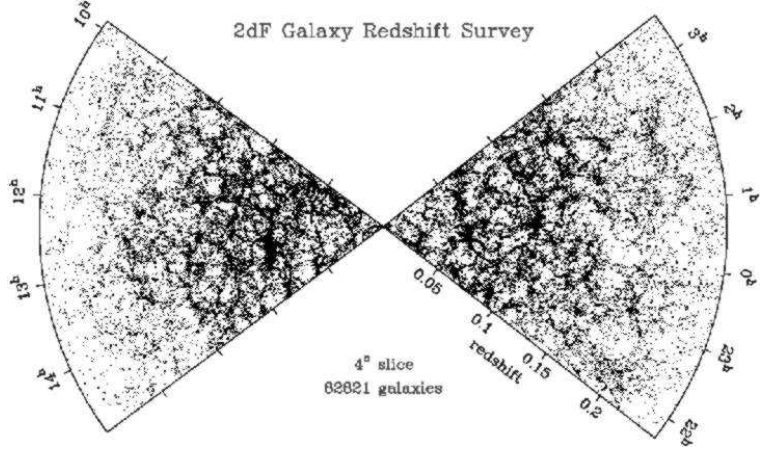}
\caption{The 2 degree Field Galaxy Redshift Survey contains some
250,000 galaxies, cove\-ring a large fraction of the sky up to 
redshifts of $z\leq0.25$. From Ref.~\cite{2dFGRS}.}
\end{center}
\end{figure}
%%%%%%%%%%%%%%%%%%%%%%%%%

Since the primordial spectrum is very approximately represented by a
scale-invariant {\em Gaussian random field}, the best way to present
the results of structure formation is by working with the 2-point
correlation function in Fourier space, the so-called {\em power
spectrum}. If the reprocessed spectrum of inhomogeneities remains
Gaussian, the power spectrum is all we need to describe the galaxy
distribution. Non-Gaussian effects are expected to arise from the
non-linear gravitational collapse of structure, and may be important
at small scales.  The power spectrum measures the degree of
inhomogeneity in the mass distribution on different scales, see
Fig.~16. It depends upon a few basic ingredientes: a) the primordial
spectrum of inhomogeneities, whether they are Gaussian or
non-Gaussian, whether {\em adiabatic} (perturbations in the energy
density) or {\em isocurvature} (perturbations in the entropy density),
whether the primordial spectrum has {\em tilt} (deviations from
scale-invariance), etc.; b) the recent creation of inhomogeneities,
whether {\em cosmic strings} or some other topological defect from an
early phase transition are responsible for the formation of structure
today; and c) the cosmic evolution of the inhomogeneity, whether the
universe has been dominated by cold or hot dark matter or by a
cosmological constant since the beginning of structure formation, and
also depending on the rate of expansion of the universe.

%%%%%%%%%%%%%%%%%%%%%%%%%
\begin{figure}\label{fig:Pk}
\begin{center}
\includegraphics[width=10cm]{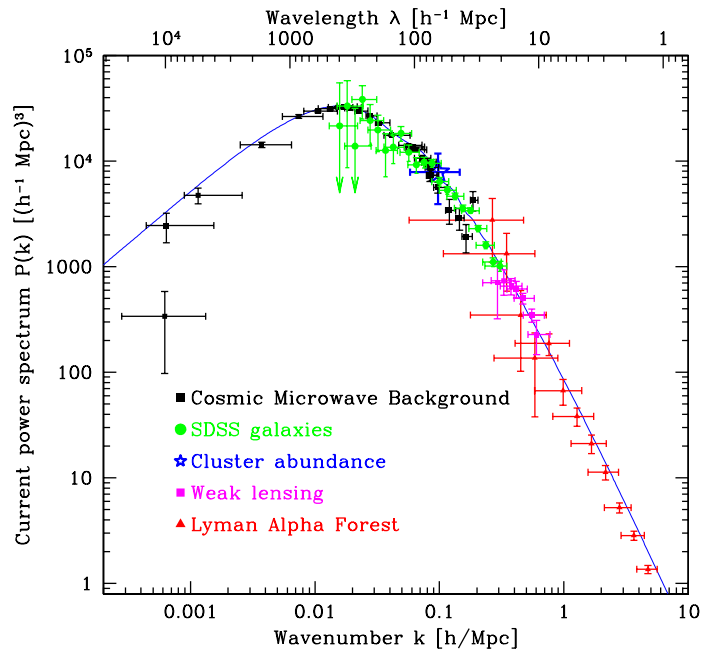}
\caption{The measured power spectrum $P(k)$ as a function of
wavenumber $k$. From observations of the Sloan Digital Sky Survey,
CMB anisotropies, cluster abundance, gravitational lensing and
Lyman-$\alpha$ forest. From Ref.~\cite{SDSS}.}
\end{center}
\end{figure}
%%%%%%%%%%%%%%%%%%%%%%%%%

The working tools used for the comparison between the observed power
spectrum and the predicted one are very precise N-body numerical
simulations and theoretical models that predict the {\em shape} but not
the {\em amplitude} of the present power spectrum. Even though a large
amount of work has gone into those analyses, we still have large
uncertainties about the nature and amount of matter necessary for
structure formation.  A model that has become a working paradigm is a
flat cold dark matter model with a cosmological constant and
$\Omega_M \sim 0.3$. This model is now been confronted with the recent
very precise measurements from 2dFGRS~\cite{2dFGRS} and SDSS~\cite{SDSS}.

\subsubsection{The new redshift catalogs, 2dF and Sloan Digital Sky Survey}

Our view of the large-scale distribution of luminous objects in the
universe has changed dramatically during the last 25 years: from the
simple pre-1975 picture of a distribution of field and cluster
galaxies, to the discovery of the first single superstructures and
voids, to the most recent results showing an almost regular web-like
network of interconnected clusters, filaments and walls, separating
huge nearly empty volumes. The increased efficiency of redshift
surveys, made possible by the development of spectrographs and --
specially in the last decade -- by an enormous increase in
multiplexing gain (i.e. the ability to collect spectra of several
galaxies at once, thanks to fibre-optic spectrographs), has allowed us
not only to do {\em cartography} of the nearby universe, but also to
statistically characterize some of its properties. At the same time,
advances in theoretical modeling of the development of structure, with
large high-resolution gravitational simulations coupled to a deeper
yet limited understanding of how to form galaxies within the dark
matter halos, have provided a more realistic connection of the models
to the observable quantities. Despite the large uncertainties that
still exist, this has transformed the study of cosmology and
large-scale structure into a truly quantitative science, where theory
and observations can progress together.

%%%%%%%%%%%%%%%%%%%%%%%%%
\begin{figure}\label{SDM}
\begin{center}
\includegraphics[width=8cm]{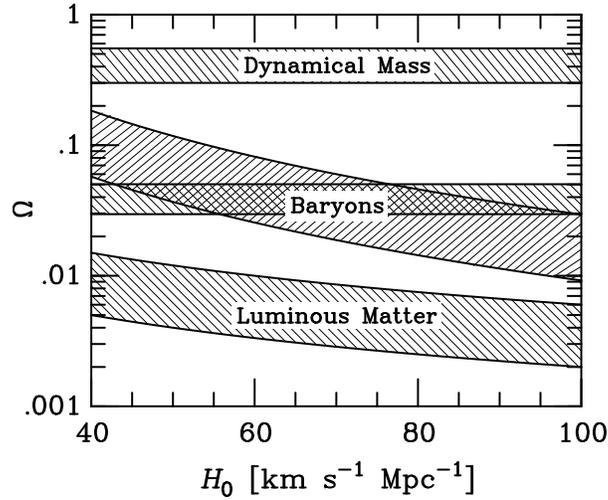} 
\caption{The observed cosmic matter components as functions of the
Hubble expansion parameter. The luminous matter component is given by
$0.002\leq\Omega_{\rm lum}\leq0.006$; the galactic halo component is
the horizontal band, $0.03\leq\Omega_{\rm halo}\leq0.05$, crossing the
baryonic component from BBN, $\Omega_B\,h^2=0.0244\pm0.0024$; and the
dynamical mass component from large scale structure analysis is given
by $\Omega_M=0.3\pm0.1$. Note that in the range $H_0 = 70\pm7$
km/s/Mpc, there are {\em three} dark matter problems, see the
text. From Ref.~\cite{Raffelt}.}
\end{center}
\end{figure}
%%%%%%%%%%%%%%%%%%%%%%%%%

\subsubsection{Summary of the matter content}

We can summarize the present situation with Fig.~17, for
$\Omega_M$ as a function of $H_0$. There are four bands, the luminous
matter $\Omega_{\rm lum}$; the baryon content $\Omega_B$, from BBN;
the galactic halo component $\Omega_{\rm halo}$, and the dynamical
mass from clusters, $\Omega_M$. From this figure it is clear that
there are in fact {\em three} dark matter problems: The first one is
where are 90\% of the baryons? Between the fraction predicted by BBN
and that seen in stars and diffuse gas there is a huge fraction which
is in the form of dark baryons. They could be in small clumps of
hydrogen that have not started thermonuclear reactions and perhaps
constitute the dark matter of spiral galaxies' halos. Note that
although $\Omega_B$ and $\Omega_{\rm halo}$ coincide at $H_0\simeq70$
km/s/Mpc, this could be just a coincidence.  The second problem is
what constitutes 90\% of matter, from BBN baryons to the mass inferred
from cluster dynamics? This is the standard dark matter problem and
could be solved in the future by direct detection of a weakly
interacting massive particle in the laboratory.  And finally, since we
know from observations of the CMB that the universe is flat, the rest,
up to $\Omega_0=1$, must be a diffuse vacuum energy, which affects the
very large scales and late times, and seems to be responsible for the
present acceleration of the universe, see Section 3. Nowadays, multiple
observations seem to converge towards a common determination of
$\Omega_M = 0.25\pm0.08$ (95\% c.l.), see Fig.~18.

%%%%%%%%%%%%%%%%%%%%%%%%%
\begin{figure}\label{CDM}
\begin{center}
\includegraphics[width=7cm,angle=-90]{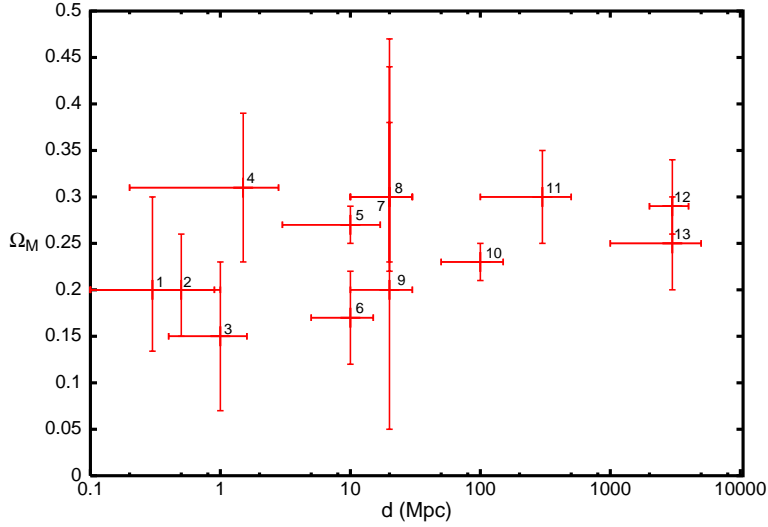} 
\caption{Different determinations of $\Omega_M$ as a function of distance,
from various sources: 1. peculiar velocities; 2. weak gravitacional
lensing; 3. shear autocorrelation function; 4. local group of
galaxies; 5. baryon mass fraction; 6. cluster mass function;
7. virgocentric flow; 8. mean relative velocities; 9. redshift space
distortions; 10. mass power spectrum; 11. integrated Sachs-Wolfe
effect; 12. angular diameter distance: SNe; 13. cluster baryon
fraction.  While a few years ago the dispersion among observed values
was huge and strongly dependent on scale, at present the observed
value of the matter density parameter falls well within a narrow
range, $\Omega_M = 0.25\pm0.07$ (95\% c.l.) and is essentially
independent on scale, from 100 kpc to 5000 Mpc. Adapted from
Ref.~\cite{Peebles2}.}
\end{center}
\end{figure}
%%%%%%%%%%%%%%%%%%%%%%%%%

\subsubsection{Massive neutrinos}

One of the `usual suspects' when addressing the problem of dark matter
are neutrinos. They are the only candidates known to exist. If
neutrinos have a mass, could they constitute the missing matter? We
know from the Big Bang theory, see Section~2.6.5, that there is a
cosmic neutrino background at a temperature of approximately 2K. This
allows one to compute the present number density in the form of
neutrinos, which turns out to be, for massless neutrinos,
$n_\nu(T_\nu) = {3\over11}\, n_\gamma(T_\gamma) = 112\ {\rm cm}^{-3}$,
per species of neutrino. If neutrinos have mass, as recent experiments
seem to suggest,\footnote{For a review on Neutrino properties, see
Gonz\'alez-Garc\ii a's lectures on these Proceedings.} see Fig.~19,
the cosmic energy density in massive neutrinos would be $\rho_\nu =
\sum n_\nu m_\nu = {3\over11}\,n_\gamma\,\sum m_\nu$, and therefore
its contribution today,
\begin{equation}\label{OmegaNeutrinos}
\Omega_\nu h^2 = {\sum m_\nu\over93.2\ {\rm eV}}\,.
\end{equation}
The discussion in the previous Sections suggest that $\Omega_{\rm M}
\leq 0.4$, and thus, for any of the three families of neutrinos, $m_\nu
\leq 40$ eV. Note that this limit improves by six orders of magnitude
the present bound on the tau-neutrino mass~\cite{PDG}. Supposing that
the missing mass in non-baryonic cold dark matter arises from a single
particle dark matter (PDM) component, its contribution to the critical
density is bounded by \ $0.05 \leq \Omega_{\rm PDM}h^2 \leq 0.4$, see
Fig.~17.

%%%%%%%%%%%%%%%%%%%%%%%%%
\begin{figure}[htb]
\begin{center}%\vspace*{-1cm}
\includegraphics[width=12cm]{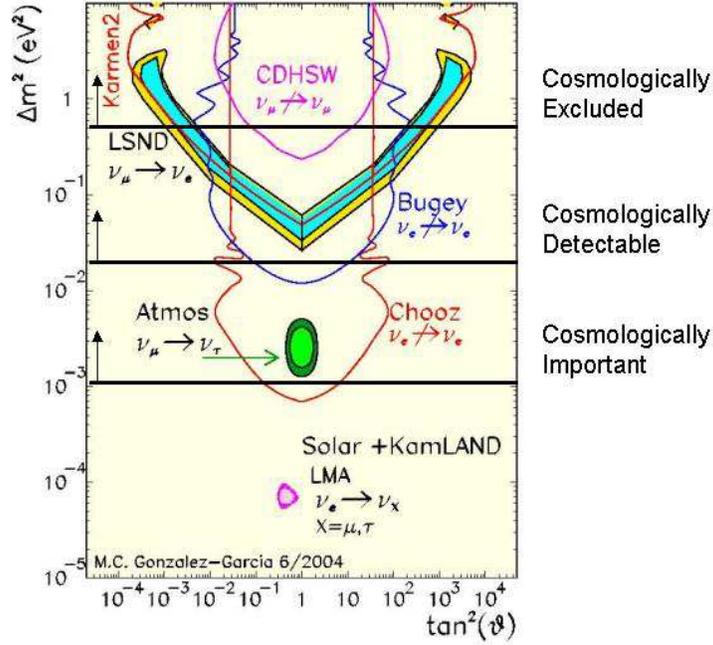} 
%\vspace*{-1cm}
\caption{The neutrino parameter space, mixing angle against $\Delta
m^2$, including the results from the different solar and atmospheric
neutrino oscillation experiments. Note the threshold of cosmologically
important masses, cosmologically detectable neutrinos (by CMB and LSS
observations), and cosmologically excluded range of masses. Adapted
from Refs.~\cite{CGG} and \cite{KK}.}
\label{NuSummary}
\end{center}
\end{figure}
%%%%%%%%%%%%%%%%%%%%%%%%%

I will now go through the various logical arguments that exclude
neutrinos as the {\em dominant} component of the missing dark matter in
the universe. Is it possible that neutrinos with a mass \ $4\ {\rm eV}
\leq m_\nu \leq 40$ eV be the non-baryonic PDM component? For instance,
could massive neutrinos constitute the dark matter halos of galaxies?
For neutrinos to be gravitationally bound to galaxies it is necessary
that their velocity be less that the escape velocity $v_{\rm esc}$, and
thus their maximum momentum is $p_{\rm max} = m_\nu\,v_{\rm esc}$. How
many neutrinos can be packed in the halo of a galaxy? Due to the Pauli
exclusion principle, the maximum number density is given by that of a
completely degenerate Fermi gas with momentum $p_{\rm F} = p_{\rm max}$,
i.e. $n_{\rm max} = p_{\rm max}^3/3\pi^2$. Therefore, the maximum local
density in dark matter neutrinos is $\rho_{\rm max} = n_{\rm max}m_\nu =
m_\nu^4\, v_{\rm esc}^3/3\pi^2$, which must be greater than the typical
halo density $\rho_{\rm halo} = 0.3$ GeV\,cm$^{-3}$. For a typical
spiral galaxy, this constraint, known as the Tremaine-Gunn limit, gives
$m_\nu \geq 40$ eV, see Ref.~\cite{TG}. However, this mass, even for a
single species, say the tau-neutrino, gives a value for $\Omega_\nu
h^2=0.5$, which is far too high for structure formation. Neutrinos of
such a low mass would constitute a relativistic hot dark matter
component, which would wash-out structure below the supercluster scale,
against evidence from present observations, see Fig.~19.
Furthermore, applying the same phase-space argument to the neutrinos as
dark matter in the halo of dwarf galaxies gives $m_\nu \geq 100$ eV,
beyond closure density (\ref{OmegaNeutrinos}). We must conclude that the
simple idea that light neutrinos could constitute the particle dark
matter on all scales is ruled out. They could, however, still play a
role as a sub-dominant hot dark matter component in a flat CDM model. In
that case, a neutrino mass of order 1 eV is not cosmological excluded,
see Fig.~19.

Another possibility is that neutrinos have a large mass, of order a
few GeV. In that case, their number density at decoupling, see Section
2.5.1, is suppressed by a Boltzmann factor, $\sim \exp(-m_\nu/T_{\rm
dec})$.  For masses $m_\nu > T_{\rm dec} \simeq 0.8$ MeV, the present
energy density has to be computed as a solution of the corresponding
Boltzmann equation. Apart from a logarithmic correction, one finds
$\Omega_\nu h^2 \simeq 0.1 (10\ {\rm GeV}/m_\nu)^2$ for Majorana
neutrinos and slightly smaller for Dirac neutrinos. In either case,
neutrinos could be the dark matter only if their mass was a few
GeV. Laboratory limits for $\nu_\tau$ of around 18 MeV~\cite{PDG}, and
much more stringent ones for $\nu_\mu$ and $\nu_e$, exclude the known
light neutrinos. However, there is always the possibility of a fourth
unknown heavy and stable (perhaps sterile) neutrino. If it couples to
the Z boson and has a mass below 45 GeV for Dirac neutrinos (39.5 GeV
for Majorana neutrinos), then it is ruled out by measurements at LEP
of the invisible width of the Z. There are two logical alternatives,
either it is a sterile neutrino (it does not couple to the Z), or it
does couple but has a larger mass. In the case of a Majorana neutrino
(its own antiparticle), their abundance, for this mass range, is too
small for being cosmologically relevant, $\Omega_\nu h^2 \leq
0.005$. If it were a Dirac neutrino there could be a lepton asymmetry,
which may provide a higher abundance (similar to the case of
baryogenesis). However, neutrinos scatter on nucleons via the weak
axial-vector current (spin-dependent) interaction. For the small
momentum transfers imparted by galactic WIMPs, such collisions are
essentially coherent over an entire nucleus, leading to an enhancement
of the effective cross section. The relatively large detection rate in
this case allowes one to exclude fourth-generation Dirac neutrinos for
the galactic dark matter~\cite{WIMP}. Anyway, it would be very
implausible to have such a massive neutrino today, since it would have
to be stable, with a life-time greater than the age of the universe,
and there is no theoretical reason to expect a massive sterile
neutrino that does not oscillate into the other neutrinos.

Of course, the definitive test to the possible contribution of neutrinos
to the overall density of the universe would be to measure {\em
directly} their mass in laboratory experiments. There are at present two
types of experiments: neutrino oscillation experiments, which measure
only {\em differences} in squared masses, and direct mass-searches
experiments, like the tritium $\beta$-spectrum and the neutrinoless
double-$\beta$ decay experiments, which measure directly the mass of the
electron neutrino. The former experiments give a bound $m_{\nu_e} \lsim
$ 2.3 eV (95\% c.l.)~\cite{Tritium}, while the latter
claim~\cite{Klapdor} they have a positive evidence for a Majorana
neutrino of mass $m_\nu = 0.05 - 0.89$ eV (95\% c.l.), although this
result still awaits confirmation by other experiments.  Neutrinos with
such a mass could very well constitute the HDM component of the
universe, $\Omega_{\rm HDM} \lsim 0.15$.  The oscillation experiments
give a range of possibilities for $\Delta\, m_\nu^2 = 0.3 - 3\ {\rm
eV}^2$ from LSND (not yet confirmed by Miniboone), to the atmospheric
neutrino oscillations from SuperKamiokande ($\Delta\, m_\nu^2 \simeq
2.2\pm0.5\times 10^{-3}\ {\rm eV}^2\,, \ \ \tan^2\theta=1.0\pm0.3$) and
the solar neutrino oscillations from KamLAND and the Sudbury Neutrino
Observatory ($\Delta\, m_\nu^2 \simeq 8.2\pm0.3\times 10^{-5}\ {\rm
eV}^2\,, \ \ \tan^2\theta=0.39\pm0.05$), see Ref.~\cite{CGG}. Only the
first two possibilities would be cosmologically relevant, see
Fig.~19. Thanks to recent observations by WMAP, 2dFGRS and SDSS, we can
put stringent limits on the absolute scale of neutrino masses, see below
(Section 3.4).

\subsubsection{Weakly Interacting Massive Particles}

Unless we drastically change the theory of gravity on large scales,
baryons cannot make up the bulk of the dark matter. Massive neutrinos
are the only alternative among the known particles, but they are
essentially ruled out as a universal dark matter candidate, even if
they may play a subdominant role as a hot dark matter component. There
remains the mystery of what is the physical nature of the dominant
cold dark matter component. Something like a heavy stable neutrino, a
generic Weakly Interacting Massive Particle (WIMP), could be a
reasonable candidate because its present abundance could fall within
the expected range,
\begin{equation}\label{OmegaPDM}
\Omega_{\rm PDM} h^2 \sim {G^{3/2}T_0^3h^2\over H_0^2\langle
\sigma_{\rm ann}v_{\rm rel}\rangle} =
{3\times10^{-27}\ {\rm cm}^3{\rm s}^{-1}\over \langle
\sigma_{\rm ann}v_{\rm rel}\rangle}\,.
\end{equation}
Here $v_{\rm rel}$ is the relative velocity of the two incoming dark
matter particles and the brackets $\langle\cdot\rangle$ denote a thermal
ave\-rage at the freeze-out temperature, $T_{\rm f}\simeq m_{\rm
PDM}/20$, when the dark matter particles go out of equilibrium with
radiation. The value of $\langle\sigma_{\rm ann}v_{\rm rel}\rangle$
needed for $\Omega_{\rm PDM} \approx 1$ is remarkably close to what one
would expect for a WIMP with a mass $m_{\rm PDM}=100$ GeV, \
$\langle\sigma_{\rm ann}v_{\rm rel} \rangle\sim \alpha^2/8\pi\,m_{\rm
PDM}\sim 3\times10^{-27}\ {\rm cm}^3 {\rm s}^{-1}$. We still do not know
whether this is just a coincidence or an important hint on the nature of
dark matter.

%%%%%%%%%%%%%%%%%%%%%%%%%
\begin{figure}[htb]
\begin{center}%\vspace*{1cm}
\includegraphics[width=6cm]{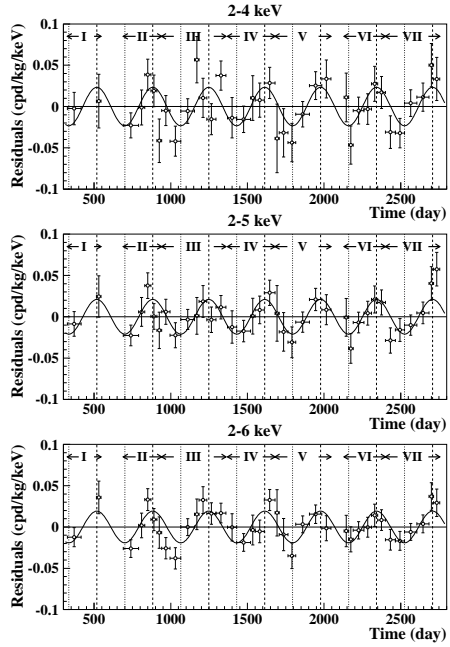}
%\vspace*{-2.8cm}
\caption{The annual-modulation signal accumulated over 7 years is
consistent with a neutralino of mass of $\ m_\chi = 59^{\ +17}_{\
-14}$ GeV and a proton cross section of $\ \xi\sigma_p = 7.0 ^{\
+0.4}_{\ -1.2} \times10^{-6}$ pb, according to DAMA. 
From Ref.~\cite{DAMA}.}
\label{Dama}
\end{center}
\end{figure}
%%%%%%%%%%%%%%%%%%%%%%%%%

There are a few theoretical candidates for WIMPs, like the neutralino,
coming from supersymme\-tric extensions of the standard model of
particle physics,\footnote{For a review of Supersymmetry (SUSY), see
Kazakov's contribution to these Proceedings.} but at present there is no
empirical evidence that such extensions are indeed realized in
nature. In fact, the non-observation of supersymmetric particles at
current accelerators places stringent limits on the neutralino mass and
interaction cross section~\cite{neutralino}. If WIMPs constitute the
dominant component of the halo of our galaxy, it is expected that some
may cross the Earth at a reasonable rate to be detected. The direct
experimental search for them rely on elastic WIMP collisions with the
nuclei of a suitable target. Dark matter WIMPs move at a typical
galactic ``virial'' velocity of around $200-300$ km/s, depending on the
model. If their mass is in the range $10-100$ GeV, the recoil energy of
the nuclei in the elastic collision would be of order 10 keV. Therefore,
one should be able to identify such energy depositions in a macroscopic
sample of the target. There are at present three different methods:
First, one could search for scintillation light in NaI crystals or in
liquid xenon; second, search for an ionization signal in a
semiconductor, typically a very pure germanium crystal; and third, use a
cryogenic detector at 10 mK and search for a measurable temperature
increase of the sample. The main problem with such a type of experiment
is the low expected signal rate, with a typical number below 1
event/kg/day. To reduce natural radioactive contamination one must use
extremely pure substances, and to reduce the background caused by cosmic
rays requires that these experiments be located deeply underground.

The best limits on WIMP scattering cross sections come from some
germanium experiments, like the Criogenic Dark Matter Search (CDMS)
collaboration at Stanford and the Soudan mine~\cite{CDMS}, as well as
from the NaI scintillation detectors of the UK dark matter collaboration
(UKDMC) in the Boulby salt mine in England~\cite{UKDMC}, and the DAMA
experiment in the Gran Sasso laboratory in Italy~\cite{DAMA}. Current
experiments already touch the parameter space expected from
supersymmetric particles, see Fig.~21, and therefore there is a chance
that they actually discover the nature of the missing dark matter. The
problem, of course, is to attribute a tentative signal unambiguously to
galactic WIMPs rather than to some unidentified radioactive background.

%%%%%%%%%%%%%%%%%%%%%%%%%
\begin{figure}[htb]
\begin{center}%\vspace*{1cm}
\includegraphics[width=8cm]{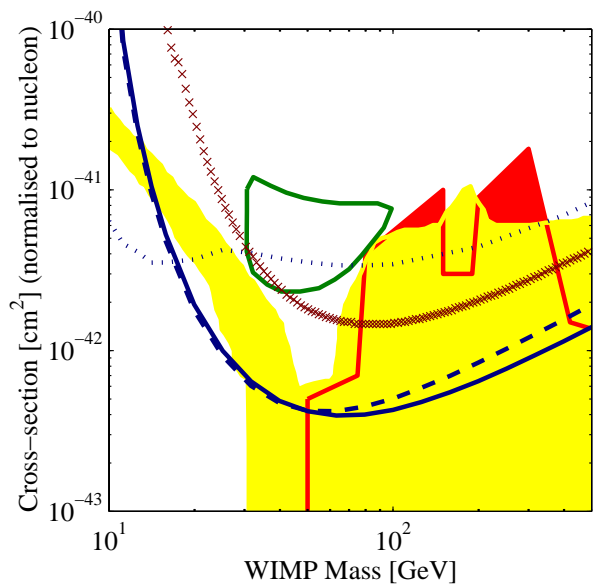}
%\vspace*{-2.8cm}
\caption{Exclusion range for the spin-independent WIMP scattering
cross section per nucleon from the NaI experiments and the Ge
detectors.  The blue lines come from the CDMS experiment, which
exclude the DAMA region at more than 3 sigma. Also shown in yellow and
red is the range of expected counting rates for neutralinos in the
MSSM. From Ref.~\cite{CDMS}.}
\label{fig25}
\end{center}
\end{figure}
%%%%%%%%%%%%%%%%%%%%%%%%%

One specific signature is the annual modulation which arises as the
Earth moves around the Sun.\footnote{The time scale of the Sun's orbit
around the center of the galaxy is too large to be relevant in the
analysis.} Therefore, the net speed of the Earth relative to the
galactic dark matter halo varies, causing a modulation of the expected
counting rate. The DAMA/NaI experiment has actually reported such a
modulation signal, from the combined analysis of their 7-year data,
see Fig.~20 and Ref.~\cite{DAMA}, which provides a confidence level of
99.6\% for a neutralino mass of $\ m_\chi = 52^{\ +10}_{\ -8}$ GeV and
a proton cross section of $\ \xi\sigma_p = 7.2 ^{\ +0.4}_{\ -0.9}
\times10^{-6}$ pb, where $\xi = \rho_\chi/0.3$ GeV\,cm$^{-3}$ is the
local neutralino energy density in units of the galactic halo
density. There has been no confirmation yet of this result from other
dark matter search groups. In fact, the CDMS collaboration claims an
exclusion of the DAMA region at the 3 sigma level, see Fig.~21.
Hopefully in the near future we will have much better sensitivity at
low masses from the Cryogenic Rare Event Search with Superconducting
Thermometers (CRESST) experiment at Gran Sasso. The CRESST
experiment~\cite{CRESST} uses sapphire crystals as targets and a new
method to simultaneously measure the phonons and the scintillating
light from particle interactions inside the crystal, which allows
excellent background discrimination.  Very recently there has been
also the proposal of a completely new method based on a Superheated
Droplet Detector (SDD), which claims to have already a similar
sensitivity as the more standard methods described above, see
Ref.~\cite{SDD}.

There exist other {\em indirect} methods to search for galactic
WIMPs~\cite{Griest}. Such particles could self-annihilate at a certain
rate in the galactic halo, producing a potentially detectable
background of high energy photons or antiprotons. The absence of such
a background in both gamma ray satellites and the Alpha Matter
Spectrometer~\cite{AMS} imposes bounds on their density in the
halo. Alternatively, WIMPs traversing the solar system may interact
with the matter that makes up the Earth or the Sun so that a small
fraction of them will lose energy and be trapped in their cores,
building up over the age of the universe. Their annihilation in the
core would thus produce high energy neutrinos from the center of the
Earth or from the Sun which are detectable by neutrino telescopes. In
fact, SuperKamiokande already covers a large part of SUSY parameter
space. In other words, neutrino telescopes are already competitive
with direct search experiments. In particular, the AMANDA experiment
at the South Pole~\cite{AMANDA}, which has approximately $10^3$
Cherenkov detectors several km deep in very clear ice, over a volume
$\sim 1$ km$^3$, is competitive with the best direct searches
proposed. The advantages of AMANDA are also directional, since the
arrays of Cherenkov detectors will allow one to reconstruct the
neutrino trajectory and thus its source, whether it comes from the
Earth or the Sun. AMANDA recently reported the detection of TeV
neutrinos~\cite{AMANDA}.

\subsection{The age of the universe $t_0$}

The universe must be older than the oldest objects it contains. Those
are believed to be the stars in the oldest clusters in the Milky Way,
globular clusters. The most reliable ages come from the application of
theoretical models of stellar evolution to observations of old stars in
globular clusters. For about 30 years, the ages of globular clusters
have remained reasonable stable, at about 15 Gyr~\cite{Vandenberg}.
However, recently these ages have been revised downward~\cite{Krauss}. 

During the 1980s and 1990s, the globular cluster age estimates have
improved as both new observations have been made with CCDs, and since
refinements to stellar evolution models, including opacities,
consideration of mixing, and different chemical abundances have been
incorporated~\cite{Chaboyer}. From the theory side, uncertainties in
globular cluster ages come from uncertainties in convection models,
opacities, and nuclear reaction rates. From the observational side,
uncertainties arise due to corrections for dust and chemical
composition. However, the dominant source of systematic errors in the
globular cluster age is the uncertainty in the cluster distances.
Fortunately, the Hipparcos satellite recently provided geometric
parallax measurements for many nearby old stars with low metallicity,
typical of glubular clusters, thus allowing for a new calibration of the
ages of stars in globular clusters, leading to a downward revision to
$10-13$ Gyr~\cite{Chaboyer}. Moreover, there were very few stars in the
Hipparcos catalog with both small parallax erros and low metal
abundance. Hence, an increase in the sample size could be critical in
reducing the statatistical uncertaintites for the calibration of the
globular cluster ages.  There are already proposed two new parallax
satellites, NASA's Space Interferometry Mission (SIM) and ESA's mission,
called GAIA, that will give 2 or 3 orders of magnitude more accurate
parallaxes than Hipparcos, down to fainter magnitude limits, for several
orders of magnitude more stars. Until larger samples are available,
however, distance errors are likely to be the largest source of
systematic uncertainty to the globular cluster age~\cite{Freedman}.

%%%%%%%%%%%%%%%%%%%%%%%%%
\begin{figure}[htb]
%\vspace*{-6cm}
\begin{center}
\includegraphics[width=8cm]{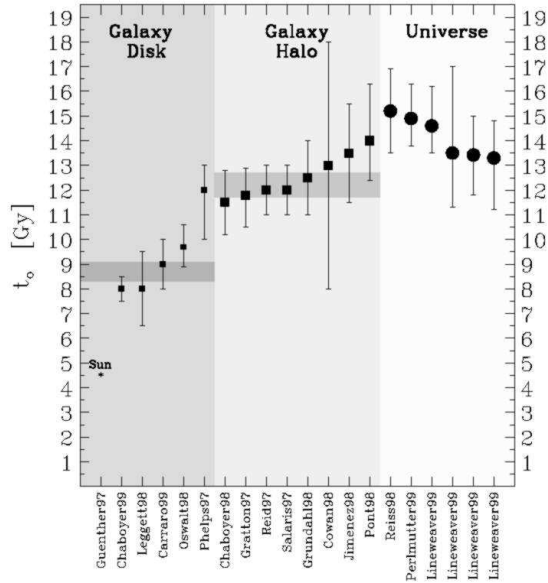}
\end{center}\vspace*{-.5cm}
\caption{The recent estimates of the age of the universe and that of the
oldest objects in our galaxy. The last three points correspond to the
combined analysis of 8 different measurements, for $h=$ 0.64, 0.68 and
7.2, which indicates a relatively weak dependence on $h$. The age of the
Sun is accurately known and is included for reference. Error bars
indicate 1$\sigma$ limits. The averages of the ages of the Galactic Halo
and Disk are shaded in gray. Note that there isn't a single age estimate
more than 2$\sigma$ away from the average. The result $t_0 > t_{\rm
gal}$ is logically inevitable, but the standard EdS model does not
satisfy this unless $h<0.55$. From Ref.~\cite{Charley}.}
\label{fig29}
\end{figure}
%%%%%%%%%%%%%%%%%%%%%%%%%

The supernovae groups can also determine the age of the universe from
their high redshift observations. The high confidence regions in the
$(\Omega_{\rm M}, \Omega_\Lambda)$ plane are almost parallel to the
contours of constant age. For any value of the Hubble constant less
than $H_0 = 70$ km/s/Mpc, the implied age of the universe is greater
than 13 Gyr, allowing enough time for the oldest stars in globular
clusters to evolve~\cite{Chaboyer}. Integrating over $\Omega_{\rm M}$
and $\Omega_\Lambda$, the best fit value of the age in Hubble-time
units is $H_0t_0=0.93\pm0.06$ or equivalently $t_0 = 14.1 \pm 1.0 \
(0.65\, h^{-1})$ Gyr, see Ref.~\cite{SCP}. Furthermore, a combination
of 8 independent recent measurements: CMB anisotropies, type Ia SNe,
cluster mass-to-light ratios, cluster abundance evolution, cluster
baryon fraction, deuterium-to-hidrogen ratios in quasar spectra,
double-lobed radio sources and the Hubble constant, can be used to
determine the present age of the universe~\cite{Charley}. The result
is shown in Fig.~\ref{fig29}, compared to other recent
determinations. The best fit value for the age of the universe is,
according to this analysis, $t_0 = 13.4 \pm 1.6$ Gyr, about a billion
years younger than other recent estimates~\cite{Charley}.

%%%%%%%%%%%%%%%%%%%%%%%%%
\begin{figure}[htb]
\label{fig:WMAP}
\begin{center}
\includegraphics[width=10cm]{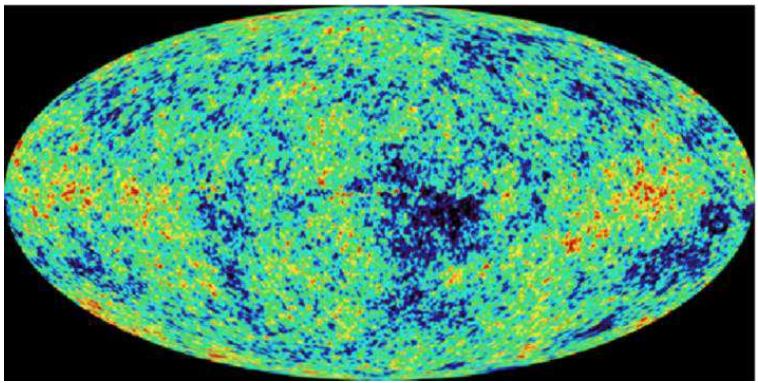} 
\caption{The anisotropies of the microwave background measured by the
WMAP satellite with 10 arcminute resolution. It shows the intrinsic
CMB anisotropies at the level of a few parts in $10^5$.  The galactic
foreground has been properly subtracted. The amount of information
contained in this map is enough to determine most of the cosmological
parameters to few percent accuracy. From Ref.~\cite{WMAP}.}
\end{center}
\end{figure}
%%%%%%%%%%%%%%%%%%%%%%%%%

\subsection{Cosmic Microwave Background Anisotropies}

The cosmic microwave background has become in the last five years the
Holy Grail of Cosmology, since precise observations of the temperature
and polarization anisotropies allow in principle to determine the
parameters of the Standard Model of Cosmology with very high
accuracy. Recently, the WMAP satellite has provided with a very
detailed map of the microwave anisotropies in the sky, see Fig.~23,
and indeed has fulfilled our expectations, see Table~2.

%%%%%%%%%%%%%%%%%%%%%%%%%
\begin{figure}[htb]
\label{fig:Power}
\begin{center}
\includegraphics[width=8cm]{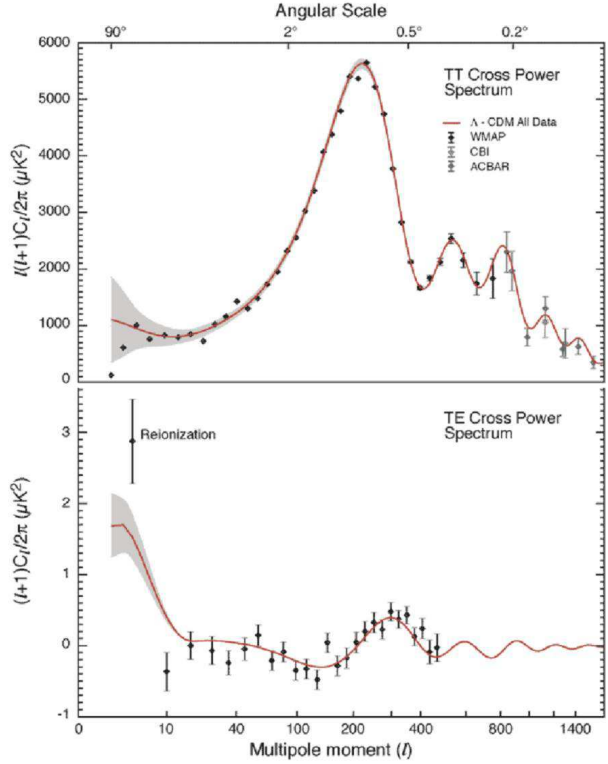} 
\caption{The Angular Power Spectrum of CMB temperature anisotropies,
compared with the cross-correlation of temperature-polarization
anisotropies. From Ref.~\cite{WMAP}.}
\end{center}
\end{figure}
%%%%%%%%%%%%%%%%%%%%%%%%%

The physics of the CMB anisotropies is relatively simple~\cite{CMB}.
The universe just before recombination is a very tightly coupled
fluid, due to the large electromagnetic Thomson cross section
$\sigma_T = 8\pi\alpha^2/3m_e^2\simeq0.7$ barn. Photons scatter off
charged particles (protons and electrons), and carry energy, so they
feel the gravitational potential associated with the perturbations
imprinted in the metric during inflation. An overdensity of baryons
(protons and neutrons) does not collapse under the effect of gravity
until it enters the causal Hubble radius. The perturbation continues
to grow until radiation pressure opposes gravity and sets up acoustic
oscillations in the plasma, very similar to sound waves. Since
overdensities of the same size will enter the Hubble radius at the
same time, they will oscillate in phase. Moreover, since photons
scatter off these baryons, the acoustic oscillations occur also in the
photon field and induces a pattern of peaks in the temperature
anisotropies in the sky, at different angular scales, see
Fig.~24. There are three different effects that determine the
temperature anisotropies we observe in the CMB. First, {\em gravity}:
photons fall in and escape off gravitational potential wells,
characterized by $\Phi$ in the comoving gauge, and as a consequence
their frequency is gravitationally blue- or red-shifted,
$\delta\nu/\nu = \Phi$. If the gravitational potential is not
constant, the photons will escape from a larger or smaller potential
well than they fell in, so their frequency is also blue- or
red-shifted, a phenomenon known as the Rees-Sciama effect.  Second,
{\em pressure}: photons scatter off baryons which fall into
gravitational potential wells and the two competing forces create
acoustic waves of compression and rarefaction. Finally, {\em
velocity}: baryons accelerate as they fall into potential wells. They
have minimum velocity at maximum compression and rarefaction. That is,
their velocity wave is exactly $90^\circ$ off-phase with the acoustic
waves. These waves induce a Doppler effect on the frequency of the
photons.  The temperature anisotropy induced by these three effects is
therefore given by~\cite{CMB}
\begin{equation}
{\delta T\over T}({\bf r}) = \Phi({\bf r},t_{\rm dec}) + 
2\int_{t_{\rm dec}}^{t_0} \dot\Phi({\bf r},t) dt \ + \
{1\over3}{\delta\rho\over\rho} \ - \ {{\bf r}\cdot{\bf v}\over c}\,.
\label{TemperatureAnisotropy}
\end{equation}
Metric perturbations of different wavelengths enter the horizon at
different times. The largest wavelengths, of size comparable to our
present horizon, are entering now. There are perturbations with
wavelengths comparable to the size of the horizon at the time of last
scattering, of projected size about $1^\circ$ in the sky today, which
entered precisely at decoupling. And there are perturbations with
wavelengths much smaller than the size of the horizon at last
scattering, that entered much earlier than decoupling, all the way to
the time of radiation-matter equality, which have gone through several
acoustic oscillations before last scattering. All these perturbations
of different wavelengths leave their imprint in the CMB anisotropies.

%%%%%%%%%%%%%%%%%%%%%%%%%%%%%%%%%%%%%%%%%%%%%%%%%%
\begin{table}[htb]
\label{table1}
\caption{{\bf The parameters of the standard cosmological model}. The
standard model of cosmology has about 20 different parameters, needed
to describe the background space-time, the matter content and the
spectrum of metric perturbations. We include here the present range of
the most relevant parameters (with 1$\sigma$ errors), as recently
determined by WMAP, and the error with which the Planck satellite will
be able to determine them in the near future.  The rate of expansion
is written in units of $H=100\,h$ km/s/Mpc.}
\begin{center}
\begin{tabular}{lllll}
\hline
physical quantity & 
\multicolumn{1}{c}{symbol} &
\multicolumn{1}{c}{WMAP} &
\multicolumn{1}{c}{Planck} \\
\hline
\hline
total density & 
\multicolumn{1}{c}{$\Omega_0$} &
\multicolumn{1}{c}{$1.02\pm0.02$} &
\multicolumn{1}{c}{0.7\%} \\
\hline
baryonic matter & 
\multicolumn{1}{c}{$\Omega_{\rm B}$} &
\multicolumn{1}{c}{$0.044\pm0.004$} &
\multicolumn{1}{c}{0.6\%} \\
\hline
cosmological constant & 
\multicolumn{1}{c}{$\Omega_\Lambda$} &
\multicolumn{1}{c}{$0.73\pm0.04$} &
\multicolumn{1}{c}{0.5\%} \\
\hline
cold dark matter & 
\multicolumn{1}{c}{$\Omega_{\rm M}$} &
\multicolumn{1}{c}{$0.23\pm0.04$} &
\multicolumn{1}{c}{0.6\%} \\
\hline
hot dark matter & 
\multicolumn{1}{c}{$\Omega_\nu h^2$} &
\multicolumn{1}{c}{$<0.0076$ (95\% c.l.)} &
\multicolumn{1}{c}{1\%} \\
\hline
sum of neutrino masses \hspace{0.7cm} & 
\multicolumn{1}{c}{$\sum m_\nu$ (eV)} &
\multicolumn{1}{c}{$<0.23$ (95\% c.l.)} &
\multicolumn{1}{c}{1\%} \\
\hline
CMB temperature & 
\multicolumn{1}{c}{$T_0\ (K)$} &
\multicolumn{1}{c}{$2.725\pm0.002$} &
\multicolumn{1}{c}{0.1\%} \\
\hline
baryon to photon ratio & 
\multicolumn{1}{c}{$\eta\times10^{10}$} &
\multicolumn{1}{c}{$6.1\pm0.3$} &
\multicolumn{1}{c}{0.5\%} \\
\hline
baryon to matter ratio & 
\multicolumn{1}{c}{$\Omega_{\rm B}/\Omega_{\rm M}$} &
\multicolumn{1}{c}{$0.17\pm0.01$} &
\multicolumn{1}{c}{1\%} \\
\hline
spatial curvature & 
\multicolumn{1}{c}{$\Omega_K$} &
\multicolumn{1}{c}{$<0.02$ (95\% c.l.)} &
\multicolumn{1}{c}{0.5\%} \\
\hline
rate of expansion & 
\multicolumn{1}{c}{$h$} &
\multicolumn{1}{c}{$0.71\pm0.03$} &
\multicolumn{1}{c}{0.8\%} \\
\hline
age of the universe & 
\multicolumn{1}{c}{$t_0$ (Gyr)} &
\multicolumn{1}{c}{$13.7\pm0.2$} &
\multicolumn{1}{c}{0.1\%} \\
\hline
age at decoupling & 
\multicolumn{1}{c}{$t_{\rm dec}$ (kyr)} &
\multicolumn{1}{c}{$379\pm8$} &
\multicolumn{1}{c}{0.5\%} \\
\hline
age at reionization & 
\multicolumn{1}{c}{$t_{\rm r}$ (Myr)} &
\multicolumn{1}{c}{$180\pm100$} &
\multicolumn{1}{c}{5\%} \\
\hline
spectral amplitude & 
\multicolumn{1}{c}{$A$} &
\multicolumn{1}{c}{$0.833\pm0.085$} &
\multicolumn{1}{c}{0.1\%} \\
\hline
spectral tilt & 
\multicolumn{1}{c}{$n_{\rm s}$} &
\multicolumn{1}{c}{$0.98\pm0.03$} &
\multicolumn{1}{c}{0.2\%} \\
\hline
spectral tilt variation& 
\multicolumn{1}{c}{$dn_{\rm s}/d\ln k$} &
\multicolumn{1}{c}{$-0.031\pm0.017$} &
\multicolumn{1}{c}{0.5\%} \\
\hline
tensor-scalar ratio & 
\multicolumn{1}{c}{$r$} &
\multicolumn{1}{c}{$<0.71$ (95\% c.l.)} &
\multicolumn{1}{c}{5\%} \\
\hline
reionization optical depth & 
\multicolumn{1}{c}{$\tau$} &
\multicolumn{1}{c}{$0.17\pm0.04$} &
\multicolumn{1}{c}{5\%} \\
\hline
redshift of equality & 
\multicolumn{1}{c}{$z_{\rm eq}$} &
\multicolumn{1}{c}{$3233\pm200$} &
\multicolumn{1}{c}{5\%} \\
\hline
redshift of decoupling & 
\multicolumn{1}{c}{$z_{\rm dec}$} &
\multicolumn{1}{c}{$1089\pm1$} &
\multicolumn{1}{c}{0.1\%} \\
\hline
width of decoupling & 
\multicolumn{1}{c}{$\Delta z_{\rm dec}$} &
\multicolumn{1}{c}{$195\pm2$} &
\multicolumn{1}{c}{1\%} \\
\hline
redshift of reionization & 
\multicolumn{1}{c}{$z_{\rm r}$} &
\multicolumn{1}{c}{$20\pm10$} &
\multicolumn{1}{c}{2\%} \\
\hline
\end{tabular}
\end{center}
\end{table}
%%%%%%%%%%%%%%%%%%%%%%%%%%%%%%%%%%%%%%%%%%%%%%%%%%

The baryons at the time of decoupling do not feel the gravitational
attraction of perturbations with wavelength greater than the size of
the horizon at last scattering, because of causality.  Perturbations
with exactly that wavelength are undergoing their first contraction,
or acoustic compression, at decoupling. Those perturbations induce a
large peak in the temperature anisotropies power spectrum, see
Fig.~24. Perturbations with wavelengths smaller than these will have
gone, after they entered the Hubble scale, through a series of
acoustic compressions and rarefactions, which can be seen as secondary
peaks in the power spectrum. Since the surface of last scattering is
not a sharp discontinuity, but a region of $\Delta z \sim 100$, there
will be scales for which photons, travelling from one energy
concentration to another, will erase the perturbation on that scale,
similarly to what neutrinos or HDM do for structure on small scales.
That is the reason why we don't see all the acoustic oscillations with
the same amplitude, but in fact they decay exponentialy towards
smaller angular scales, an effect known as Silk damping, due to photon
diffusion~\cite{Silk,CMB}.

From the observations of the CMB anisotropies it is possible to
determine most of the parameters of the Standard Cosmological Model
with few percent accuracy, see Table~2. However, there are many
degeneracies between parameters and it is difficult to disentangle one
from another.  For instance, as mentioned above, the first peak in the
photon distribution corresponds to overdensities that have undergone
half an oscillation, that is, a compression, and appear at a scale
associated with the size of the horizon at last scattering, about
$1^\circ$ projected in the sky today. Since photons scatter off
baryons, they will also feel the acoustic wave and create a peak in
the correlation function. The height of the peak is proportional to
the amount of baryons: the larger the baryon content of the universe,
the higher the peak. The position of the peak in the power spectrum
depends on the geometrical size of the particle horizon at last
scattering. Since photons travel along geodesics, the projected size
of the causal horizon at decoupling depends on whether the universe is
flat, open or closed. In a flat universe the geodesics are straight
lines and, by looking at the angular scale of the first acoustic peak,
we would be measuring the actual size of the horizon at last
scattering. In an open universe, the geodesics are inward-curved
trajectories, and therefore the projected size on the sky appears
smaller. In this case, the first acoustic peak should occur at higher
multipoles or smaller angular scales. On the other hand, for a closed
universe, the first peak occurs at smaller multipoles or larger
angular scales. The dependence of the position of the first acoustic
peak on the spatial curvature can be approximately given by $l_{\rm
peak} \simeq 220\,\Omega_0^{-1/2}$, where $\Omega_0=\Omega_{\rm M} +
\Omega_\Lambda = 1-\Omega_K$. Present observations by WMAP and other
experiments give $\Omega_0=1.00\pm0.02$ at one standard
deviation~\cite{WMAP}.

%%%%%%%%%%%%%%%%%%%%%%%%%
\begin{figure}[htb]
\label{Future}
\begin{center}
\includegraphics[width=9cm]{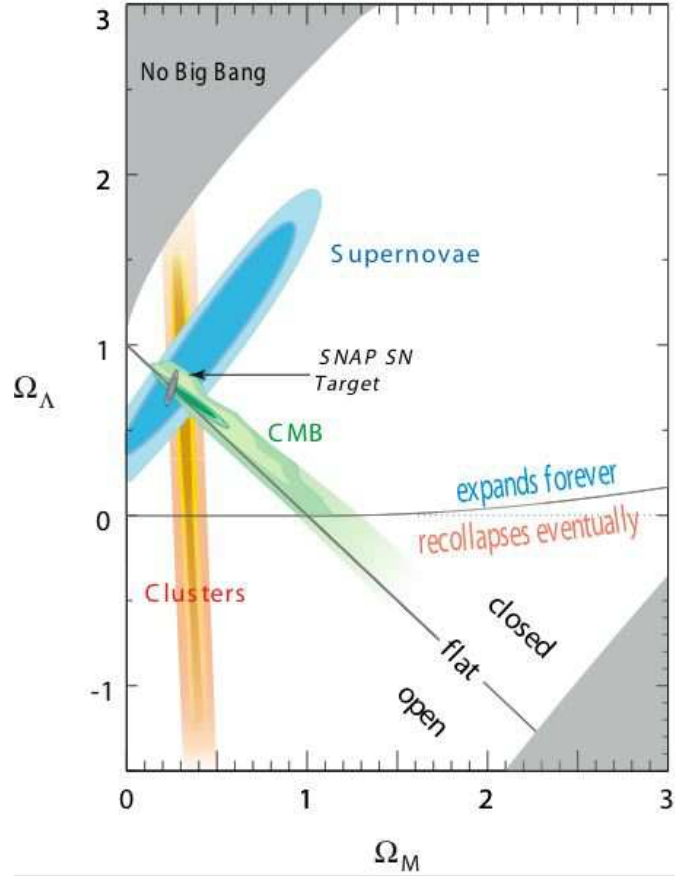} 
\caption{The $(\Omega_M,\,\Omega_\Lambda)$ plane with the present data
set of cosmological observations $-$ the acceleration of the universe,
the large scale structure and the CMB anisotropies $-$ as well as the
future determinations by SNAP and Planck of the fundamental parameters 
which define our Standard Model of Cosmology.}
\end{center}
\end{figure}
%%%%%%%%%%%%%%%%%%%%%%%%%

The other acoustic peaks occur at harmonics of this, corresponding to
smaller angular scales. Since the amplitude and position of the
primary and secondary peaks are directly determined by the sound speed
(and, hence, the equation of state) and by the geometry and expansion
of the universe, they can be used as a powerful test of the density of
baryons and dark matter, and other cosmological parameters. With the
joined data from WMAP, VSA, CBI and ACBAR, we have rather good
evidence of the existence of the second and third acoustic peaks,
which confirms one of the most important predictions of inflation $-$
the non-causal origin of the primordial spectrum of perturbations $-$,
and rules out cosmological defects as the dominant source of structure
in the universe~\cite{Turok}. Moreover, since the observations of CMB
anisotropies now cover almost three orders of magnitude in the size of
perturbations, we can determine the much better accuracy the value of
the spectral tilt, $n=0.98\pm0.03$, which is compatible with the
approximate scale invariant spectrum needed for structure formation,
and is a prediction of the simplest models of inflation.  Soon after
the release of data from WMAP, there was some expectation at the claim
of a scale-dependent tilt. Nowadays, with better resolution in the
linear matter power spectrum from SDSS~\cite{Seljak}, we can not
conclude that the spectral tilt has any observable dependence on
scale.

The microwave background has become also a testing ground for theories
of particle physics. In particular, it already gives stringent
constraints on the mass of the neutrino, when analysed together with
large scale structure observations. Assuming a flat $\Lambda$CDM model,
the 2-sigma upper bounds on the sum of the masses of light neutrinos is
$\sum m_\nu < 1.0$ eV for degenerate neutrinos (i.e. without a large
hierachy between them) if we don't impose any priors, and it comes down
to $\sum m_\nu < 0.6$ eV if one imposes the bounds coming from the HST
measurements of the rate of expansion and the supernova data on the
present acceleration of the universe~\cite{Nu}. The final bound on the
neutrino density can be expressed as $\Omega_\nu\,h^2 = \sum m_\nu/93.2$
eV $\leq 0.01$. In the future, both with Planck and with the 
Atacama Cosmology Telescope (ACT) we will be able to put constraints
on the neutrino masses down to the 0.1 eV level.

Moreover, the present data is good enough that we can start to put
constraints on the models of inflation that give rise to structure.  In
particular, multifield models of inflation predict a mixture of
adiabatic and isocurvature perturbations,\footnote{This mixture is
generic, unless all the fields thermalize simultaneously at reheating,
just after inflation, in which case the entropy perturbations that would
give rise to the isocurvature modes disappear.} and their signatures in
the cosmic microwave background anisotropies and the matter power
spectrum of large scale structure are specific and perfectly
distinguishable. Nowadays, thanks to precise CMB, LSS and SNIa data, one
can put rather stringent limits on the relative fraction and correlation
of the isocurvature modes to the dominant adiabatic 
perturbations~\cite{iso}.

We can summarize this Section by showing the region in parameter space
where we stand nowadays, thanks to the recent cosmological
observations.  We have plotted that region in Fig.~25. One
could also superimpose the contour lines corresponding to equal
$t_0H_0$ lines, as a cross check.  It is extraordinary that only in
the last few months we have been able to reduce the concordance region
to where it stands today, where all the different observations seem to
converge. There are still many uncertainties, mainly systematic;
however, those are quickly decreasing and becoming predominantly
statistical. In the near future, with precise observations of the
anisotropies in the microwave background temperature and polarization
anisotropies, thanks to Planck satellite, we will be able to reduce
those uncertainties to the level of one percent. This is the reason
why cosmologists are so excited and why it is claimed that we live in
the Golden Age of Cosmology.

\section{THE INFLATIONARY PARADIGM}

The hot Big Bang theory is nowadays a very robust edifice, with many
independent observational checks: the expansion of the universe; the
abundance of light elements; the cosmic microwave background; a
predicted age of the universe compatible with the age of the oldest
objects in it, and the formation of structure via gravitational collapse
of initially small inhomogeneities. Today, these observations are
confirmed to within a few percent accuracy, and have helped establish
the hot Big Bang as the preferred model of the universe. All the physics
involved in the above observations is routinely tested in the laboratory
(atomic and nuclear physics experiments) or in the solar system
(general relativity).

However, this theory leaves a range of crucial questions unanswered,
most of which are initial conditions' problems. There is the reasonable
assumption that these cosmological problems will be solved or explained
by {\em new physical principles} at high energies, in the early
universe. This assumption leads to the natural conclusion that accurate
observations of the present state of the universe may shed light onto
processes and physical laws at energies above those reachable by
particle accelerators, present or future. We will see that this is a
very optimistic approach indeed, and that there are many unresolved
issues related to those problems. However, there might be in the near
future reasons to be optimistic.

\subsection{Shortcomings of Big Bang Cosmology}

The Big Bang theory could not explain the origin of matter and structure
in the universe; that is, the origin of the matter--antimatter
asymmetry, without which the universe today would be filled by a uniform
radiation continuosly expanding and cooling, with no traces of matter,
and thus without the possibility to form gravitationally bound systems
like galaxies, stars and planets that could sustain life. Moreover, the
standard Big Bang theory assumes, but cannot explain, the origin of the
extraordinary smoothness and flatness of the universe on the very large
scales seen by the microwave background probes and the largest galaxy
catalogs. It cannot explain the origin of the primordial density
perturbations that gave rise to cosmic structures like galaxies,
clusters and superclusters, via gravitational collapse; the quantity and
nature of the dark matter that we believe holds the universe together;
nor the origin of the Big Bang itself.

A summary~\cite{JGB} of the problems that the Big Bang theory cannot
explain is:

\begin{itemize}

\item The global structure of the universe.

- Why is the universe so close to spatial flatness?

- Why is matter so homogeneously distributed on large scales?

\item The origin of structure in the universe.

- How did the primordial spectrum of density perturbations originate?

\item The origin of matter and radiation.

- Where does all the energy in the universe come from?

- What is the nature of the dark matter in the universe?

- How did the matter-antimatter asymmetry arise?

%\newpage

\item The initial singularity.

- Did the universe have a beginning?

- What is the global structure of the universe beyond our observable 
patch?

\end{itemize}

\noindent
Let me discuss one by one the different issues:

\subsubsection{The Flatness Problem}

The Big Bang theory assumes but cannot explain the extraordinary spatial
flatness of our local patch of the universe. In the general FRW metric
(\ref{FRWmetric}) the parameter $K$ that characterizes spatial curvature
is a free parameter. There is nothing in the theory that determines this
parameter a priori. However, it is directly related, via the Friedmann
equation (\ref{FriedmannEquation}), to the dynamics, and thus the matter
content, of the universe,
\begin{equation}\label{SpatialK}
K = {8\pi G\over3} \rho a^2 - H^2a^2 \ = \ {8\pi G\over3} \rho a^2
\Big({\Omega-1\over\Omega}\Big)\,.
\end{equation}
We can therefore define a new variable,
\begin{equation}\label{NewVariableX}
x \equiv {\Omega-1\over\Omega} = {{\rm const.}\over\rho a^2}\,,
\end{equation}
whose time evolution is given by
\begin{equation}\label{Stability}
x' = {dx\over dN} = (1+3\omega)\,x\,,
\end{equation}
where $N=\ln(a/a_i)$ characterizes the {\em number of $e$-folds} of
universe expansion ($dN=H dt$) and where we have used
Eq.~(\ref{EnergyConservation}) for the time evolution of the total
energy, $\rho a^3$, which only depends on the barotropic ratio
$\omega$. It is clear from Eq.~(\ref{Stability}) that the phase-space
diagram $(x,x')$ presents an unstable critical (saddle) point at $x=0$
for $\omega > -1/3$, i.e. for the radiation ($\omega = 1/3$) and matter
($\omega = 0$) eras. A small perturbation from $x=0$ will drive the
system towards $x=\pm\infty$. Since we know the universe went through
both the radiation era (because of primordial nucleosynthesis) and the
matter era (because of structure formation), tiny deviations from
$\Omega=1$ would have grown since then, such that today
\begin{equation}\label{x0}
x_0 = {\Omega_0 - 1\over\Omega_0} = x_{\rm in}\,
\Big({T_{\rm in}\over T_{\rm eq}}\Big)^2 (1+z_{\rm eq})\,.
\end{equation}
In order that today's value be in the range $0.1 < \Omega_0 < 1.2$, or
$x_0 \approx {\cal O}(1)$, it is required that at, say, primordial
nucleosynthesis
($T_{_{\rm NS}} \simeq 10^6\, T_{\rm eq}$) \ its value be
\begin{equation}\label{OmegaNucleosynthesis}
\Omega(t_{_{\rm NS}}) = 1 \pm 10^{-15}\,,
\end{equation}
which represents a tremendous finetuning. Perhaps the universe indeed
started with such a peculiar initial condition, but it is
epistemologically more satisfying if we give a fundamental dynamical
reason for the universe to have started so close to spatial flatness.
These arguments were first used by Robert Dicke in the 1960s, much
before inflation. He argued that the most natural initial condition for
the spatial curvature should have been the Planck scale curvature,
$^{(3)}\!R = 6K/l_{\rm P}^2$, where the Planck length is $l_{\rm P} =
(\hbar G/c^3)^{1/2} = 1.62\times 10^{-33}$ cm, that is, 60 orders of
magnitude smaller than the present size of the universe, $a_0 =
1.38\times 10^{28}$ cm. A universe with this immense curvature would
have collapsed within a Planck time, $t_{\rm P} = (\hbar G/c^5)^{1/2} =
5.39\times 10^{-44}$ s, again 60 orders of magnitude smaller than the
present age of the universe, $t_0 = 4.1\times 10^{17}$ s. Therefore, the
flatness problem is also related to the Age Problem, why is it that the
universe is so old and flat when, under ordinary circumstances (based on
the fundamental scale of gravity) it should have lasted only a Planck
time and reached a size of order the Planck length? As we will see,
inflation gives a dynamical reason to such a peculiar initial condition.

\subsubsection{The Homogeneity Problem}

An expanding universe has {\em particle horizons}, that is, spatial
regions beyond which causal communication cannot occur. The horizon
distance can be defined as the maximum distance that light could have
travelled since the origin of the universe~\cite{KT},
\begin{equation}\label{HorizonDistance}
d_{\rm H}(t) \equiv a(t)\int_0^t {dt'\over a(t')} \sim 
H^{-1}(t)\,,
\end{equation}
which is proportional to the Hubble scale.\footnote{For the radiation
era, the horizon distance is equal to the Hubble scale. For the
matter era it is twice the Hubble scale.} For
instance, at the beginning of nucleosynthesis the horizon distance is a
few light-seconds, but grows {\em linearly} with time and by the end of
nucleosynthesis it is a few light-minutes, i.e. a factor 100 larger,
while the scale factor has increased {\em only} a factor of 10. The fact
that the causal horizon increases faster, $d_{\rm H}\sim t$, than the
scale factor, $a\sim t^{1/2}$, implies that at any given time the
universe contains regions within itself that, according to the Big Bang
theory, were {\em never} in causal contact before. For instance, the
number of causally disconnected regions at a given redshift $z$ present 
in our causal volume today, $d_{\rm H}(t_0)\equiv a_0$, is
\begin{equation}\label{DisconnectedRegions}
N_{\rm CD}(z) \sim \left({a(t)\over d_{\rm H}(t)}\right)^3 \simeq
(1+z)^{3/2}\,,
\end{equation}
which, for the time of decoupling, is of order $N_{\rm CD}(z_{\rm dec}) 
\sim 10^5 \gg1$.

%%%%%%%%%%%%%%%%%%%%%%%%%
\begin{figure}[htb]
\vspace*{-3cm}
\begin{center}
\includegraphics[width=10cm]{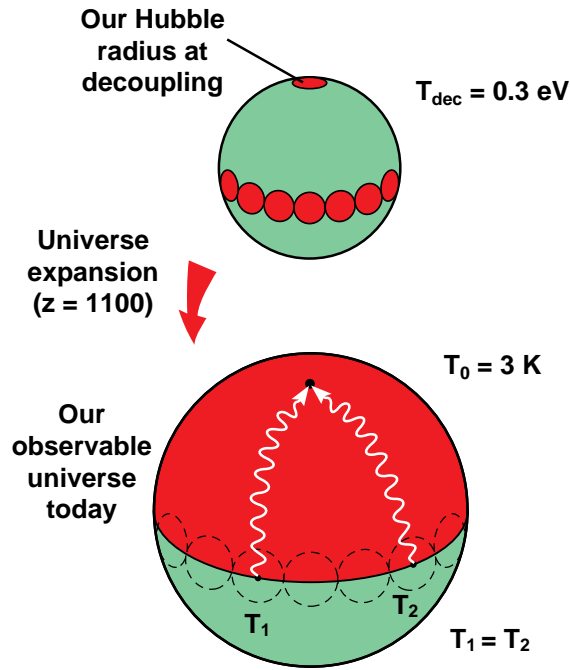}
\vspace*{-2cm}
\caption{Perhaps the most acute problem of the Big Bang theory is
explaining the extraordinary homogeneity and isotropy of the microwave
background, see Fig.~\ref{fig6}. At the time of decoupling, the volume
that gave rise to our present universe contained many causally
disconnected regions (top figure). Today we observe a blackbody spectrum
of photons coming from those regions and they appear to have the same
temperature, $T_1=T_2$, to one part in $10^5$. Why is the universe so
homogeneous?  This constitutes the so-called horizon problem, which is
spectacularly solved by inflation. From Ref.~\cite{Bellido}.}
\label{fig31}
\end{center}
\end{figure}
%%%%%%%%%%%%%%%%%%%%%%%%%

This phenomenon is particularly acute in the case of the observed
microwave background. Information cannot travel faster than the speed of
light, so the causal region at the time of photon decoupling could not
be larger than $d_{\rm H}(t_{\rm dec}) \sim 3\times 10^5$ light years
across, or about $1^\circ$ projected in the sky today. So why should
regions that are separated by more than $1^\circ$ in the sky today have
exactly the same temperature, to within 10 ppm, when the photons that
come from those two distant regions could not have been in causal
contact when they were emitted? This constitutes the so-called horizon
problem, see Fig.~\ref{fig31}, and was first discussed by Robert Dicke
in the 1970s as a profound inconsistency of the Big Bang theory.

\subsection{Cosmological Inflation}

\begin{figure}[htb]
%\vspace*{1cm}
\begin{center}%\hspace*{-7cm}
\includegraphics[width=8cm]{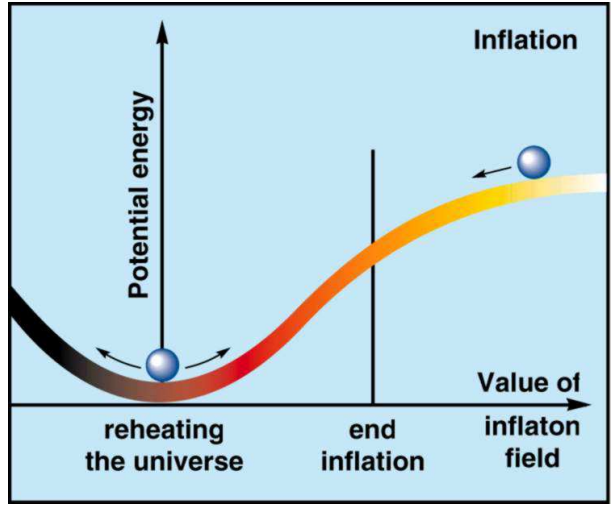}
%\vspace*{-6cm}
\caption{The inflaton field can be represented as a ball rolling
down a hill.  During inflation, the energy density is approximately
constant, driving the tremendous expansion of the universe. When the
ball starts to oscillate around the bottom of the hill, inflation ends
and the inflaton energy decays into particles. In certain cases, the
coherent oscillations of the inflaton could generate a resonant
production of particles which soon thermalize, reheating the universe.
From Ref.~\cite{Bellido}.}
\label{fig32}
\end{center}
\end{figure}

In the 1980s, a new paradigm, deeply rooted in fundamental physics, was
put forward by Alan H. Guth~\cite{Guth}, Andrei D. Linde~\cite{Linde}
and others~\cite{Andy,Guthbook,LindeBook}, to address these fundamental
questions. According to the inflationary paradigm, the early universe
went through a period of exponential expansion, driven by the
approximately constant energy density of a scalar field called the
inflaton. In modern physics, elementary particles are represented by
quantum fields, which resemble the familiar electric, magnetic and
gravitational fields. A field is simply a function of space and time
whose quantum oscillations are interpreted as particles. In our case,
the inflaton field has, associated with it, a large potential energy
density, which drives the exponential expansion during inflation, see
Fig.~\ref{fig32}. We know from general relativity that the density of
matter determines the expansion of the universe, but a constant energy
density acts in a very peculiar way: as a repulsive force that makes any
two points in space separate at exponentially large speeds. (This does
not violate the laws of causality because there is no information
carried along in the expansion, it is simply the stretching of
space-time.)

%%%%%%%%%%%%%%%%%%%%%%%%%
%%%%%%%%%%%%%%%%%%%%%%%%%

This superluminal expansion is capable of explaining the large scale
homogeneity of our observable universe and, in particular, why the
microwave background looks so isotropic: regions separated today by more
than $1^\circ$ in the sky were, in fact, in causal contact before
inflation, but were stretched to cosmological distances by the
expansion. Any inhomogeneities present before the tremendous expansion
would be washed out. This explains why photons from supposedly causally
disconneted regions have actually the same spectral distribution with
the same temperature, see Fig.~\ref{fig31}. 

\begin{figure}[htb]
%\vspace*{-1cm}
\begin{center}\hspace*{-4cm}
\includegraphics[width=8cm]{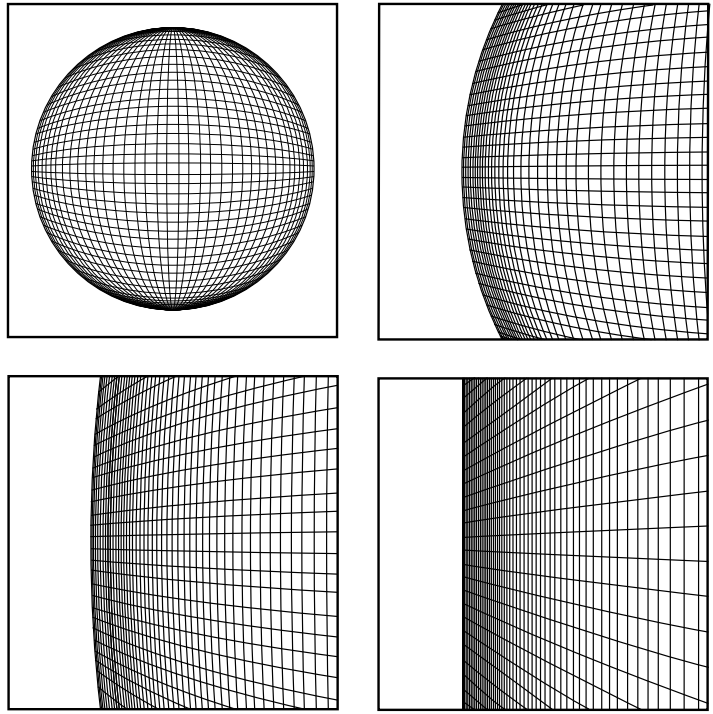}
\vspace*{-5.5cm}
\caption{The exponential expansion during inflation made the
radius of curvature of the universe so large that our observable patch
of the universe today appears essentialy flat, analogous (in three
dimensions) to how the surface of a balloon appears flatter and flatter
as we inflate it to enormous sizes. This is a crucial prediction of
cosmological inflation that will be tested to extraordinary accuracy in
the next few years. From Ref.~\cite{Guthbook,Bellido}.}
\label{fig33}
\end{center}
\end{figure}

Moreover, in the usual Big Bang scenario a flat universe, one in which
the gravitational attraction of matter is exactly balanced by the cosmic
expansion, is unstable under perturbations: a small deviation from
flatness is amplified and soon produces either an empty universe or a
collapsed one. As we discussed above, for the universe to be nearly flat
today, it must have been extremely flat at nucleosynthesis, deviations
not exceeding more than one part in $10^{15}$. This extreme fine tuning
of initial conditions was also solved by the inflationary paradigm, see
Fig.~\ref{fig33}. Thus inflation is an extremely elegant hypothesis that
explains how a region much, much greater that our own observable
universe could have become smooth and flat without recourse to {\em ad
hoc} initial conditions. Furthermore, inflation dilutes away any
``unwanted'' relic species that could have remained from early universe
phase transitions, like monopoles, cosmic strings, etc., which are
predicted in grand unified theories and whose energy density could be so
large that the universe would have become unstable, and collapsed, long
ago. These relics are diluted by the superluminal expansion, which
leaves at most one of these particles per causal horizon, making them
harmless to the subsequent evolution of the universe.

%%%%%%%%%%%%%%%%%%%%%%%%%
%%%%%%%%%%%%%%%%%%%%%%%%%

The only thing we know about this peculiar scalar field, the {\em
inflaton}, is that it has a mass and a self-interaction potential
$V(\phi)$ but we ignore everything else, even the scale at which its
dynamics determines the superluminal expansion. In particular, we still
do not know the nature of the inflaton field itself, is it some new {\em
fundamental} scalar field in the electroweak symmetry breaking sector,
or is it just some {\em effective} description of a more fundamental
high energy interaction?  Hopefully, in the near future, experiments in
particle physics might give us a clue to its nature. Inflation had its
original inspiration in the Higgs field, the scalar field supposed to be
responsible for the masses of elementary particles (quarks and leptons)
and the breaking of the electroweak symmetry. Such a field has not been
found yet, and its discovery at the future particle colliders would help
understand one of the truly fundamental problems in physics, the origin
of masses. If the experiments discover something completely new and
unexpected, it would automatically affect the idea of inflation at a
fundamental level.

\subsubsection{Homogeneous scalar field dynamics}

In this subsection I will describe the theoretical basis for the
phenomenon of inflation. Consider a scalar field $\phi$, a singlet under
any given interaction, with an effective potential $V(\phi)$. The
Lagrangian for such a field in a curved background is 
\begin{equation}\label{ScalarLagrangian}
{\cal L}_{\rm inf} = {1\over2}\,g^{\mu\nu}\partial_\mu\phi
\partial_\nu\phi - V(\phi)\,, 
\end{equation}
whose evolution equation in a Friedmann-Robertson-Walker metric
(\ref{FRWmetric}) and for a {\em homogeneous} field $\phi(t)$ is given by
\begin{equation}\label{ScalarEvolution}
\ddot\phi + 3H\dot\phi + V'(\phi)=0\,,
\end{equation}
where $H$ is the rate of expansion, together with the Einstein equations,
\begin{eqnarray}\label{ScalarEinstein}
H^2 &=& {\kappa^2\over3}\Big({1\over2}\,\dot\phi^2 + V(\phi)\Big)\,,\\
\dot H &=& - {\kappa^2\over2}\,\dot\phi^2\,,
\end{eqnarray}
where $\kappa^2\equiv 8\pi G$. The dynamics of inflation can be
described as a perfect fluid (\ref{PerfectFluid}) with a time dependent
pressure and energy density given by
\begin{eqnarray}\label{ScalarFluid}
\rho &=& {1\over2}\,\dot\phi^2 + V(\phi)\,,\\
p &=& {1\over2}\,\dot\phi^2 - V(\phi)\,.
\end{eqnarray}
The field evolution equation (\ref{ScalarEvolution}) can then be written
as the energy conservation equation, 
\begin{equation}\label{ScalarEnergyCons}
\dot\rho + 3H(\rho+p) = 0\,. 
\end{equation}
If the potential energy density of the scalar field dominates the kinetic
energy, $V(\phi) \gg \dot\phi^2$, then we see that
\begin{equation}\label{ConstantRate}
p\simeq -\rho \hspace{5mm} \Rightarrow \hspace{5mm} 
\rho\simeq {\rm const.} \hspace{5mm}
\Rightarrow \hspace{5mm} H(\phi) \simeq {\rm const.} \,,
\end{equation}
which leads to the solution
\begin{equation}\label{ExponentialSolution}
a(t) \sim \exp(Ht) \hspace{5mm} \Rightarrow \hspace{5mm} 
{\ddot a\over a} > 0 \hspace{5mm} {\rm accelerated\ expansion} \,.
\end{equation}
Using the definition of the number of $e$-folds, $N=\ln(a/a_i)$, we see
that the scale factor grows exponentially, $a(N) = a_i \exp(N)$. This
solution of the Einstein equations solves immediately the flatness
problem. Recall that the problem with the radiation and matter eras is
that $\Omega=1$ ($x=0$) is an unstable critical point in phase-space.
However, during inflation, with $p\simeq -\rho \ \Rightarrow \
\omega\simeq -1$, we have that $1+3\omega\geq0$ and therefore $x=0$ is a
stable {\em attractor} of the equations of motion, see
Eq.~(\ref{Stability}). As a consequence, what seemed an {\em ad hoc}
initial condition, becomes a natural {\em prediction} of inflation.
Suppose that during inflation the scale factor increased $N$ $e$-folds,
then
\begin{equation}\label{SolutionFlatness}
x_0 = x_{\rm in}\,e^{-2N}\,%\Big({T_{\rm end}\over T_{\rm rh}}\Big)^2 
\Big({T_{\rm rh}\over T_{\rm eq}}\Big)^2 (1+z_{\rm eq}) \
\simeq \ e^{-2N}\,10^{56} \leq 1 \hspace{5mm} 
\Rightarrow \hspace{5mm} N \geq 65\,,
\end{equation}
where we have assumed that inflation ended at the scale $V_{\rm end}$,
and the transfer of the inflaton energy density to thermal radiation at
reheating occurred almost instantaneously\footnote{There could be a
small delay in thermalization, due to the intrinsic inefficiency of
reheating, but this does not change significantly the required number of
$e$-folds.} at the temperature \ $T_{\rm rh} \sim V_{\rm end}^{1/4} \sim
10^{15}$ GeV. Note that we can now have initial conditions with a large
uncertainty, $x_{\rm in} \simeq 1$, and still have today $x_0 \simeq 1$,
thanks to the inflationary attractor towards $\Omega=1$.  This can be
understood very easily by realizing that the three curvature evolves
during inflation as
\begin{equation}\label{R3Flatness}
^{(3)}R = {6K\over a^2} = \ ^{(3)}\!R_{\rm in}\,e^{-2N} \hspace{5mm} 
\longrightarrow \hspace{5mm} 0\,, \hspace{5mm} {\rm for} \ \ N\gg 1\,. 
\end{equation}
Therefore, if cosmological inflation lasted over 65 $e$-folds, as most
models predict, then today the universe (or at least our local patch)
should be exactly flat, see Fig.~\ref{fig33}, a prediction that can be
tested with great accuracy in the near future and for which already
seems to be some evidence from observations of the microwave
background~\cite{Boomerang}.

Furthermore, inflation also solves the homogeneity problem in a
spectacular way. First of all, due to the superluminal expansion, 
any inhomogeneity existing prior to inflation will be washed out,
\begin{equation}\label{Homogeneity}
\delta_k \sim \ \left({k\over aH}\right)^2\,
\Phi_k \hspace{2mm} \propto \hspace{2mm} e^{-2N} \hspace{5mm} 
\longrightarrow \hspace{5mm} 0\,, \hspace{5mm} {\rm for} \ \ N\gg 1\,. 
\end{equation}
Moreover, since the scale factor grows exponentially, while the horizon
distance remains essentially constant, $d_H(t) \simeq H^{-1} = $ const.,
any scale within the horizon during inflation will be stretched by the
superluminal expansion to enormous distances, in such a way that at
photon decoupling all the causally disconnected regions that encompass
our present horizon actually come from a single region during inflation,
about 65 $e$-folds before the end. This is the reason why two points
separated more than $1^\circ$ in the sky have the same backbody
temperature, as observed by the COBE satellite: they were actually in
causal contact during inflation. There is at present no other proposal
known that could solve the homogeneity problem without invoquing an
acausal mechanism like inflation.

Finally, any relic particle species (relativistic or not) existing prior
to inflation will be diluted by the expansion,
\begin{eqnarray}\label{RelicSpecies}
\rho_{\rm M} &\propto& a^{-3} \ \sim \ e^{-3N} \hspace{5mm} 
\longrightarrow \hspace{5mm} 0\,, \hspace{5mm} {\rm for} \ \ N\gg 1\,,\\
\rho_{\rm R} &\propto& a^{-4} \ \sim \ e^{-4N} \hspace{5mm} 
\longrightarrow \hspace{5mm} 0\,, \hspace{5mm} {\rm for} \ \ N\gg 1\,.
\end{eqnarray}
Note that the vacuum energy density $\rho_v$ remains constant under the
expansion, and therefore, very soon it is the only energy density
remaining to drive the expansion of the universe.

\subsubsection{The slow-roll approximation}

In order to simplify the evolution equations during inflation, we will
consider the slow-roll approximation (SRA). Suppose that, during
inflation, the scalar field evolves very slowly down its effective
potential, then we can define the slow-roll parameters~\cite{LL93},
\begin{eqnarray}\label{SlowRollParameters}
\epsilon&\equiv&-\,{\dot H\over H^2} \ = \ 
{\kappa^2\over2}\,{\dot\phi^2\over H^2} \ \ll \ 1\,,\\
\delta &\equiv& -\,{\ddot\phi\over H\dot\phi} \ \ll \ 1\,,\\
\xi &\equiv& {\stackrel{\dots}{\phi}\over H^2\dot\phi} - \delta^2 \ \ll \ 1\,.
\end{eqnarray}
It is easy to see that the condition 
\begin{equation}\label{Inflation}
\epsilon < 1 \hspace{5mm} \Longleftrightarrow \hspace{5mm} 
{\ddot a\over a} > 0
\end{equation}
characterizes inflation: it is all you need for superluminal expansion,
i.e. for the horizon distance to grow more slowly than the scale factor,
in order to solve the homogeneity problem, as well as for the spatial
curvature to decay faster than usual, in order to solve the flatness
problem.

The number of $e$-folds during inflation can be written with the help of
Eq.~(\ref{SlowRollParameters}) as
\begin{equation}\label{Nefolds}
N \ = \ \ln{a_{\rm end}\over a_i} \ = \ \int_{t_i}^{t_e} H dt \ = \
\int_{\phi_i}^{\phi_e} {\kappa d\phi\over\sqrt{2\epsilon(\phi)}} \,,
\end{equation}
which is an exact expression in terms of $\epsilon(\phi)$.

In the limit given by Eqs.~(\ref{SlowRollParameters}), the evolution
equations (\ref{ScalarEvolution}) and (\ref{ScalarEinstein}) become
\begin{eqnarray}\label{SlowRollEvolution}
H^2\Big(1- {\epsilon\over3}\Big) &\simeq& H^2 \ = \
{\kappa^2\over3}\,V(\phi)\,,\\
3H\dot\phi\,\Big(1- {\delta\over3}\Big) &\simeq& 3H\dot\phi \ = \
-\,V'(\phi)\,.
\end{eqnarray}
Note that this corresponds to a reduction of the dimensionality of 
phase-space from two to one dimensions, $H(\phi,\dot\phi) \rightarrow
H(\phi)$.  In fact, it is possible to prove a theorem, for single-field
inflation, which states that the slow-roll approximation is an attractor
of the equations of motion, and thus we can always evaluate the
inflationary trajectory in phase-space within the SRA, therefore reducing
the number of initial conditions to just one, the initial value of the
scalar field. If $H(\phi)$ only depends on $\phi$, then $H'(\phi) =
-\kappa^2\dot\phi/2$ and we can rewrite the slow-roll parameters 
(\ref{SlowRollParameters}) as
\begin{eqnarray}\label{neweps}
\epsilon&=&{2\over\kappa^2}\left({H'(\phi)\over H(\phi)}\right)^2 \ 
\simeq \ {1\over2\kappa^2}\left({V'(\phi)\over V(\phi)}\right)^2 \ 
\equiv \ \epsilon_V \ll \ 1\,,\\
\delta &=& {2\over\kappa^2}\,{H''(\phi)\over H(\phi)} \ 
\simeq \ {1\over\kappa^2}\,{V''(\phi)\over V(\phi)} -
{1\over2\kappa^2}\left({V'(\phi)\over V(\phi)}\right)^2 \ 
\equiv \ \eta_V - \epsilon_V \ \ll \ 1\,, \label{neweta}\\
\xi &=& {4\over\kappa^4}\,{H'(\phi)H'''(\phi)\over H^2(\phi)} \ 
\simeq \ {1\over\kappa^4}\,{V'(\phi)V'''(\phi)\over V^2(\phi)} -
{3\over2\kappa^4}{V''(\phi)\over V(\phi)}
\left({V'(\phi)\over V(\phi)}\right)^2 \nonumber \\
&+& {3\over4\kappa^4}\left({V'(\phi)\over V(\phi)}\right)^4 \ 
\equiv \ \xi_V - 3\eta_V\epsilon_V + 3\epsilon_V^2 \ \ll \ 1\,. 
\label{newxi}
\end{eqnarray}
These expressions define the new slow-roll parameters $\epsilon_V,
\ \eta_V$ and $\xi_V$. The number of
$e$-folds can also be rewritten in this approximation as
\begin{equation}\label{newefolds}
N \ \simeq \
\int_{\phi_i}^{\phi_e} {\kappa d\phi\over\sqrt{2\epsilon_V(\phi)}} \ 
= \ \kappa^2\, \int_{\phi_i}^{\phi_e} {V(\phi)\,d\phi\over V'(\phi)} \,,
\end{equation}
a very useful expression for evaluating $N$ for a given effective
scalar potential $V(\phi)$.

\subsection{The origin of density perturbations}

If cosmological inflation made the universe so extremely flat and
homogeneous, where did the galaxies and clusters of galaxies come from?
One of the most astonishing predictions of inflation, one that was not
even expected, is that quantum fluctuations of the inflaton field are
stretched by the exponential expansion and generate large-scale
perturbations in the metric. Inflaton fluctuations are small wave
packets of energy that, according to general relativity, modify the
space-time fabric, creating a whole spectrum of curvature
perturbations. The use of the word spectrum here is closely related to
the case of light waves propagating in a medium: a spectrum
characterizes the amplitude of each given wavelength. In the case of
inflation, the inflaton fluctuations induce waves in the space-time
metric that can be decomposed into different wavelengths, all with
approximately the same amplitude, that is, corresponding to a
scale-invariant spectrum. These patterns of perturbations in the metric
are like fingerprints that unequivocally characterize a period of
inflation. When matter fell in the troughs of these waves, it created
density perturbations that collapsed gravitationally to form galaxies,
clusters and superclusters of galaxies, with a spectrum that is also
scale invariant. Such a type of spectrum was proposed in the early 1970s
(before inflation) by Harrison and Zel'dovich~\cite{HZ}, to explain
the distribution of galaxies and clusters of galaxies on very large
scales in our observable universe. Perhaps the most interesting aspect
of structure formation is the possibility that the detailed knowledge of
what seeded galaxies and clusters of galaxies will allow us to test the
idea of inflation.

\subsubsection{Reparametrization invariant perturbation theory}

Until now we have considered only the unperturbed FRW metric described
by a scale factor $a(t)$ and a homogeneous scalar field $\phi(t)$,
\begin{eqnarray}
ds^2 &=& a^2(\eta) [-\,d\eta^2 + \gamma_{ij}\,dx^i dx^j]\,,
\label{HomogeneousBackground}\\ \phi &=& \phi(\eta)\,,
\end{eqnarray}
where $\eta=\int dt/a(t)$ is the conformal time, under which the
background equations of motion can be written as
\begin{eqnarray}
{\cal H}^2 &=& {\kappa^2\over3}\left({1\over2}{\phi'}^2 + 
a^2 V(\phi)\right)\,,\\
{\cal H}'-{\cal H}^2 &=& -\,{\kappa^2\over2}{\phi'}^2 \,,\\[1mm]
\phi''+2{\cal H}\phi'+ a^2V'(\phi)&=&0\,,\label{HomogeneousEquations}
\end{eqnarray}
where ${\cal H}=aH$ and $\phi'=a\dot\phi$.

During inflation, the quantum fluctuations of the scalar field will
induce metric perturbations which will backreact on the scalar field.
Let us consider, in linear perturbation theory, the most general line
element with both scalar and tensor metric 
perturbations~\cite{Bardeen},\footnote{Note
that inflation cannot generate, to linear order, a vector perturbation.}
together with the scalar field perturbations
\begin{eqnarray}\label{MetricPerturbations}
ds^2 &=& a^2(\eta)\left[-\,(1+2A)d\eta^2 + 2B_{|i} dx^i d\eta + \Big\{
(1+2{\cal R})\gamma_{ij}+2E_{|ij}+2h_{ij}\Big\} dx^i dx^j\right]\,,\\
\phi&=&\phi(\eta)+\delta\phi(\eta,x^i)\,.\label{PerturbedBackground}
\end{eqnarray}
The indices $\{i,j\}$ label the three-dimensional spatial coordinates
with metric $\gamma_{ij}$, and the $|i$ denotes covariant derivative
with respect to that metric. The gauge invariant tensor perturbation
$h_{ij}$ corresponds to a transverse traceless gravitational wave,
$\nabla^ih_{ij} = h^i_i=0$. The four scalar perturbations $(A, B, {\cal
R}, E)$ are {\em gauge dependent} functions of $(\eta,x^i)$.
Under a general coordinate (gauge) transformation~\cite{Bardeen,MFB}
\begin{eqnarray}
&&\tilde\eta = \eta + \xi^0(\eta,x^i)\,,\\
&&\tilde x^i = x^i + \gamma^{ij}\xi_{|j}(\eta,x^i)\,,
\label{GaugeTransformation}
\end{eqnarray}
with arbitrary functions $(\xi^0,\xi)$, the scalar and tensor 
perturbations transform, to linear order, as
\begin{eqnarray}
&&\tilde A = A - {\xi^0}' - {\cal H}\xi^0\,,\hspace{1cm}
\tilde B = B + \xi^0 - \xi'\,,\\
&&\tilde {\cal R} = {\cal R} - {\cal H}\xi^0\,,\hspace{1.9cm}
\tilde E = E - \xi\,,\\
&&\hspace{3cm} \tilde h_{ij} = h_{ij}\,, \label{TransformedPerturbations}
\end{eqnarray}
where a prime denotes derivative with respect to conformal time. It is
possible to construct, however, two gauge-invariant gravitational
potentials~\cite{Bardeen,MFB},
\begin{eqnarray}
&& \Phi = A + (B-E')' + {\cal H}(B-E')\,,\\
&& \Psi = {\cal R} + {\cal H}(B-E')\,, 
\label{GravitationalPotentials}
\end{eqnarray}
which are related through the perturbed Einstein equations,
\begin{eqnarray}
 \Phi &=& \Psi\,, \\
 {k^2-3K\over a^2}\Psi &=& {\kappa^2\over2} \delta\rho\,,
\label{PertEinsteinEqs}
\end{eqnarray}
where $\delta\rho$ is the gauge-invariant density perturbation,
and the latter expression is nothing but the Poisson equation for 
the gravitational potential, written in relativistic form.

During inflation, the energy density is given in terms of a scalar
field, and thus the gauge-invariant equations for the perturbations on
comoving hypersurfaces (constant energy density hypersurfaces) are
\begin{eqnarray}
\Phi''+3{\cal H}\Phi'+({\cal H}'+2{\cal H}^2)\Phi &=& 
{\kappa^2\over2} [\phi'\delta\phi'- a^2V'(\phi)\delta\phi]\,,\\
-\nabla^2\Phi+3{\cal H}\Phi'+({\cal H}'+2{\cal H}^2)\Phi &=& 
- {\kappa^2\over2} [\phi'\delta\phi'+ a^2V'(\phi)\delta\phi]\,,\\
\Phi'+{\cal H}\Phi &=& {\kappa^2\over2} \phi'\delta\phi\,,\\[1mm]
\delta\phi''+2{\cal H}\delta\phi'-\nabla^2\delta\phi &=& 4\phi'\Phi'
-2a^2V'(\phi)\Phi-a^2V''(\phi)\delta\phi\,.
\label{PerturbedEquations}
\end{eqnarray}

This system of equations seem too difficult to solve at first sight. 
However, there is a gauge invariant combination of variables that
allows one to find exact solutions. Let us define~\cite{MFB}
\begin{eqnarray}
&& u \equiv a\delta\phi + z\Phi\,,\\
&& z \equiv a{\phi'\over{\cal H}}\,.
\label{RedefinitionVariables}
\end{eqnarray}
Under this redefinition, the above equations simplify enormously to
just three independent equations,
\begin{eqnarray}
&&u'' - \nabla^2u - {z''\over z}u = 0\,,\label{EquationU}\\
&&\nabla^2\Phi = {\kappa^2\over2} {{\cal H}\over a^2} (zu'-z'u)\,,
\label{NablaPhi}\\
&& \Big({a^2\Phi\over{\cal H}}\Big)' = {\kappa^2\over2} zu\,.
\label{EquationPhi}
\end{eqnarray}
From Equation (\ref{EquationU}) we can find a solution $u(z)$,
which substituted into (\ref{EquationPhi}) can be integrated to give
$\Phi(z)$, and together with $u(z)$ allow us to obtain
$\delta\phi(z)$. 

\subsubsection{Quantum Field Theory in curved space-time}

Until now we have treated the perturbations as classical, but we should
in fact consider the perturbations $\Phi$ and $\delta\phi$ as quantum
fields. Note that the perturbed action for the scalar mode $u$
can be written as
\begin{equation}
\delta S = {1\over2} \int d^3x\,d\eta\,\Big[(u')^2 - (\nabla u)^2
+ {z''\over z}u^2\Big]\,.
\end{equation}
In order to quantize the field $u$ in the curved background defined
by the metric (\ref{HomogeneousBackground}), we can write the operator
\begin{equation}
\hat u(\eta,{\bf x}) = \int {d^3{\bf k}\over(2\pi)^{3/2}}
\Big[u_k(\eta)\,\hat a_{\bf k}\,e^{i{\bf k}\cdot{\bf x}} + 
u_k^*(\eta)\,\hat a_{\bf k}^\dagger\,e^{-i{\bf k}\cdot{\bf x}} \Big]\,,
\label{OperatorU}
\end{equation}
where the creation and annihilation operators satisfy the commutation
relation of bosonic fields, and the scalar field's Fock space is defined
through the vacuum condition,
\begin{eqnarray}\label{CommutationRelation}
[\hat a_{\bf k}, \hat a_{{\bf k}'}^\dagger] &=& \delta^3({\bf k - k}')
\,,\\ \hat a_{\bf k}|0\rangle &=& 0\,.
\end{eqnarray}
Note that we are not assuming that the inflaton is a fundamental scalar
field, but that is can be written as a quantum field with its
commutation relations (as much as a pion can be described as a quantum
field).

The equations of motion for each mode $u_k(\eta)$ are decoupled in
linear perturbation theory,
\begin{equation}
u''_k +\Big(k^2-{z''\over z}\Big)u_k = 0\,.
\label{EquationModek}
\end{equation}
The ratio $z''/z$ acts like a time-dependent potential for this
Schr\"odinger like equation. In order to find exact solutions to the
mode equation, we will use the slow-roll parameters
(\ref{SlowRollParameters}), see Ref.~\cite{LL93}
\begin{eqnarray}
&&\epsilon = 1 - {{\cal H}'\over{\cal H}^2} = {\kappa^2\over2}
{z^2\over a^2}\,,\label{ConfEpsilon}\\
&&\delta = 1 - {\phi''\over{\cal H}\phi'} = 1 + \epsilon - 
{z'\over{\cal H}z}\,,\label{ConfDelta}\\
&&\xi = - \left(2-\epsilon-3\delta+\delta^2 - 
{\phi'''\over{\cal H}^2\phi'}\right)\,.
\label{ConfXi}
\end{eqnarray}
In terms of these parameters, the conformal time and the effective
potential for the $u_k$ mode can be written as
\begin{eqnarray}
&&\eta = {-1\over{\cal H}} + \int {\epsilon da\over a{\cal H}}\,,\\
&&{z''\over z} = {\cal H}^2\,\Big[(1+\epsilon-\delta)(2-\delta) +
{\cal H}^{-1}(\epsilon'-\delta')\Big]\,.\label{PotentialZ}
\end{eqnarray}
Note that the slow-roll parameters, (\ref{ConfEpsilon}) and
(\ref{ConfDelta}), can be taken as {\em constant},\footnote{For
instance, there are models of inflation, like power-law inflation,
$a(t)\sim t^p$, where $\epsilon=\delta=1/p<1$, that give constant 
slow-roll parameters.} to order $\epsilon^2$,
\begin{equation}\label{ConstantSlowRoll}
\begin{array}{l}
\epsilon' = 2{\cal H}\,\Big(\epsilon^2-\epsilon\delta\Big) = 
{\cal O}(\epsilon^2)\,,\\[2mm]
\delta' = {\cal H}\,\Big(\epsilon\delta-\xi\Big) = 
{\cal O}(\epsilon^2)\,.
\end{array}
\end{equation}
In that case, for constant slow-roll parameters, we can write
\begin{eqnarray}
&&\eta = {-1\over{\cal H}}{1\over1-\epsilon}\,,\label{EtaEpsilon}\\
{z''\over z} = {1\over\eta^2}\Big(\nu^2-{1\over4}\Big)\,, && 
\hspace{5mm} {\rm where} \hspace{7mm} \nu =
{1+\epsilon-\delta\over1-\epsilon} + {1\over2}\,.\label{NuParam}
\end{eqnarray}

We are now going to search for approximate solutions of the mode
equation (\ref{EquationModek}), where the effective potential
(\ref{PotentialZ}) is of order $z''/z \simeq 2{\cal H}^2$ in the
slow-roll approximation. In quasi-de Sitter there is a characteristic
scale given by the (event) horizon size or Hubble scale during
inflation, $H^{-1}$. There will be modes $u_k$ with physical wavelengths
much smaller than this scale, $k/a\gg H$, that are well within the
de Sitter horizon and therefore do not feel the curvature of space-time.
On the other hand, there will be modes with physical wavelengths much
greater than the Hubble scale, $k/a\ll H$. In these two asymptotic
regimes, the solutions can be written as
\begin{eqnarray}
&&u_k = {1\over\sqrt{2k}}\,e^{-ik\eta} \hspace{1cm} k\gg aH\,, 
\label{MinkowskyModes}\\
&&u_k = C_1\,z \hspace{2cm} k\ll aH\,.\label{deSitterModes}
\end{eqnarray}
In the limit $k\gg aH$ the modes behave like ordinary quantum modes in
Minkowsky space-time, appropriately normalized, while in the opposite
limit, $u/z$ becomes constant on superhorizon scales. For approximately
constant slow-roll parameters one can find exact solutions to
(\ref{EquationModek}), with the effective potential given by
(\ref{NuParam}), that interpolate between the two asymptotic solutions,
\begin{equation}
u_k(\eta) = {\sqrt\pi\over2}\,e^{i(\nu+{1\over2}){\pi\over2}}\,
(-\eta)^{1/2}\,H_\nu^{_{(1)}}(-k\eta)\,,
\label{ExactSolution}
\end{equation}
where $H_\nu^{_{(1)}}(z)$ is the Hankel function of the first
kind~\cite{AS}, and $\nu$ is given by (\ref{NuParam}) in terms of the
slow-roll parameters. In the limit $k\eta\to0$, the solution becomes
\begin{eqnarray}
&&|u_k| = {2^{\nu-{3\over2}}\over\sqrt{2k}}\,
{\Gamma(\nu)\over\Gamma({3\over2})}\,(-k\eta)^{{1\over2}-\nu} \equiv
{C(\nu)\over\sqrt{2k}}\,\Big({k\over aH}\Big)^{{1\over2}-\nu}\,, 
\label{LimitSolution}\\[2mm]
&&C(\nu) = 2^{\nu-{3\over2}}\,
{\Gamma(\nu)\over\Gamma({3\over2})}\,(1-\epsilon)^{\nu-{1\over2}}
\ \simeq 1 \hspace{1cm} {\rm for} \hspace{.5cm} \epsilon, \delta\ll1\,.
\end{eqnarray}

We can now compute $\Phi$ and $\delta\phi$ from the super-Hubble-scale
mode solution (\ref{deSitterModes}), for $k\ll aH$. Substituting into
Eq.~(\ref{EquationPhi}), we find
\begin{eqnarray}
&&\Phi = C_1\,\Big(1 - {{\cal H}\over a^2}\,\int a^2 d\eta\Big) + 
C_2\,{{\cal H}\over a^2}\,, 
\label{PhiSolution}\\
&&\delta\phi = {C_1\over a^2}\,\int a^2 d\eta - {C_2\over a^2}\,.
\label{DeltaphiSolution}
\end{eqnarray}
The term proportional to $C_1$ corresponds to the growing solution,
while that proportional to $C_2$ corresponds to the decaying solution,
which can soon be ignored. These quantities are gauge invariant but
evolve with time outside the horizon, during inflation, and before
entering again the horizon during the radiation or matter eras. We would
like to write an expression for a gauge invariant quantity that is also
{\em constant} for superhorizon modes. Fortunately, in the case of
adiabatic perturbations, there is such a quantity:
\begin{equation}
\zeta \equiv \Phi + {1\over\epsilon{\cal H}}\,(\Phi'+
{\cal H}\Phi) = {u\over z}\,,\label{Zeta}
\end{equation}
which is constant, see Eq.~(\ref{deSitterModes}), for $k\ll aH$. In
fact, this quantity $\zeta$ is identical, for superhorizon modes, to the
gauge invariant curvature metric perturbation ${\cal R}_c$ on comoving
(constant energy density) hypersurfaces, see Ref.~\cite{Bardeen,GBW},
\begin{equation}
\zeta = {\cal R}_c + {1\over\epsilon{\cal H}^2}\,
\nabla^2\Phi\,.\label{ZetaRc}
\end{equation}
Using Eq.~(\ref{NablaPhi}) we can write the evolution equation for
$\zeta={u\over z}$ as \ $\zeta' = {1\over\epsilon{\cal
H}}\,\nabla^2\Phi$, which confirms that $\zeta$ is constant for
(adiabatic\footnote{This conservation fails for entropy or isocurvature
perturbations, see Ref.~\cite{GBW}.}) superhorizon modes, $k\ll
aH$. Therefore, we can evaluate the Newtonian potential $\Phi_k$ when
the perturbation reenters the horizon during radiation/matter eras in
terms of the curvature perturbation ${\cal R}_k$ when it left the Hubble
scale during inflation,
\begin{equation}
\Phi_k = \Big(1 - {{\cal H}\over a^2}\,\int a^2 d\eta\Big)\,
{\cal R}_k = {3+3\omega\over5+3\omega}\,{\cal R}_k = 
\hspace{1mm}\left\{\begin{array}{cl}
{2\over3}\,{\cal R}_k&\hspace{5mm}{\rm radiation 
\hspace{2mm} era}\,,\\[3mm]
{3\over5}\,{\cal R}_k&\hspace{5mm}{\rm matter 
\hspace{2mm} era}\,.\end{array}\right.
\label{PhiRc}
\end{equation}

Let us now compute the tensor or gravitational wave metric perturbations
generated during inflation. The perturbed action for the tensor mode can
be written as
\begin{equation}
\delta S = {1\over2} \int d^3x\,d\eta\,{a^2\over2\kappa^2}\Big[
(h_{ij}')^2 - (\nabla h_{ij})^2\Big]\,,
\end{equation}
with the tensor field $h_{ij}$ considered as a quantum field,
\begin{equation}
\hat h_{ij}(\eta,{\bf x}) = \int {d^3{\bf k}\over(2\pi)^{3/2}}
\sum_{\lambda=1,2}\Big[h_k(\eta)\,e_{ij}({\bf k},\lambda)\,
\hat a_{\bf k,\lambda}\,e^{i{\bf k}\cdot{\bf x}} + h.c.\Big]\,,
\label{TensorField}
\end{equation}
where $e_{ij}({\bf k},\lambda)$ are the two polarization tensors, 
satisfying symmetric, transverse and traceless conditions
\begin{eqnarray}
&&e_{ij}=e_{ji}\,, \hspace{5mm} k^ie_{ij}=0\,, \hspace{5mm} e_{ii} =0\,, 
\label{PolarCons}\\[2mm]
&&e_{ij}(-{\bf k},\lambda) = e_{ij}^*({\bf k},\lambda)\,, \hspace{5mm} 
\sum_{\lambda}e_{ij}^*({\bf k},\lambda)e^{ij}({\bf k},\lambda)=4\,,
\label{polarSum}
\end{eqnarray}
while the creation and annihilation operators satisfy the usual
commutation relation of bosonic fields, Eq.~(\ref{CommutationRelation}).
We can now redefine our gauge invariant tensor amplitude as
\begin{equation}
v_k(\eta)={a\over\sqrt2\kappa}\,h_k(\eta)\,,
\label{FieldV}
\end{equation}
which satisfies the following evolution equation, decoupled for each 
mode $v_k(\eta)$ in linear perturbation theory,
\begin{equation}
v''_k +\Big(k^2-{a''\over a}\Big)v_k = 0\,.
\label{ModeVk}
\end{equation}
The ratio $a''/a$ acts like a time-dependent potential for this
Schr\"odinger like equation, analogous to the term $z''/z$ for the
scalar metric perturbation. For constant slow-roll parameters, the
potential becomes
\begin{eqnarray}\label{Potentialappa}
&&{a''\over a} = 2{\cal H}^2\Big(1-{\epsilon\over2}\Big) =
{1\over\eta^2}\Big(\mu^2-{1\over4}\Big)\,,\\[1mm]
&&\mu = {1\over1-\epsilon} + {1\over2}\,.\label{MuParam}
\end{eqnarray}
We can solve equation (\ref{ModeVk}) in the two asymptotic regimes,
\begin{eqnarray}
&&v_k = {1\over\sqrt{2k}}\,e^{-ik\eta} \hspace{1cm} k\gg aH\,, 
\label{MinkowskyVkModes}\\[1mm]
&&v_k = C\,a \hspace{2.2cm} k\ll aH\,.\label{deSitterVkModes}
\end{eqnarray}
In the limit $k\gg aH$ the modes behave like ordinary quantum modes in
Minkowsky space-time, appropriately normalized, while in the opposite
limit, the metric perturbation $h_k$ becomes {\em constant} on
superhorizon scales. For constant slow-roll parameters one can find
exact solutions to (\ref{ModeVk}), with effective potential given by
(\ref{Potentialappa}), that interpolate between the two asymptotic
solutions. These are identical to Eq.~(\ref{ExactSolution}) except for
the substitution $\nu\to\mu$.  In the limit $k\eta\to0$, the solution
becomes
\begin{equation}
|v_k| = {C(\mu)\over\sqrt{2k}}\,\Big({k\over aH}\Big)^{{1\over2}-\mu}\,.
\label{LimitVkSolution}
\end{equation}
Since the mode $h_k$ becomes constant on superhorizon scales, we can
evaluate the tensor metric perturbation when it reentered during the
radiation or matter era directly in terms of its value during inflation.

\subsubsection{Power spectrum of scalar and tensor metric perturbations}

Not only do we expect to measure the amplitude of the metric
perturbations generated during inflation and responsible for the
anisotropies in the CMB and density fluctuations in LSS, but we should
also be able to measure its power spectrum, or two-point correlation
function in Fourier space. Let us consider first the scalar metric
perturbations ${\cal R}_k$, which enter the horizon at $a=k/H$. Its
correlator is given by~\cite{LL93}
\begin{eqnarray}
&&\langle0|{\cal R}_k^*{\cal R}_{k'}|0\rangle = {|u_k|^2\over z^2}\,
\delta^3({\bf k - k'}) \equiv {{\cal P}_{\cal R}(k)\over4\pi k^3}\,
(2\pi)^3\,\delta^3({\bf k - k'})\,, 
\label{CorrelatorRc}\\[2mm]
&&{\cal P}_{\cal R}(k)={k^3\over2\pi^2}\,{|u_k|^2\over z^2}=
{\kappa^2\over2\epsilon}\,\Big({H\over2\pi}\Big)^2\,
\Big({k\over aH}\Big)^{3-2\nu}\equiv A_S^2\,
\Big({k\over aH}\Big)^{n_s-1}\,,
\label{PowerSpectrumRc}
\end{eqnarray}
where we have used \ ${\cal R}_k=\zeta_k={u_k\over z}$ and
Eq.~(\ref{LimitSolution}). This last equation determines the power
spectrum in terms of its amplitude at horizon-crossing, $A_S$, and a
tilt,
\begin{equation}
n_s-1\equiv{d\ln{\cal P}_{\cal R}(k)\over d\ln k} = 3-2\nu = 
2 \Big({\delta-2\epsilon\over1-\epsilon}\Big)\simeq 2\eta_V-6\epsilon_V\,,
\label{ScalarTilt}
\end{equation}
see Eqs.~(\ref{neweps}), (\ref{neweta}).  Note from this equation that
it is possible, in principle, to obtain from inflation a scalar tilt
which is either positive ($n>1$) or negative ($n<1$). Furthermore,
depending on the particular inflationary model~\cite{LR99}, we
can have significant departures from scale invariance.

Note that at horizon entry $k\eta=-1$, and thus we can alternatively
evaluate the tilt as
\begin{equation}
n_s-1\equiv -\,{d\ln{\cal P}_{\cal R}\over d\ln\eta} = 
-2\eta{\cal H}\,\Big[(1-\epsilon)-(\epsilon-\delta)-1\Big] =
2 \Big({\delta-2\epsilon\over1-\epsilon}\Big)\simeq 2\eta_V-6\epsilon_V\,,
\label{STilt}
\end{equation}
and the running of the tilt
\begin{equation}
{d n_s\over d\ln k} = -\,{d n _s\over d\ln\eta} = 
-\eta{\cal H}\,\Big(2\xi+8\epsilon^2-10\epsilon\delta\Big) 
\simeq 2\xi_V + 24\epsilon_V^2-16\eta_V\epsilon_V\,,
\label{RunningScalarTilt}
\end{equation}
where we have used Eqs.~(\ref{ConstantSlowRoll}).

Let us consider now the tensor (gravitational wave) metric perturbation,
which enter the horizon at \ $a=k/H$,
\begin{eqnarray}
&&\sum_\lambda\langle0|h^*_{k,\lambda}h_{k',\lambda}|0\rangle = 4\,
{2\kappa^2\over a^2}\,|v_k|^2
\delta^3({\bf k - k'}) \equiv {{\cal P}_g(k)\over4\pi k^3}\,
(2\pi)^3\,\delta^3({\bf k - k'})\,, 
\label{CorrelatorHk}\\[2mm]
&&{\cal P}_g(k)=8\kappa^2\,\Big({H\over2\pi}\Big)^2\,
\Big({k\over aH}\Big)^{3-2\mu}\equiv A_T^2\,
\Big({k\over aH}\Big)^{n_T}\,,
\label{TensorSpectrum}
\end{eqnarray}
where we have used Eqs.~(\ref{FieldV}) and (\ref{LimitVkSolution}).
Therefore, the power spectrum can be approximated by a power-law
expression, with amplitude $A_T$ and tilt
\begin{equation}
n_T\equiv{d\ln{\cal P}_g(k)\over d\ln k} = 3-2\mu = 
{-2\epsilon\over1-\epsilon} \simeq -2\epsilon_V < 0\,,
\label{TensorTilt}
\end{equation}
which is always negative. In the slow-roll approximation, $\epsilon\ll1$,
the tensor power spectrum is scale invariant.

Alternatively, we can evaluate the tensor tilt by
\begin{equation}
n_T\equiv -\,{d\ln{\cal P}_g\over d\ln\eta} = 
-2\eta{\cal H}\,\Big[(1-\epsilon)-1\Big] = {-2\epsilon\over1-\epsilon}
\simeq -2\epsilon_V \,,
\label{TTilt}
\end{equation}
and its running by
\begin{equation}
{d n_T\over d\ln k} = -\,{d n _T\over d\ln\eta} = 
-\eta{\cal H}\,\Big(4\epsilon^2-4\epsilon\delta\Big) 
\simeq 8\epsilon_V^2-4\eta_V\epsilon_V\,,
\label{RunningTensorTilt}
\end{equation}
where we have used Eqs.~(\ref{ConstantSlowRoll}).

\subsection{The anisotropies of the microwave background}

The metric fluctuations generated during inflation are not only
responsible for the density perturbations that gave rise to galaxies via
gravitational collapse, but one should also expect to see such ripples
in the metric as temperature anisotropies in the cosmic microwave
background, that is, minute deviations in the temperature of the
blackbody spectrum when we look at different directions in the sky. Such
anisotropies had been looked for ever since Penzias and Wilson's
discovery of the CMB, but had eluded all detection, until COBE satellite
discovered them in 1992, see Fig.~\ref{fig6}. The reason why they took
so long to be discovered was that they appear as perturbations in
temperature of only one part in $10^5$. Soon after COBE, other groups
quickly confirmed the detection of temperature anisotropies at around
30\,$\mu$K, at higher multipole numbers or smaller angular scales. 

\subsubsection{The Sachs-Wolfe effect}

The anisotropies corresponding to large angular scales are only
generated via gravitational red-shift and density perturbations through
the Einstein equations, $\delta\rho/\rho = -2\Phi$ for adiabatic
perturbations; we can ignore the Doppler contribution, since the
perturbation is non-causal. In that case, the temperature anisotropy 
in the sky today is given by~\cite{SW}
\begin{equation}
{\delta T\over T}(\theta,\phi) = {1\over3}\Phi(\eta_{\rm LS})\,
Q(\eta_0, \theta, \phi) + 2\int_{\eta_{\rm LS}}^{\eta_0} 
dr\,\Phi'(\eta_0-r)\,Q(r, \theta, \phi)\,,
\label{SachsWolfe}
\end{equation}
where $\eta_0$ is the {\em coordinate distance} to the last scattering
surface, i.e. the present conformal time, while $\eta_{\rm LS}\simeq0$
determines that comoving hypersurface. The above expression is known as
the Sachs-Wolfe effect~\cite{SW}, and contains two parts, the intrinsic
and the Integrated Sachs-Wolfe (ISW) effect, due to integration along
the line of sight of time variations in the gravitational potential.

In linear perturbation theory, the scalar metric perturbations can be
separated into $\Phi(\eta, {\bf x}) \equiv \Phi(\eta)\,Q({\bf x})$,
where $Q({\bf x})$ are the scalar harmonics, eigenfunctions of the
Laplacian in three dimensions, $\nabla^2 Q_{klm}(r, \theta, \phi) =
-k^2\,Q_{klm}(r, \theta, \phi)$. These functions have the general 
form~\cite{Harrison}
\begin{equation}
Q_{klm}(r, \theta, \phi) = \Pi_{kl}(r)\,Y_{lm}(\theta, \phi) \,,
\label{ScalarHarmonics}
\end{equation}
where $Y_{lm}(\theta, \phi)$ are the usual spherical
harmonics~\cite{AS}.

In order to compute the temperature anisotropy associated with the
Sachs-Wolfe effect, we have to know the evolution of the metric
perturbation during the matter era,
\begin{equation}
\Phi''+3{\cal H}\,\Phi' + a^2 \Lambda\,\Phi - 2K\,\Phi = 0\,.
\label{PhiEvEq}
\end{equation}
In the case of a flat universe without cosmological constant, the
Newtonian potential remains constant during the matter era and only
the intrinsic SW effect contributes to $\delta T/T$. In case of a
non-vanishing $\Lambda$, since its contribution is negligible in the
past, most of the photon's trajectory towards us is unperturbed, and
the only difference with respect to the $\Lambda=0$ case is an overall
factor~\cite{CPT}. We will consider here the approximation $\Phi=$
constant during the matter era and ignore that factor, see
Ref.~\cite{BLW}.

In a flat universe, the radial part of the eigenfunctions
(\ref{ScalarHarmonics}) can be written as~\cite{Harrison}
\begin{equation}
\Pi_{kl}(r) = \sqrt{2\over\pi}\,k\,j_l(kr)\,,
\label{RadialEigenfunctions}
\end{equation}
where $j_l(z)$ are the spherical Bessel functions~\cite{AS}. The growing
mode solution of the metric perturbation that left the Hubble scale
during inflation contributes to the temperature anisotropies on large
scales (\ref{SachsWolfe}) as
\begin{equation}\label{alm}
{\delta T\over T}(\theta,\phi) \ = \ {1\over3}\Phi(\eta_{\rm LS})\,Q \ 
= \ {1\over5}\,{\cal R}\,Q(\eta_0, \theta, \phi) \ \equiv \
\sum_{l=2}^\infty \sum_{m=-l}^l\,a_{lm}\,Y_{lm}(\theta, \phi)\,,
\end{equation}
where we have used the fact that at reentry (at the surface of last
scattering) the gauge invariant Newtonian potential $\Phi$ is related to
the curvature perturbation ${\cal R}$ at Hubble-crossing during
inflation, see Eq.~(\ref{PhiRc}); and we have expanded $\delta T/T$ in
spherical harmonics. 

We can now compute the two-point correlation function or angular power
spectrum, $C(\theta)$, of the CMB anisotropies on large scales, defined
as an expansion in multipole number,
\begin{equation}
C(\theta)=
\left\langle{\delta T\over T}^*\!({\bf n}){\delta T\over T}({\bf n}')
\right\rangle_{{\bf n}\cdot{\bf n}'=\cos\theta} = {1\over4\pi}
\sum_{l=2}^\infty (2l+1)\,C_l\,P_l(\cos\theta)\,,
\label{Ctheta}
\end{equation}
where $P_l(z)$ are the Legendre polynomials~\cite{AS}, and we have
averaged over different universe realizations. Since the coefficients
$a_{lm}$ are isotropic (to first order), we can compute the $C_l =
\langle|a_{lm}|^2\rangle$ as
\begin{equation}
C_l^{(S)} = {4\pi\over25}\,\int_0^\infty\,{dk\over k}\,
{\cal P}_{\cal R}(k)\,j_l^2(k\eta_0)\,,
\label{ClScalar}
\end{equation}
where we have used Eqs.~(\ref{alm}) and (\ref{CorrelatorRc}). In the
case of scalar metric perturbation produced during inflation, the scalar
power spectrum at reentry is given by ${\cal P}_{\cal R}(k)=
A_S^2(k\eta_0)^{n-1}$, in the power-law approximation, see
Eq.~(\ref{PowerSpectrumRc}). In that case, one can integrate
(\ref{ClScalar}) to give
\begin{eqnarray}
&&C_l^{(S)}={2\pi\over25}\,A_S^2\,
{\Gamma[{3\over2}]\,\Gamma[1-{n-1\over2}]\,\Gamma[l+{n-1\over2}]\over
\Gamma[{3\over2}-{n-1\over2}]\,\Gamma[l+2-{n-1\over2}]}\,,\\[2mm]
&&{l(l+1)\,C_l^{(S)}\over2\pi}={A_S^2\over25} \ = \ {\rm constant}\,,
\hspace{1cm}{\rm for} \hspace{3mm} n=1\,.\label{CLS}
\end{eqnarray}
This last expression corresponds to what is known as the Sachs-Wolfe
plateau, and is the reason why the coefficients $C_l$ are always plotted
multiplied by $l(l+1)$, see Fig.~\ref{fig:Power}. 

Tensor metric perturbations also contribute with an approximately
constant angular power spectrum, $l(l+1)C_l$. The Sachs-Wolfe effect for
a gauge invariant tensor perturbation is given by~\cite{SW}
\begin{equation}
{\delta T\over T}(\theta,\phi) = \int_{\eta_{\rm LS}}^{\eta_0} 
dr\,h'(\eta_0-r)\,Q_{rr}(r, \theta, \phi)\,,
\label{SachsWolfeTensor}
\end{equation}
where $Q_{rr}$ is the $rr$-component of the tensor harmonic along the
line of sight~\cite{Harrison}. The tensor perturbation $h$ during the
matter era satisfies the following evolution equation
\begin{equation}
h''_k+3{\cal H}\,h'_k+(k^2+2K)\,h_k=0\,,
\end{equation}
which depends on the wavenumber $k$, contrary to what happens with the
scalar modes, see Eq.~(\ref{PhiEvEq}). For a flat ($K=0$) universe, the
solution to this equation is \ $h_k(\eta)=h\,G_k(\eta)$, where $h$ is
the constant tensor metric perturbation at horizon crossing and
$G_k(\eta)=3\,j_1(k\eta)/k\eta$, normalized so that $G_k(0)=1$ at the
surface of last scattering. The radial part of the tensor harmonic
$Q_{rr}$ in a flat universe can be written as~\cite{Harrison}
\begin{equation}
Q^{rr}_{kl}(r) = \left[{(l-1)l(l+1)(l+2)\over\pi k^2}\right]^{1/2}\,
{j_l(kr)\over r^2}\,.
\end{equation}
The tensor angular power spectrum can finally be expressed as
\begin{eqnarray}\label{CLTensor}
&&C_l^{(T)}={9\pi\over4}\,(l-1)l(l+1)(l+2)\,\int_0^\infty\,{dk\over k}\,
{\cal P}_g(k)\,I_{kl}^2\,,\\[1mm]
&&I_{kl} = \int_0^{x_0}\,dx\,{j_2(x_0-x)j_l(x)\over(x_0-x)x^2}\,,
\end{eqnarray}
where $x\equiv k\eta$, and ${\cal P}_g(k)$ is the primordial tensor spectrum
(\ref{TensorSpectrum}). For a scale invariant spectrum, $n_T=0$, we can
integrate (\ref{CLTensor}) to give~\cite{Staro}
\begin{equation}\label{CLT}
l(l+1)\,C_l^{(T)} = {\pi\over36}\Big(1 + {48\pi^2\over385}\Big)\,A_T^2\,
B_l\,,
\end{equation}
with \ $B_l = (1.1184, 0.8789, \dots, 1.00)$ \ for $l=2, 3, \dots,30$.
Therefore, $l(l+1)\,C_l^{(T)}$ also becomes constant for large $l$. Beyond
$l\sim30$, the Sachs-Wolfe expression is not a good approximation and
the tensor angular power spectrum decays very quickly at large $l$,
see Fig.~\ref{fig40}.

\subsubsection{The consistency relation}

In spite of the success of inflation in predicting a homogeneous and
isotropic background on which to imprint a scale-invariant spectrum of
inhomogeneities, it is difficult to test the idea of inflation. A CMB
cosmologist before the 1980s would have argued that {\em ad hoc} initial
conditions could have been at the origin of the homogeneity and flatness
of the universe on large scales, while a LSS cosmologist would have
agreed with Harrison and Zel'dovich that the most natural spectrum
needed to explain the formation of structure was a scale-invariant
spectrum. The surprise was that inflation incorporated an understanding
of {\em both} the globally homogeneous and spatially flat background, and
the approximately scale-invariant spectrum of perturbations in the same
formalism. But that could have been just a coincidence.

What is {\em unique} to inflation is the fact that inflation determines
not just one but {\em two} primordial spectra, corresponding to the
scalar (density) and tensor (gravitational waves) metric perturbations,
from a single continuous function, the inflaton potential $V(\phi)$. In
the slow-roll approximation, one determines, from $V(\phi)$, two
continuous functions, ${\cal P}_{\cal R}(k)$ and ${\cal P}_g(k)$, that
in the power-law approximation reduces to two amplitudes, $A_S$ and
$A_T$, and two tilts, $n$ and $n_T$. It is clear that there must be a
relation between the four parameters. Indeed, one can see from
Eqs.~(\ref{CLT}) and (\ref{CLS}) that the ratio of the tensor to scalar
contribution to the angular power spectrum is proportional to the tensor
tilt~\cite{LL93}, 
\begin{equation}\label{Consistency}
R\equiv {C_l^{(T)}\over C_l^{(S)}} = {25\over9}
\Big(1 + {48\pi^2\over385}\Big)\,2\epsilon \simeq - 2\pi\,n_T\,.
\end{equation}
This is a unique prediction of inflation, which could not have been
postulated a priori by any cosmologist. If we finally observe a tensor
spectrum of anisotropies in the CMB, or a stochastic gravitational wave
background in laser interferometers like LIGO or LISA, with sufficient
accuracy to determine their spectral tilt, one might have some chance to
test the idea of inflation, via the consistency relation
(\ref{Consistency}). For the moment, observations of the microwave
background anisotropies suggest that the Sachs-Wolfe plateau exists, see
Fig.~\ref{fig:Power}, but it is still premature to determine the tensor
contribution. Perhaps in the near future, from the analysis of
polarization as well as temperature anisotropies, with the CMB
satellites MAP and Planck, we might have a chance of determining the
validity of the consistency relation.

Assuming that the scalar contribution dominates over the tensor on large
scales, i.e. $R \ll 1$, one can actually give a measure of the amplitude
of the scalar metric perturbation from the observations of the
Sachs-Wolfe plateau in the angular power spectrum~\cite{WMAP},
\begin{eqnarray}\label{MeasuredPower}
\left[{l(l+1)\,C_l^{(S)}\over2\pi}\right]^{1/2}\hspace{-2mm}&=&
{A_S\over5} \ = \ (1.03\pm0.07)\times10^{-5}\,,\\[2mm]
n&=&0.97\pm0.03\,.
\label{MeasuredTilt}
\end{eqnarray}
These measurements can be used to normalize the primordial spectrum and
determine the parameters of the model of inflation~\cite{LR99}.
In the near future these parameters will be determined with much better
accuracy, as described in Section~4.4.5.

\subsubsection{The acoustic peaks}

The Sachs-Wolfe plateau is a distinctive feature of Fig.~24. These
observations confirm the existence of a primordial spectrum of scalar
(density) perturbations on all scales, otherwise the power spectrum
would have started from zero at $l=2$. However, we see that the
spectrum starts to rise around $l=20$ towards the first acoustic peak,
where the SW approximation breaks down and the above formulae are no
longer valid.

As mentioned above, the first peak in the photon distribution
corresponds to overdensities that have undergone half an oscillation,
that is, a compression, and appear at a scale associated with the size
of the horizon at last scattering, about $1^\circ$ projected in the sky
today. Since photons scatter off baryons, they will also feel the
acoustic wave and create a peak in the correlation function. The height
of the peak is proportional to the amount of baryons: the larger the
baryon content of the universe, the higher the peak. The position of the
peak in the power spectrum depends on the geometrical size of the
particle horizon at last scattering. Since photons travel along
geodesics, the projected size of the causal horizon at decoupling
depends on whether the universe is flat, open or closed. In a flat
universe the geodesics are straight lines and, by looking at the angular
scale of the first acoustic peak, we would be measuring the actual size
of the horizon at last scattering. In an open universe, the geodesics
are inward-curved trajectories, and therefore the projected size on the
sky appears smaller. In this case, the first acoustic peak should occur
at higher multipoles or smaller angular scales. On the other hand, for a
closed universe, the first peak occurs at smaller multipoles or larger
angular scales. The dependence of the position of the first acoustic
peak on the spatial curvature can be approximately given by $l_{\rm
peak} \simeq 220\,\Omega_0^{-1/2}$, where $\Omega_0=\Omega_{\rm M} +
\Omega_\Lambda = 1-\Omega_K$. Past observations from the balloon
experiment BOOMERANG~\cite{Boomerang}, suggested clearly a few years ago
that the first peak was between $l=180$ and 250 at 95\% c.l., with an
amplitude $\delta T = 80\pm10\ \mu$K, and therefore the universe was
most probably flat.  However, with the high precision WMAP data we can
now pinpoint the spatial curvature to a few percent,
\begin{equation}
\Omega_0 = 1.02\pm0.02 \hspace{5mm}(95\% \ {\rm c.l.})
\label{Omega0}
\end{equation}
That is, the universe is spatially flat (i.e. Euclidean), within 2\%
uncertainty, which is much better than we could ever do before.

%%%%%%%%%%%%%%%%%%%%%%%%%
\begin{figure}[htb]
\begin{center}\hspace{-1mm}
\includegraphics[width=6cm,angle=-90]{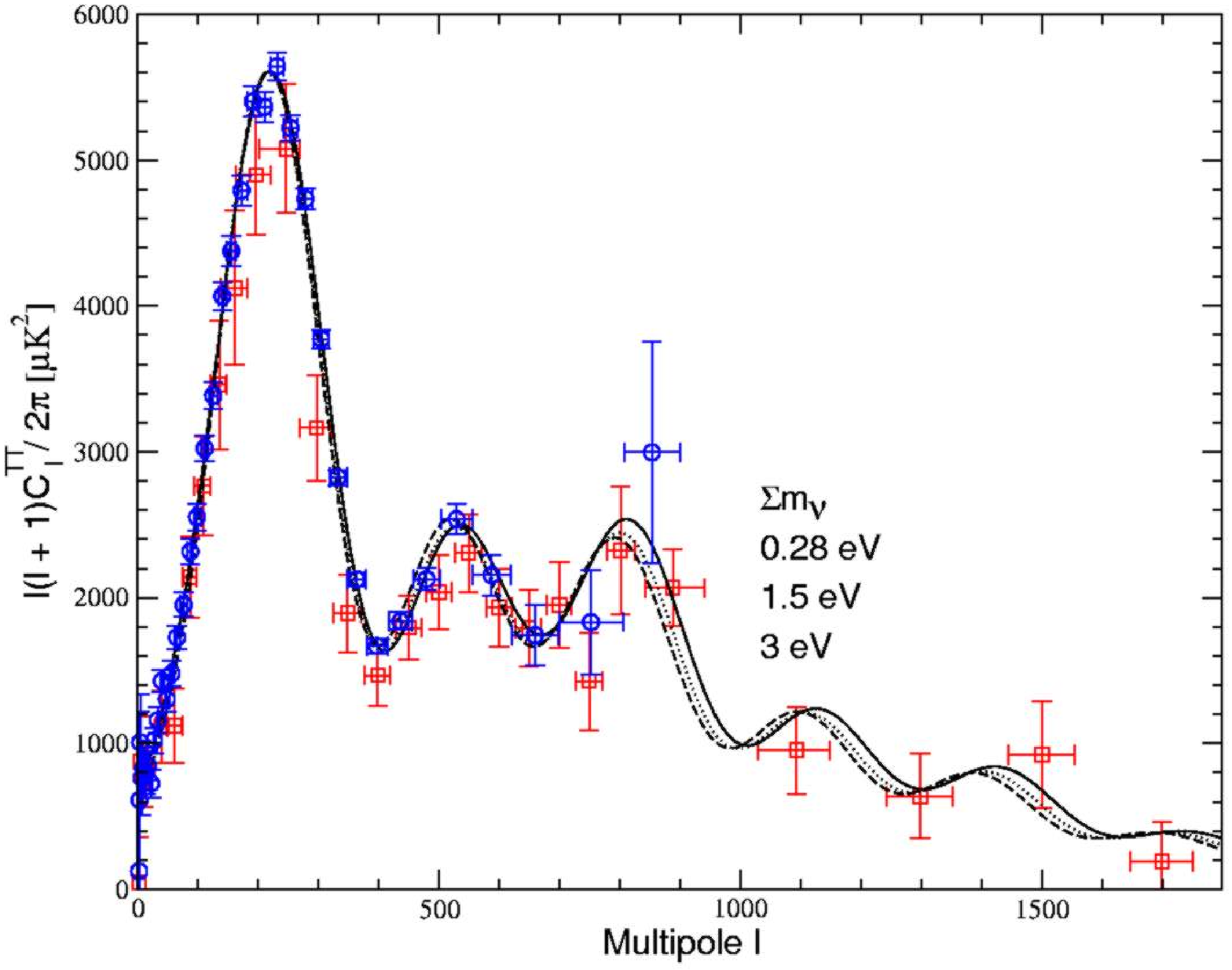}
\includegraphics[width=6cm,angle=-90]{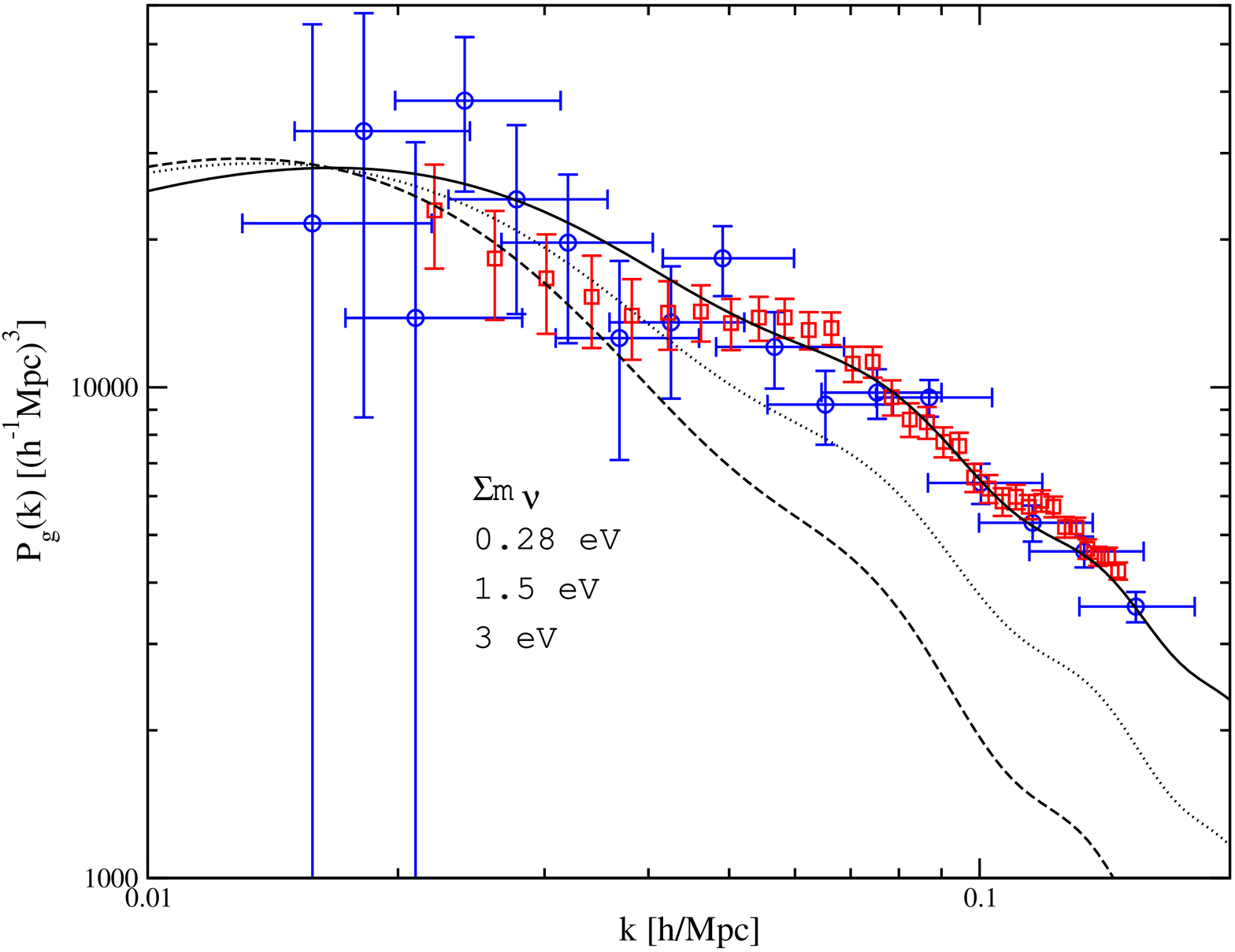}
\end{center}\vspace*{-3mm}
\caption{The dependence of CMB anisotropies and LSS power spectrum on
the sum of the mass of all neutrino species. The blue(red) data
corresponds to WMAP(Boomerang, etc.) and SDSS(2dFGRS), for the CMB and
LSS respectively.}
\label{CMBLSSNu}
\end{figure}
%%%%%%%%%%%%%%%%%%%%%%%%%

With Boomerang, CBI, VSA, and specially with WMAP, we have evidence of
at least three distinct acoustic peaks. In the near furture, even before
Planck, we may be able to distinguish anothes two. These peaks should
occur at harmonics of the first one, but are typically much lower
because of Silk damping. Since the amplitude and position of the primary
and secondary peaks are directly determined by the sound speed (and,
hence, the equation of state) and by the geometry and expansion of the
universe, they can be used as a powerful test of the density of baryons
and dark matter, and other cosmological parameters. 

By looking at these patterns in the anisotropies of the microwave
background, cosmologists can determine not only the cosmological
parameters, but also the primordial spectrum of
density perturbations produced during inflation. It turns out that the
observed temperature anisotropies are compatible with a scale-invariant
spectrum, see Eq.~(\ref{MeasuredTilt}), as predicted by inflation. This
is remarkable, and gives very strong support to the idea that inflation
may indeed be responsible for both the CMB anisotropies and the
large-scale structure of the universe. Different models of inflation
have different specific predictions for the fine details associated with
the spectrum generated during inflation. It is these minute differences
that will allow cosmologists to differentiate between alternative models
of inflation and discard those that do not agree with observations.
However, most importantly, perhaps, the pattern of anisotropies
predicted by inflation is completely different from those predicted by
alternative models of structure formation, like cosmic defects: strings,
vortices, textures, etc. These are complicated networks of energy
density concentrations left over from an early universe phase
transition, analogous to the defects formed in the laboratory in certain
kinds of liquid crystals when they go through a phase transition. The
cosmological defects have spectral properties very different from those
generated by inflation. That is why it is so important to launch more
sensitive instruments, and with better angular resolution, to determine
the properties of the CMB anisotropies.

\subsubsection{The new microwave anisotropy satellites, WMAP and Planck}

The large amount of information encoded in the anisotropies of the
microwave background is the reason why both NASA and the European Space
Agency have decided to launch two independent satellites to measure the
CMB temperature and polarization anisotropies to unprecendented
accuracy. The Wilkinson Microwave Anisotropy Probe~\cite{WMAPHP} was
launched by NASA at the end of 2000, and has fulfilled most of our
expectation, while Planck~\cite{Planck} is expected to be lanched by ESA
in 2007. There are at the moment other large proposals like CMB
Pol~\cite{CMBpol}, ACT~\cite{ACT}, etc. which will see the light in the
next few years, see Ref.~\cite{Tegmark}.

As we have emphasized before, the fact that these anisotropies have
such a small amplitude allow for an accurate calculation of the
predicted anisotropies in linear perturbation theory. A particular
cosmological model is characterized by a dozen or so parameters: the
rate of expansion, the spatial curvature, the baryon content, the cold
dark matter and neutrino contribution, the cosmological constant
(vacuum energy), the reionization parameter (optical depth to the last
scattering surface), and various primordial spectrum parameters like
the amplitude and tilt of the adiabatic and isocurvature spectra, the
amount of gravitational waves, non-Gaussian effects, etc. All these
parameters can now be fed into very fast CMB codes called
CMBFAST~\cite{CMBFAST} and CAMB~\cite{CAMB}, that compute the
predicted temperature and polarization anisotropies to better than 1\%
accuracy, and thus can be used to compare with observations.

These two satellites will improve both the sensitivity, down to $\mu$K,
and the resolution, down to arc minutes, with respect to the previous
COBE satellite, thanks to large numbers of microwave horns of various
sizes, positioned at specific angles, and also thanks to recent advances
in detector technology, with high electron mobility transistor
amplifiers (HEMTs) for frequencies below 100 GHz and bolometers for
higher frequencies. The primary advantage of HEMTs is their ease of use
and speed, with a typical sensitivity of 0.5 mKs$^{1/2}$, while the
advantage of bolometers is their tremendous sensitivity, better than 0.1
mKs$^{1/2}$, see Ref.~\cite{Page}. This will allow cosmologists to
extract information from around 3000 multipoles! Since most of the
cosmological parameters have specific signatures in the height and
position of the first few acoustic peaks, the higher the resolution, the
more peaks one is expected to see, and thus the better the accuracy with
which one will be able to measure those parameters, see Table~2.

\begin{figure}[htb]
\begin{center}%\vspace*{-4cm}
%\hspace{1cm}
\includegraphics[width=9cm]{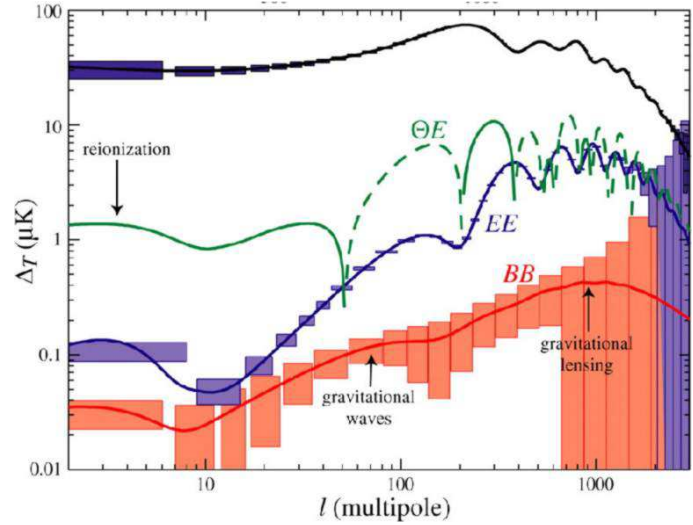}
\end{center}
\vspace*{-2mm}
\caption{Theoretical predictions for the four non-zero CMB
temperature-polarization spectra as a function of multipole moment,
together with the expectations from Planck. From Ref.~\cite{HD}.}
\label{fig39}
\end{figure}

Although the satellite probes were designed for the accurate
measurement of the CMB temperature anisotropies, there are other
experiments, like balloon-borne and ground
interferometers~\cite{Tegmark}.  Probably the most important objective
of the future satellites (beyond WMAP) will be the measurement of the
CMB polarization anisotropies, discovered by DASI in November
2002~\cite{DASI}, and confirmed a few months later by WMAP with
greater accuracy~\cite{WMAP}, see Fig.~24. These anisotropies were
predicted by models of structure formation and indeed found at the
level of microKelvin sensitivities, where the new satellites were
aiming at. The complementary information contained in the polarization
anisotropies already provides much more stringent constraints on the
cosmological parameters than from the temperature anisotropies alone.
However, in the future, Planck and CMB pol will have much better
sensitivities. In particular, the curl-curl component of the
polarization power spectra is nowadays the only means we have to
determine the tensor (gravitational wave) contribution to the metric
perturbations responsible for temperature anisotropies, see
Fig.~\ref{fig39}. If such a component is found, one could constraint
very precisely the model of inflation from its spectral properties,
specially the tilt~\cite{KK}.

%%%%%%%%%%%%%%%%%%%%%%%%%
%%%%%%%%%%%%%%%%%%%%%%%%%

\subsection{From metric perturbations to large scale structure}

If inflation is responsible for the metric perturbations that gave rise
to the temperature anisotropies observed in the microwave background,
then the primordial spectrum of density inhomogeneities induced by the
same metric perturbations should also be responsible for the present
large scale structure~\cite{LiddleLyth}. This simple connection allows
for more stringent tests on the inflationary paradigm for the generation
of metric perturbations, since it relates the large scales (of order the
present horizon) with the smallest scales (on galaxy scales). This
provides a very large lever arm for the determination of primordial
spectra parameters like the tilt, the nature of the perturbations,
whether adiabatic or isocurvature, the geometry of the universe, as well
as its matter and energy content, whether CDM, HDM or mixed CHDM.

\subsubsection{The galaxy power spectrum}

As metric perturbations enter the causal horizon during the radiation or
matter era, they create density fluctuations via gravitational
attraction of the potential wells. The density contrast $\delta$ can be
deduced from the Einstein equations in linear perturbation theory, see
Eq.~(\ref{PertEinsteinEqs}),
\begin{equation}\label{deltaRc}
\delta_k \equiv {\delta\rho_k\over\rho} = 
\left({k\over aH}\right)^2\,{2\over3}\,\Phi_k =
\left({k\over aH}\right)^2\,{2+2\omega\over5+3\omega}\,{\cal R}_k\,,
\end{equation}
where we have assumed $K=0$, and used Eq.~(\ref{PhiRc}). From this
expression one can compute the power spectrum, at horizon crossing, of
matter density perturbations induced by inflation, see
Eq.~(\ref{CorrelatorRc}),
\begin{equation}
P(k) = \langle|\delta_k|^2\rangle = A\,\left({k\over aH}\right)^n\,,
\end{equation}
with $n$ given by the scalar tilt (\ref{ScalarTilt}), $n = 1 + 2\eta -
6\epsilon$. This spectrum reduces to a Harrison-Zel'dovich spectrum 
(\ref{HarrisonZeldovich}) in the slow-roll approximation: \ $\eta, \,
\epsilon \ll 1$.

Since perturbations evolve after entering the horizon, the power
spectrum will not remain constant. For scales entering the horizon well
after matter domination ($k^{-1} \gg k^{-1}_{\rm eq} \simeq 81$ Mpc),
the metric perturbation has not changed significantly, so that ${\cal
R}_k({\rm final}) = {\cal R}_k({\rm initial})$. Then Eq.~(\ref{deltaRc})
determines the final density contrast in terms of the initial one. On
smaller scales, there is a linear transfer function $T(k)$, which may be
defined as~\cite{LL93}
\begin{equation}\label{TransferFunction}
{\cal R}_k({\rm final}) = T(k)\,{\cal R}_k({\rm initial})\,.
\end{equation}
To calculate the transfer function one has to specify the initial
condition with the relative abundance of photons, neutrinos, baryons and
cold dark matter long before horizon crossing. The most natural
condition is that the abundances of all particle species are uniform on
comoving hypersurfaces (with constant total energy density). This is
called the {\em adiabatic} condition, because entropy is conserved
independently for each particle species $X$, i.e. $\delta\rho_X =
\dot\rho_X\delta t$, given a perturbation in time from a comoving
hypersurface, so
\begin{equation}\label{AdiabaticCondition}
{\delta\rho_X\over\rho_X+p_X} = {\delta\rho_Y\over\rho_Y+p_Y}\,,
\end{equation}
where we have used the energy conservation equation for each species,
$\dot\rho_X=-3H(\rho_X+p_X)$, valid to first order in perturbations. It
follows that each species of radiation has a common density contrast
$\delta_r$, and each species of matter has also a common density
contrast $\delta_m$, with the relation \ $\delta_m={3\over4}\delta_r$.

Given the adiabatic condition, the transfer function is determined by
the physical processes occuring between horizon entry and matter
domination. If the radiation behaves like a perfect fluid, its density
perturbation oscillates during this era, with decreasing amplitude. The
matter density contrast living in this background does not grow
appreciably before matter domination because it has negligible
self-gravity. The transfer function is therefore given roughly by,
see Eq.~(\ref{PowerSpectrum}),
\begin{equation}\label{Tk}
T(k) = \left\{\begin{array}{ll}
1\,,&\hspace{1cm}k\ll k_{\rm eq}\\[2mm]
(k/k_{\rm eq})^2\,,&
\hspace{1cm}k\gg k_{\rm eq}\end{array}\right.
\end{equation}

The perfect fluid description of the radiation is far from being correct
after horizon entry, because roughly half of the radiation consists of
neutrinos whose perturbation rapidly disappears through free
streeming. The photons are also not a perfect fluid because they diffuse
significantly, for scales below the Silk scale, $k_S^{-1} \sim 1$ Mpc.
One might then consider the opposite assumption, that the radiation has
zero perturbation after horizon entry. Then the matter density
perturbation evolves according to
\begin{equation}\label{EvolutionDensity}
\ddot\delta_k + 2H\dot\delta_k + (c_s^2\,k^2_{\rm ph}-4\pi G\rho)\,
\delta_k = 0\,,
\end{equation}
which corresponds to the equation of a damped harmonic oscillator.  The
zero-frequency oscillator defines the Jeans wavenumber, $k_J =
\sqrt{4\pi G\rho/c_s^2}$. For $k\ll k_J$, $\delta_k$ grows exponentially
on the dynamical timescale, $\tau_{\rm dyn} = {\rm Im}\,\omega^{-1} =
(4\pi G\rho)^{-1/2} = \tau_{\rm grav}$, which is the time scale for
gravitational collapse. One can also define the Jeans length,
\begin{equation}\label{JeansLength}
\lambda_J = {2\pi\over k_J} = c_s\,\sqrt{\pi\over G\rho}\,,
\end{equation}
which separates gravitationally stable from unstable modes. If we define
the pressure response timescale as the size of the perturbation over the
sound speed, $\tau_{\rm pres} \sim \lambda/c_s$, then, if $\tau_{\rm pres} 
> \tau_{\rm grav}$, gravitational collapse of a perturbation can occur
before pressure forces can response to restore hydrostatic equilibrium
(this occurs for $\lambda > \lambda_J$). On the other hand, if
$\tau_{\rm pres} < \tau_{\rm grav}$, radiation pressure prevents
gravitational collapse and there are damped acoustic oscillations
(for $\lambda < \lambda_J$).

\begin{figure}[htb]
%\vspace*{-1cm}
\begin{center}
\hspace{-5mm}
\includegraphics[width=9cm]{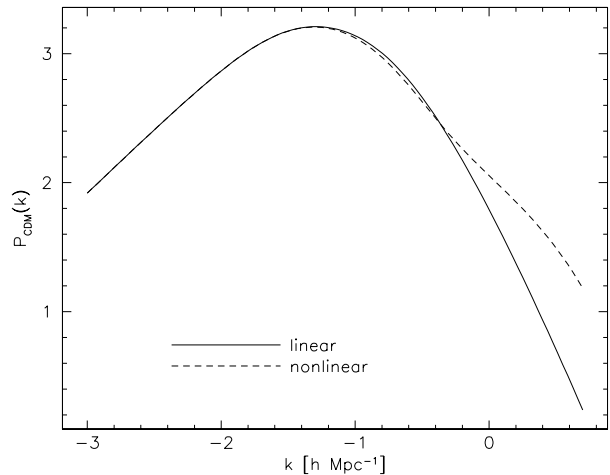}
\end{center}
\vspace*{-8mm}
\caption{The CDM power spectrum $P(k)$ as a function of wavenumber $k$,
in logarithmic scale, normalized to the local abundance of galaxy
clusters, for an Einstein-de Sitter universe with $h=0.5$. The solid
(dashed) curve shows the linear (non-linear) power spectrum. While the
linear power spectrum falls off like $k^{-3}$, the non-linear
power-spectrum illustrates the increased power on small scales due to
non-linear effects, at the expense of the large-scale structures.  From
Ref.~\cite{Bartelmann}.}
\label{fig40}
\end{figure}

We will consider now the behaviour of modes within the horizon during
the transition from the radiation ($c_s^2=1/3$) to the matter era
($c_s^2=0$). The growing mode solution increases only by a factor of 2
between horizon entry and the epoch when matter starts to dominate,
i.e. $y=1$.  The transfer function is therefore again roughly given by
Eq.~(\ref{Tk}). Since the radiation consists roughly half of
neutrinos, which free streem, and half of photons, which either form a
perfect fluid or just diffuse, neither the perfect fluid nor the
free-streeming approximation looks very sensible. A more precise
calculation is needed, including: neutrino free streeming around the
epoch of horizon entry; the diffusion of photons around the same time,
for scales below Silk scale; the diffusion of baryons along with the
photons, and the establishment after matter domination of a common
matter density contrast, as the baryons fall into the potential wells
of cold dark matter. All these effects apply separately, to first
order in the perturbations, to each Fourier component, so that a
linear transfer function is produced. There are several
parametrizations in the literature, but the one which is more widely
used is that of Ref.~\cite{Bond},
\begin{eqnarray}\label{TFk}
&&T(k) = \left[1+\Big(ak+(bk)^{3/2}+(ck)^2\Big)^\nu
\right]^{-1/\nu}\,, \hspace{1cm} \nu=1.13\,,\\[2mm]
&&a=6.4\,(\Omega_{\rm M}h)^{-1}\,h^{-1} \ {\rm Mpc}\,, \\
&&b=3.0\,(\Omega_{\rm M}h)^{-1}\,h^{-1} \ {\rm Mpc}\,, \\
&&c=1.7\,(\Omega_{\rm M}h)^{-1}\,h^{-1} \ {\rm Mpc}\,.
\end{eqnarray}
We see that the behaviour estimated in Eq.~(\ref{Tk}) is roughly
correct, although the break at $k=k_{\rm eq}$ is not at all sharp, see
Fig.~\ref{fig40}. The transfer function, which encodes the soltion to
linear equations, ceases to be valid when the density contrast becomes
of order 1. After that, the highly nonlinear phenomenon of gravitational
collapse takes place, see Fig.~\ref{fig40}.

%%%%%%%%%%%%%%%%%%%%%%%%%
%%%%%%%%%%%%%%%%%%%%%%%%%

\subsubsection{The new redshift catalogs, 2dF and Sloan Digital Sky Survey}

Our view of the large-scale distribution of luminous objects in the
universe has changed dramatically during the last 25 years:
from the simple pre-1975 picture of a distribution of field and cluster
galaxies, to the discovery of the first single superstructures and
voids, to the most recent results showing an almost regular web-like
network of interconnected clusters, filaments and walls, separating huge
nearly empty volumes. The increased efficiency of redshift surveys, made
possible by the development of spectrographs and -- specially in the
last decade -- by an enormous increase in multiplexing gain (i.e. the
ability to collect spectra of several galaxies at once, thanks to
fibre-optic spectrographs), has allowed us not only to do {\em
cartography} of the nearby universe, but also to statistically
characterize some of its properties, see Ref.~\cite{royalsoc}. At the
same time, advances in theoretical modeling of the development of
structure, with large high-resolution gravitational simulations coupled
to a deeper yet limited understanding of how to form galaxies within the
dark matter halos, have provided a more realistic connection of the
models to the observable quantities~\cite{Moore}. Despite the large
uncertainties that still exist, this has transformed the study of
cosmology and large-scale structure into a truly quantitative science,
where theory and observations can progress side by side.

I will concentrate on two of the new catalogs, which are taking data at
the moment and which have changed the field, the 2-degree-Field (2dF)
Catalog and the Sloan Digital Sky Survey (SDSS). The advantages of
multi-object fibre spectroscopy have been pushed to the extreme with the
construction of the 2dF spectrograph for the prime focus of the
Anglo-Australian Telescope~\cite{2dFGRS}. This instrument is able to
accommodate 400 automatically positioned fibres over a 2 degree in
diameter field. This implies a density of fibres on the sky of
approximately 130 deg$^{-2}$, and an optimal match to the galaxy counts
for a magnitude $b_J\simeq 19.5$, similar to that of previous surveys
like the ESP, with the difference that with such an area yield, the same
number of redshifts as in the ESP survey can be collected in about 10
exposures, or slightly more than one night of telescope time with
typical 1 hour exposures.  This is the basis of the 2dF galaxy redshift
survey. Its goal is to measure redshifts for more than 250,000 galaxies
with $b_J<19.5$.  In addition, a faint redshift survey of 10,000
galaxies brighter than $R=21$ will be done over selected fields within
the two main strips of the South and North Galactic Caps.  The survey
has now finished, with a quarter of a million redshifts. The final
result can be seen in Ref.~\cite{2dFGRS}.

The most ambitious and comprehensive galaxy survey currently in
progress is without any doubt the Sloan Digital Sky
Survey~\cite{SDSS}. The aim of the project is, first of all, to
observe photometrically the whole Northern Galactic Cap, 30$^\circ$
away from the galactic plane (about $10^4$ deg$^2$) in five bands, at
limiting magnitudes from 20.8 to 23.3. The expectation is to detect
around 50 million galaxies and around $10^8$ star-like sources. This
has already led to the discovery of several high-redshift ($z>4$)
quasars, including the highest-redshift quasar known, at $z=5.0$, see
Ref.~\cite{SDSS}. Using two fibre spectrographs carrying 320 fibres
each, the spectroscopic part of the survey will then collect spectra
from about $10^6$ galaxies with $r'<18$ and $10^5$ AGNs with
$r'<19$. It will also select a sample of about $10^5$ red luminous
galaxies with $r'<19.5$, which will be observed spectroscopically,
providing a nearly volume-limited sample of early-type galaxies with a
median redshift of $z\simeq0.5$, that will be extremely valuable to
study the evolution of clustering. The data that is coming from these
catalogs is so outstanding that already cosmologists are using them
for the determination of the cosmological parameters of the standard
model of cosmology. The main outcome of these catalogs is the linear
power spectrum of matter fluctuations that give rise to galaxies, and
clusters of galaxies. It covers from the large scales of order
Gigaparsecs, the realm of the unvirialised superclusters, to the small
scales of hundreds of kiloparsecs, where the Lyman-$\alpha$ systems
can help reconstruct the linear power spectrum, since they are less
sensitive to the nonlinear growth of perturbations.

As often happens in particle physics, not always are observations from
a single experiment sufficient to isolate and determine the precise
value of the parameters of the standard model. We mentioned in the
previous Section that some of the cosmological parameters created
similar effects in the temperature anisotropies of the microwave
background. We say that these parameters are {\em degenerate} with
respect to the observations.  However, often one finds combinations of
various experiments/observations which break the degeneracy, for
example by depending on a different combination of parameters. This is
precisely the case with the cosmological parameters, as measured by a
combination of large-scale structure observations, microwave
background anisotropies, Supernovae Ia observations and Hubble Space
Telescope measurements. It is expected that in the near future we will
be able to determine the parameters of the standard cosmological model
with great precision from a combination of several different
experiments.

\section{CONCLUSION}

In the last five years we have seen a true revolution in the quality
and quantity of cosmological data that has allowed cosmologists to
determine most of the cosmological parameters with a few percent
accuracy and thus fix a Standard Model of Cosmology. The art of
measuring the cosmos has developed so rapidly and efficiently that one
may be temped of renaming this science as Cosmonomy, leaving the word
Cosmology for the theories of the Early Universe. In summary, we now
know that the stuff we are made of $-$ baryons $-$ constitutes just
about 4\% of all the matter/energy in the Universe, while 25\% is dark
matter $-$ perhaps a new particle species related to theories beyond
the Standard Model of Particle Physics $-$, and the largest fraction,
70\%, some form of diffuse tension also known as dark energy $-$
perhaps a cosmological constant. The rest, about 1\%, could be in the
form of massive neutrinos. 

Nowadays, a host of observations $-$ from CMB anisotropies and large
scale structure to the age and the acceleration of the universe $-$
all converge towards these values, see Fig.~25. Fortunately,
we will have, within this decade, new satellite experiments like
Planck, CMBpol, SNAP as well as deep galaxy catalogs from Earth, to
complement and precisely pin down the values of the Standard Model
cosmological parameters below the percent level, see Table~1.

All these observations would not make much sense without the
encompassing picture of the inflationary paradigm that determines the
homogeneous and isotropic background on top of which it imprints an
approximately scale invariant gaussian spectrum of adiabatic
fluctuations. At present all observations are consistent with the
predictions of inflation and hopefully in the near future we may have
information, from the polarization anisotropies of the microwave
background, about the scale of inflation, and thus about the physics
responsible for the early universe dynamics.

%\vskip1cm
\noindent

\section*{ACKNOWLEDGEMENTS}

I would like to thank the organizers of the CERN-JINR European School of
High Energy Physics 2004, and very specially Matteo Cavalli-Sforza,
without whom this wonderful school would not have been the success it
was. This work was supported in part by a CICYT project FPA2003-04597.


\begin{thebibliography}{99}

\bibitem{Einstein} A. Einstein, Sitz. Preuss. Akad. Wiss. Phys. 
{\bf 142} (1917) (\S 4); Ann. Phys. {\bf 69} (1922) 436.

\bibitem{Friedmann} A. Friedmann, Z. Phys. {\bf 10} (1922) 377.

\bibitem{Hubble} E.P. Hubble, Publ. Nat. Acad. Sci. {\bf 15} (1929) 168.

\bibitem{Gamow} G. Gamow, Phys. Rev. {\bf 70} (1946) 572; Phys. Rev. 
{\bf 74} (1948) 505. 

\bibitem{Wilson} A.A. Penzias and R.W. Wilson, Astrophys. J. {\bf 142} 
(1965) 419.

\bibitem{Weinberg} S. Weinberg, {\em Gravitation and Cosmology} 
(John Wiley \& Sons, San Francisco, 1972).

\bibitem{SCP} S. Perlmutter {\it et al.} [Supernova Cosmology Project], 
Astrophys. J. {\bf 517} (1999) 565.
Home Page {\tt http://scp.berkeley.edu/}

\bibitem{HSTKP}
W.~L.~Freedman {\it et al.}, Astrophys.\ J.\  {\bf 553} (2001) 47. 

\bibitem{HRS} A.~G.~Riess {\it et al.} [High-z Supernova Search],
Astron. J. {\bf 116} (1998) 1009. \\
Home Page {\tt http://cfa-www.harvard.edu/cfa/oir/Research/supernova/}

\bibitem{JGB}J. Garc\'{\i}a-Bellido in {\em European School of 
High Energy Physics}, ed. A. Olchevski (CERN report 2000-007);
e-print Archive: hep-ph/0004188.

\bibitem{SCP2003} R.~A.~Knop {\it et al.}, e-print Archive: astro-ph/0309368.

\bibitem{Riess2004} A.~G.~Riess {\it et al.}, e-print Archive: 
astro-ph/0402512.

\bibitem{WeinbergCosmo} S. Weinberg, Rev. Mod. Phys. {\bf 61} (1989) 1;
S.~M.~Carroll, Living Rev.\ Rel.\  {\bf 4} (2001) 1; T.~Padmanabhan,
Phys.\ Rept.\  {\bf 380} (2003) 235; P.~J.~E.~Peebles and B.~Ratra,
Rev.\ Mod.\ Phys.\  {\bf 75} (2003) 559.

\bibitem{SNAP} The SuperNova/Acceleration Probe Home page:
{\tt http://snap.lbl.gov/}

\bibitem{KT} E.W. Kolb and M.S. Turner, ``The Early Universe'',
Addison Wesley (1990).

\bibitem{QSO} R. Srianand, P. Petitjean and C. Ledoux, Nature {\bf
408} (2000) 931.


\bibitem{Burles} S. Burles, K.M. Nollett, J.N. Truran, M.S. Turner,
Phys. Rev. Lett. {\bf 82} (1999) 4176; S. Burles, K.M. Nollett,
M.S. Turner, ``Big-Bang Nucleosynthesis: Linking Inner Space and Outer
Space'', e-print Archive: \ astro-ph/9903300.

\bibitem{BBN} K.~A.~Olive, G.~Steigman and T.~P.~Walker,
Phys.\ Rept.\ {\bf 333} (2000) 389; J.~P.~Kneller and G.~Steigman,
``BBN For Pedestrians,'' New J.\ Phys.\  {\bf 6} (2004) 117.

\bibitem{PDG} Particle Data Group Home Page, \ 
{\tt http://pdg.web.cern.ch/pdg/}

\bibitem{WMAP} D.~N.~Spergel {\it et al.},
Astrophys.\ J.\ Suppl.\  {\bf 148} (2003) 175.

\bibitem{FIRAS} J.C. Mather et al., Astrophys. J. {\bf 512} (1999) 511.

\bibitem{Dicke} R.H. Dicke, P.J.E. Peebles, P.G. Roll and D.T.
Wilkinson, Astrophys. J. {\bf 142} (1965) 414.

\bibitem{DMR} C.L. Bennett et al., Astrophys. J. {\bf 464} (1996) L1.

\bibitem{Peebles} P.J.E. Peebles, ``Principles of Physical Cosmology'',
Princeton U.P. (1993).

\bibitem{Padmanabhan} T. Padmanabhan, ``Structure Formation in the
Universe'', Cambridge U.P. (1993).

\bibitem{HZ} E.R. Harrison, Phys. Rev. D {\bf 1} (1970) 2726;
Ya. B. Zel'dovich, Astron. Astrophys. {\bf 5} (1970) 84.

\bibitem{PSCz} The IRAS Point Source Catalog Web page:\\
{\tt http://www-astro.physics.ox.ac.uk/\~{}wjs/pscz.html}

\bibitem{PJS} P.J. Steinhardt, in Particle and Nuclear
Astrophysics and Cosmology in the Next Millennium, ed. by E.W. kolb and
R. Peccei (World Scientific, Singapore, 1995).

\bibitem{Freedman} W.L. Freedman, ``Determination of cosmological
parameters'', Nobel Symposium (1998), e-print Archive: \ hep-ph/9905222.

\bibitem{Refsdael} S. Refsdael, Mon. Not. R. Astr. Soc. {\bf 128} (1964)
295; {\bf 132} (1966) 101.

\bibitem{Blandford} R.D. Blandford and T. Kundi\'c, ``Gravitational 
Lensing and the Extragalactic Distance Scale'', e-print Archive: \ 
astro-ph/9611229.

\bibitem{Grogin} N.A. Grogin and R. Narayan, Astrophys. J. {\bf 464}
(1996) 92.

\bibitem{Birkinshaw} M. Birkinshaw, Phys. Rep. {\bf 310} (1999) 97.

\bibitem{Chandra} The Chandra X-ray observatory Home Page: \
{\tt http://chandra.harvard.edu/}

\bibitem{HST} W.~L.~Freedman {\it et al.},
Astrophys.\ J.\  {\bf 553} (2001) 47

\bibitem{Zwicky} F. Zwicky, Helv. Phys. Acata {\bf 6} (1933) 110.

\bibitem{Freeman} K.C. Freeman, Astrophys. J. {\bf 160} (1970) 811.

\bibitem{Frenk} C.M. Baugh et al., ``Ab initio galaxy formation'',
e-print Archive: \ astro-ph/9907056; Astrophys. J. {\bf 498} 
(1998) 405.

\bibitem{Klypin} F.~Prada {\it et al.},
Astrophys.\ J.\  {\bf 598} (2003) 260.

\bibitem{Sarazin} C.L. Sarazin, Rev. Mod. Phys. {\bf 58} (1986) 1.

\bibitem{Bartelmann} M. Bartelmann et al., Astron. \& Astrophys.  {\bf
330} (1998) 1; M.~Bartelmann and P.~Schneider,
Phys.\ Rept.\  {\bf 340} (2001) 291


\bibitem{2dFGRS} M.~Colless {\it et al.}  [2dFGRS Coll.],
``The 2dF Galaxy Redshift Survey: Final Data Release,''
Archive: \ astro-ph/0306581.  The 2dFGRS Home Page: \ 
{http://www.mso.anu.edu.au/2dFGRS/}

\bibitem{SDSS} M.~Tegmark {\it et al.}  [SDSS Collaboration],
Astrophys.\ J.\  {\bf 606} (2004) 702; Phys.\ Rev.\ D {\bf 69} 
(2004) 103501. The SDSS Home Page: \
{\tt http://www.sdss.org/sdss.html}

\bibitem{Raffelt} G.G. Raffelt, ``Dark Matter: Motivation, Candidates
and Searches'', European Summer School of High Energy Physics 1997. 
CERN Report pp. 235-278, e-print Archive: \ hep-ph/9712538.

\bibitem{Peebles2} P.J.E. Peebles, ``Testing GR on the Scales of
Cosmology,'' e-print Archive: \ astro-ph/0410284.

\bibitem{CGG} M.~C.~Gonzalez-Garcia,
``Global analysis of neutrino data,''
e-print Archive: \ hep-ph/0410030.

\bibitem{TG} S.D. Tremaine and J.E. Gunn, Phys. Rev. Lett. {\bf 42}
(1979) 407; J. Madsen, Phys. Rev. D {\bf 44} (1991) 999.

\bibitem{WIMP} J. Primack, D. Seckel and B. Sadoulet, Ann. Rev. Nucl.
Part. Sci. {\bf 38} (1988) 751; N.E. Booth, B. Cabrera and E. Fiorini, 
Ann. Rev. Nucl. Part. Sci. {\bf 46} (1996) 471.

\bibitem{Tritium} C.~Kraus {\it et al.}, ``Final results from phase II
of the Mainz neutrino mass search in tritium beta decay,''
e-print Archive: \ hep-ex/0412056.

\bibitem{Klapdor} H.~V.~Klapdor-Kleingrothaus et al.,
Mod.\ Phys.\ Lett.\ A {\bf 16} (2001) 2409; 
Mod.\ Phys.\ Lett.\ A {\bf 18} (2003) 2243. 

\bibitem{DAMA} R.~Bernabei {\it et al.}, ``Dark matter search,''
Riv.\ Nuovo Cim.\  {\bf 26N1} (2003) 1. DAMA Home Page, 
{\tt http://www.lngs.infn.it/lngs/htexts/dama/welcome.html}

\bibitem{neutralino} K.~A.~Olive,
``Dark matter candidates in supersymmetric models,''
e-print Archive: \ hep-ph/0412054.

\bibitem{CDMS} D.~S.~Akerib {\it et al.}  [CDMS Collaboration],
Phys.\ Rev.\ Lett.\  {\bf 93} (2004) 211301; D.~S.~Akerib {\it et al.},
Phys.\ Rev.\ D {\bf 68} (2003) 082002.

\bibitem{UKDMC} B.~Ahmed {\it et al.},
Nucl.\ Phys.\ Proc.\ Suppl.\  {\bf 124} (2003) 193; 
Astropart.\ Phys.\  {\bf 19} (2003) 691.
UKDMC Home Page at {\tt http://hepwww.rl.ac.uk/ukdmc/}

\bibitem{CRESST} M. Bravin et al., Astropart. Phys. {\bf 12} (1999) 107.

\bibitem{SDD} J.~I.~Collar et al.,
Phys.\ Rev.\ Lett.\  {\bf 85} (2000) 3083.

\bibitem{Griest} G. Jungman, M. Kamionkowski and K. Griest,
Phys. Rep. {\bf 267} (1996) 195.

\bibitem{AMS} The Alpha Magnetic Spectrometer Home Page: \
{\tt http://ams.cern.ch/AMS/}

\bibitem{AMANDA} F. Halzen et al., Phys. Rep. {\bf 307} (1998) 243;
M.~Ackermann {\it et al.}  [The AMANDA Collaboration], ``Search for
extraterrestrial point sources of high energy neutrinos with AMANDA-II
using data collected in 2000-2002,'' e-print Archive: \
astro-ph/0412347.


\bibitem{Vandenberg} D.A. Vandenberg, M. Bolte and P.B. Stetson,
Ann. Rev. Astron. Astrophys. {\bf 34} (1996) 461;
e-print Archive: \ astro-ph/9605064.

\bibitem{Krauss} L.~M.~Krauss, Phys.\ Rept.\  {\bf 333} (2000) 33.


\bibitem{Chaboyer} B. Chaboyer, P. Demarque, P.J. Kernan and
L.M. Krauss, Science {\bf 271} (1996) 957; Astrophys. J. {\bf 494}
(1998) 96.

\bibitem{Charley} C.H. Lineweaver, 
Science {\bf 284} (1999) 1503.

\bibitem{CMB} D. Scott, J. Silk and M. White, Science {\bf 268} (1995)
829; W. Hu, N. Sugiyama and J. Silk, Nature {\bf 386} (1997) 37;
E.~Gawiser and J.~Silk, Phys.\ Rept.\ {\bf 333} (2000) 245.

\bibitem{Silk} J. Silk, Nature {\bf 215} (1967) 1155; 

\bibitem{Turok} A.~Albrecht, R.~A.~Battye and J.~Robinson, 
Phys.\ Rev.\ Lett.\ {\bf 79} (1997) 4736;
N.~Turok, U.~L.~Pen, U.~Seljak, 
Phys.\ Rev.\ D {\bf 58} (1998) 023506; L.~Pogosian, Int.\ J.\ Mod.\
Phys.\ A {\bf 16S1C} (2001) 1043.

\bibitem{Seljak} U.~Seljak et al., e-print Archive: \ astro-ph/0407372.

\bibitem{Nu} V.~Barger, D.~Marfatia and A.~Tregre,
Phys.\ Lett.\ B {\bf 595} (2004) 55;
P.~Crotty, J.~Lesgourgues and S.~Pastor,
Phys.\ Rev.\ D {\bf 69} (2004) 123007;
S.~Hannestad, ``Neutrino mass bounds from cosmology,''
e-print Archive: \ hep-ph/0412181.

\bibitem{iso} H.~V.~Peiris {\it et al.}, Astrophys.\ J.\ Suppl.\ {\bf
148} (2003) 213; P.~Crotty, J.~Garc\'\i a-Bellido, J.~Lesgourgues and
A.~Riazuelo, Phys.\ Rev.\ Lett.\ {\bf 91} (2003) 171301; J.~Valiviita
and V.~Muhonen, Phys.\ Rev.\ Lett.\ {\bf 91} (2003) 131302;
M.~Beltr\'an, J.~Garc\'\i a-Bellido, J.~Lesgourgues and A.~Riazuelo,
Phys.\ Rev.\ D {\bf 70} (2004) 103530; K.~Moodley, M.~Bucher,
J.~Dunkley, P.~G.~Ferreira and C.~Skordis, Phys.\ Rev.\ D {\bf 70}
(2004) 103520; C.~Gordon and K.~A.~Malik, Phys.\ Rev.\ D {\bf 69} (2004)
063508; F.~Ferrer, S.~Rasanen and J.~Valiviita, JCAP {\bf 0410} (2004)
010; H.~Kurki-Suonio, V.~Muhonen and J.~Valiviita, e-print Archive: \
astro-ph/0412439; M.~Beltr\'an, J.~Garc\'\i a-Bellido, J.~Lesgourgues,
A.~R.~Liddle and A.~Slosar, ``Bayesian model selection and isocurvature
perturbations,'' e-print Archive: \ astro-ph/0501477.

\bibitem{Bellido} J. Garc\'\i a-Bellido, Phil. Trans. R. Soc. Lond. {\bf
A 357} (1999) 3237.

\bibitem{Guth} A. Guth, Phys. Rev. D {\bf 23} (1981) 347.

\bibitem{Linde} A.D. Linde, Phys. Lett. {\bf 108B} (1982) 389. 

\bibitem{Andy} A. Albrecht and P.J. Steinhardt, Phys. Rev. Lett. 
{\bf 48} (1982) 1220.

\bibitem{Guthbook} For a personal historical account, see A. Guth, ``The
Inflationary Universe'', Perseus Books (1997).

\bibitem{LindeBook} A.D. Linde, ``Particle Physics and Inflationary
Cosmology'', Harwood Academic Press (1990).

\bibitem{LL93} A.R. Liddle and D.H. Lyth, Phys. Rep. {\bf 231} (1993) 1.


\bibitem{Bardeen} J.M. Bardeen, Phys. Rev. D {\bf 22} (1980) 1882.

\bibitem{MFB} V.F. Mukhanov, H.A. Feldman and R.H. Brandenberger, 
Phys. Rep. {\bf 215} (1992) 203.

\bibitem{AS} M. Abramowitz and I. Stegun, ``Handbook of Mathematical
Functions'', Dover (1972).

\bibitem{GBW} J. Garc\'\i a-Bellido and D. Wands, Phys. Rev. D {\bf 53} 
(1996) 5437; D. Wands, K.~A.~Malik, D.~H.~Lyth and A.~R.~Liddle, 
Phys. Rev. D {\bf 62} (2000) 043527.


\bibitem{LR99} D.H. Lyth and A. Riotto, Phys. Rep. {\bf 314} (1999) 1.

\bibitem{SW} R.K. Sachs and A.M. Wolfe, Astrophys. J. {\bf 147} (1967)
73.

\bibitem{Harrison} E.R. Harrison, Rev. Mod. Phys. {\bf 39} (1967) 862;
L.F. Abbott and R.K. Schaefer, Astrophys. J. {\bf 308} (1986) 546.

\bibitem{BLW} E.F. Bunn, A.R. Liddle and M. White,
Phys. Rev. D {\bf 54} (1996) 5917.

\bibitem{Staro} A.A. Starobinsky, Sov. Astron. Lett. {\bf 11} (1985) 133.

%\bibitem{LIGO} LIGO Home Page: \ {\tt http://www.ligo.caltech.edu/}\\
%VIRGO Home Page: \ {\tt http://www.virgo.infn.it/}

\bibitem{CPT} S.M. Carroll, W.H. Press and E.L. Turner,
Ann. Rev. Astron. Astrophys. {\bf 30} (1992) 499.

\bibitem{Boomerang} P.~de Bernardis {\it et al.},
New Astron.\ Rev.\  {\bf 43} (1999) 289;
P.~D.~Mauskopf {\it et al.}  [Boomerang Coll.],
Astrophys.\ J.\  {\bf 536} (2000) L59. \ Boomerang Home Page: \ 
{ http://oberon.roma1.infn.it/boomerang/}

\bibitem{WMAPHP} Microwave Anisotropy Probe Home Page: \ 
{\tt http://map.gsfc.nasa.gov/}

\bibitem{Planck} Planck Surveyor Home Page: \ 
{\tt http://astro.estec.esa.nl/Planck/}

\bibitem{Tegmark} M. Tegmark Home Page: \
{http://www.hep.upenn.edu/\~{}max/cmb/experiments.html}


\bibitem{KK} M. Kamionkowski and A. Kosowsky, Ann. Rev. Nucl. Part. Sci. 
{\bf 49} (1999) 77.

\bibitem{HD} W.~Hu and S.~Dodelson,
Ann.\ Rev.\ Astron.\ Astrophys.\  {\bf 40} (2002) 171.


\bibitem{CMBFAST} U.~Seljak and M.~Zaldarriaga, CMBFAST code Home Page: \
{\tt http://www.cmbfast.org/}

\bibitem{CAMB} A.~Lewis and A.~Challinor, 
CAMB code Home Page: {\tt http://camb.info/}

\bibitem{CMBpol} CMB Polarization experiment Home Page: \\
{\tt http://www.mssl.ucl.ac.uk/www$_{-}$astro/submm/CMBpol1.html}

\bibitem{ACT} ACT experiment Home Page: \
{\tt http://www.hep.upenn.edu/act/}

\bibitem{Page} L.A. Page, ``Measuring the anisotropy in the CMB'', 
e-print Archive: \ astro-ph/9911199.

\bibitem{DASI} J.~Kovac et al., Nature {\bf 420} (2002) 772; DASI
Home Page: \ http://astro.uchicago.edu/dasi/

\bibitem{LiddleLyth} A.R. Liddle and D.H. Lyth, ``Cosmological Inflation
and Large Scale Structure'', Cambridge University Press (2000).

\bibitem{Bond} J.R. Bond and G. Efstathiou, Astrophys. J. {\bf 285}
(1984) L45.

\bibitem{royalsoc} G. Efstathiou {\em et al.} (Eds.) ``Large-scale
structure in the universe'', Phil. Trans. R. Soc. Lond. {\bf A 357}
(1999) 1-198.

\bibitem{Moore} B. Moore, 
Phil. Trans. R. Soc. Lond. {\bf A 357} (1999) 3259.


\end{thebibliography}
\end{document}